\documentclass[11pt]{article}
\pdfoutput=1
\usepackage{jheppub}
\usepackage{amssymb,amsmath,amstext,amsfonts}
\usepackage{mathtools}
\usepackage{hyperref}
\usepackage{color}

\usepackage{graphicx}
\usepackage{mwe}
\usepackage{mathrsfs}
\usepackage{multirow}
\usepackage{array}

\usepackage[caption=false]{subfig}
\usepackage{bm}
\usepackage{braket}
\usepackage{listings}
\usepackage{cases}
\usepackage{comment}
\usepackage{soul}
\usepackage{cancel}
\usepackage{cases}
\usepackage[utf8]{inputenc}
\usepackage{url}
\usepackage{longtable}
\usepackage{empheq}
\usepackage{aas_macros}
\usepackage{here}

\newcommand{\dif}[2]{\frac{\mathrm{d} #1}{\mathrm{d} #2}}
\newcommand{\pdif}[2]{\frac{\partial #1}{\partial #2}}
\newcommand{\var}[2]{\frac{\delta #1}{\delta #2}}
\newcommand{\dd}{\mathrm{d}}

\newcommand{\ee}{\mathrm{e}}
\newcommand{\diag}{\mathrm{diag}}

\newcommand{\Det}{\mathrm{Det}}

\newcommand{\Mpl}{M_\text{Pl}}

\newcommand{\cs}{c_{{}_\mathrm{S}}}
\newcommand{\IR}{\text{IR}}
\newcommand{\UV}{\text{UV}}
\renewcommand{\Re}{\mathrm{Re}}
\renewcommand{\Im}{\mathrm{Im}}
\newcommand{\dk}{\frac{\dd^3 \mathbf{k}}{(2\pi)^3}}

\newcommand{\kdx}{\mathbf{k}\cdot\mathbf{x}}

\newcommand{\efolds}{$e$-folds\xspace}

\newcommand{\calU}{\mathcal{U}}

\newcommand{\scrD}{\mathscr{D}}

\newcommand{\calH}{\mathcal{H}}

\newcommand{\bfk}{\mathbf{k}}
\newcommand{\calL}{\mathcal{L}}

\newcommand{\calN}{\mathcal{N}}

\newcommand{\calO}{\mathcal{O}}

\newcommand{\calP}{\mathcal{P}}

\newcommand{\scrQ}{\mathscr{Q}}
\newcommand{\calR}{\mathcal{R}}

\newcommand{\calS}{\mathcal{S}}
\newcommand{\calV}{\mathcal{V}}
\newcommand{\calX}{\mathcal{X}}
\newcommand{\bfx}{\mathbf{x}}

\newcommand{\bae}[1]{\begin{align} #1 \end{align}}
\newcommand{\bce}[1]{\begin{cases} #1 \end{cases}}
\newcommand{\dps}{\displaystyle}
\newcommand{\bfe}[4]{
\begin{figure} 
	\centering
	\includegraphics[#1]{#2}
	\caption{#3}
	\label{#4}
\end{figure}}
\newcommand{\bpme}[1]{\begin{pmatrix} #1 \end{pmatrix}}

\newcommand{\phifull}{\phi}
\newcommand{\pifull}{\pi}
\newcommand{\phiIR}{\varphi}
\newcommand{\piIR}{\varpi}
\newcommand{\dphi}{\delta\phi}
\newcommand{\dpi}{\delta\pi}

\newcommand{\bg}{0}
\newcommand{\phibg}{{\phi_\bg}}
\newcommand{\pibg}{{\pi_\bg}}
\newcommand{\dotphi}{{\phifull^\prime}}
\newcommand{\dotphiIR}{{\phiIR^\prime}}

\newcommand{\covP}{{\tilde{P}}}
\newcommand{\UVE}{E}
\newcommand{\bb}{b}

\newcommand{\ksigma}{k_\sigma}
\newcommand{\dotksigma}{{k_\sigma{}^\prime}}
\newcommand{\tsigma}{N_\sigma}
\newcommand{\sg}{\sigma}
\newcommand{\Lag}{\calL}
\newcommand{\Ham}{\calH}
\newcommand{\Ricci}{\calR}
\newcommand{\lapse}{\calN}

\newcommand{\lapseIR}{\calN_\IR}
\newcommand{\lapseUV}{\calN_\UV}
\newcommand{\shift}{\beta}
\newcommand{\shiftUV}{\psi}
\newcommand{\spmet}{\gamma}
\newcommand{\eN}{N}
\newcommand{\excurv}{K}
\newcommand{\momv}{v}

\newcommand{\dotH}{H^\prime}

\newcommand{\covD}{{\cal D}}
\newcommand{\ItoD}{\mathfrak{D}}
\newcommand{\phasecovD}{D}
\newcommand{\SHam}{S}
\newcommand{\So}{\SHam^{(0)}}
\newcommand{\Sl}{\SHam^{(1)}}
\newcommand{\Sz}{\SHam^{(2)}}
\newcommand{\SE}{\SHam^{(3)}}
\newcommand{\Sint}{\SHam^{(\mathrm{int})}}
\newcommand{\Seff}{\SHam_\mathrm{eff}}
\newcommand{\Steff}{\tilde{\SHam}_\mathrm{eff}}
\newcommand{\SIA}{\SHam_\mathrm{IA}}
\newcommand{\Sren}{\SHam_\mathrm{ren}}

\newcommand{\JUV}{J}
\newcommand{\mD}{\scrD}

\newcommand{\discAlpha}{\alpha}
\newcommand{\genS}{\calS}

\newcommand{\genX}{\calX}
\newcommand{\genU}{\calU}

\newcommand{\Xt}{\bar{\genX}}
\newcommand{\Gammat}{\bar{\Gamma}}
\newcommand{\genV}{\calV}
\newcommand{\Vt}{\bar{\genV}}
\newcommand{\Ut}{\bar{\genU}}
\newcommand{\covU}{{\tilde{\genU}}}
\newcommand{\covV}{{\tilde{\genV}}}

\newcommand{\Ps}{P_\mathrm{s}}

\newcommand{\indI}{I}
\newcommand{\indJ}{J}
\newcommand{\indK}{K}
\newcommand{\indL}{L}
\newcommand{\indM}{M}

\newcommand{\indR}{R}
\newcommand{\indS}{S}
\newcommand{\indIt}{{\bar{\indI}}}
\newcommand{\indJt}{{\bar{\indJ}}}
\newcommand{\indKt}{{\bar{\indK}}}

\newcommand{\indX}{X}
\newcommand{\indY}{Y}
\newcommand{\covX}{\tilde{X}}
\newcommand{\covY}{\tilde{Y}}

\newcommand{\inda}{a}
\newcommand{\indb}{b}
\newcommand{\indc}{c}
\newcommand{\indd}{d}

\newcommand{\KelC}{\mathrm{cl}}
\newcommand{\KelD}{\mathrm{q}}
\newcommand{\Kela}{\mathfrak{a}}
\newcommand{\Kelb}{\mathfrak{b}}

\newcommand{\indA}{A}
\newcommand{\indB}{B}
\newcommand{\indC}{C}
\newcommand{\At}{{\bar{\indA}}}
\newcommand{\Bt}{{\bar{\indB}}}
\newcommand{\Ct}{{\bar{\indC}}}
\newcommand{\gt}{{\bar{g}}}

\newcommand{\indn}{n}
\newcommand{\indm}{m}

\newcommand{\indi}{i}
\newcommand{\indj}{j}
\newcommand{\indk}{k}

\newcommand{\N}{\nabla}
\newcommand{\Laplacian}{\partial_i^2}
\newcommand{\Rot}{R}
\newcommand{\U}{U}
\newcommand{\barQ}{\bar{Q}}
\newcommand{\fsigma}{f_\sigma}
\newcommand{\D}{D}

\newcommand{\Hs}{H_\star}

\begin{document}
\title{
A manifestly covariant theory of multifield stochastic inflation in phase space:\\
\Large Solving the discretisation ambiguity in stochastic inflation
}

\date{\today}

\author[a]{Lucas Pinol,}
\affiliation[a]{Institut d'Astrophysique de Paris, UMR 7095 du CNRS et Sorbonne Universit\'e, 98 bis bd Arago, 75014 Paris, France}

\author[a]{S\'ebastien Renaux-Petel,}

\author[b]{Yuichiro Tada}
\affiliation[b]{Department of Physics, Nagoya University, Nagoya 464-8602, Japan}

\emailAdd{pinol@iap.fr,renaux@iap.fr}
\emailAdd{tada.yuichiro@e.mbox.nagoya-u.ac.jp}

\abstract{Stochastic inflation is an effective theory describing the super-Hubble, coarse-grained, scalar fields driving inflation, by a set of Langevin equations.
We previously highlighted the difficulty of deriving a theory of stochastic inflation that is invariant under field redefinitions, and the link with the ambiguity of discretisation schemes defining stochastic differential equations.
In this paper, we solve the issue of these “inflationary stochastic anomalies” by using the Stratonovich discretisation satisfying general covariance, and identifying that the quantum nature of the fluctuating fields entails the existence of a preferred frame defining independent stochastic noises. Moreover, we derive physically equivalent It\^o-Langevin equations that are manifestly covariant and well suited for numerical computations.
These equations are formulated in the general context of multifield inflation with curved field space, taking into account the coupling to gravity as well as the full phase space in the Hamiltonian language, but this resolution is also relevant in simpler single-field setups.
We also develop a path-integral derivation of these equations, which solves conceptual issues of the heuristic approach made at the level of the classical equations of motion, and allows in principle to compute corrections to the stochastic formalism.
Using the Schwinger-Keldysh formalism, we integrate out small-scale fluctuations, derive the influence action that describes their effects on the coarse-grained fields, and show how the resulting coarse-grained effective Hamiltonian action can be interpreted to derive Langevin equations with manifestly real noises. Although the corresponding dynamics is not rigorously Markovian, we show the covariant, phase-space Fokker-Planck equation for the Probability Density Function of fields and momenta when the Markovian approximation is relevant, and we give analytical approximations for the noises' amplitudes in multifield scenarios.}

\keywords{physics of the early universe, multifield inflation, stochastic inflation, Langevin equations, Schwinger-Keldysh formalism}

\maketitle

\section{Introduction}

\subsection*{Inflation and High-Energy Physics}

Inflation, an era of accelerated expansion of the early universe, currently provides us with the best understanding of the initial conditions for the subsequent cosmological eras. The simplest mechanism to explain this quasi-exponential, de Sitter like expansion, is to assume that the energy density of the universe was then dominated by the one of a scalar field, the inflaton, endowed with a very flat potential in Planck units, so that it slowly rolls down its potential.
This results in a homogeneous, isotropic and spatially flat Universe on cosmological scales, as required by observations of the cosmic microwave background (CMB).
Moreover, it naturally comes with a mechanism by which the quantum fluctuations of the inflaton are stretched to cosmological scales to give rise to primordial density fluctuations at the origin of the CMB anisotropies and of the large scale structure of the universe that we observe nowadays, a scenario in perfect agreement with the latest CMB data from the Planck satellite~\cite{Akrami:2018odb,Akrami:2019izv}. 

Despite its success at explaining data in a simple manner, single-field slow-roll inflation is usually seen only as a phenomenological description that emerges from a more realistic physical framework to be determined (see, e.g., Ref.~\cite{Baumann:2014nda}). One of the main reasons behind this is the peculiar ultraviolet (UV) sensitivity of inflation: order-one changes in the strengths of the interactions of the field(s) responsible for inflation with Planck-scale degrees of freedom generically have significant effects on the inflationary dynamics, to the point sometimes of ruining inflation itself. Addressing this UV sensitivity implies justifying in a controllable setup that high-energy interactions are innocuous, which can be done either by specifying the physics at the Planck scale, typically in string theory constructions, or at least by taking it into account using the methods of effective field theory (EFT). Either way, this naturally leads one to consider the impact of the existence of several degrees of freedom during inflation, and indeed the UV sensitivity of inflation provides us with a formidable opportunity to use the early universe as a giant particle detector. In this respect, looking for new physics in cosmological data, for instance through non-Gaussianities or/and features of the primordial fluctuations, can be seen as looking for multifield effects (see, e.g., Refs.~\cite{Wands:2010af,Chen:2010xka,Wang:2013eqj,Renaux-Petel:2015bja,Meerburg:2019qqi} for reviews). 
Typical UV embeddings of inflation include several scalar fields interacting through their potential as well as through their kinetic terms, with a Lagrangian of the type
\bae{\label{S-intro}
	{\cal L}=-\frac{1}{2}g^{\mu\nu}G_{\indI\indJ}(\phifull)\partial_\mu\phifull^\indI\partial_\nu\phifull^\indJ-V(\phifull).
}
This general class of so-called non-linear sigma models have been studied for a long time
(see, e.g., the review~\cite{Lyth:1998xn}), but recent years have seen a flurry of activity concerning them (see, e.g., Refs.~\cite{Cremonini:2010ua,
Turzynski:2014tza,
Carrasco:2015uma,
Renaux-Petel:2015mga,
Hetz:2016ics,
Achucarro:2016fby,
Tada:2016pmk,
Brown:2017osf,
Renaux-Petel:2017dia,
Mizuno:2017idt,
Achucarro:2017ing,
Krajewski:2018moi,
Christodoulidis:2018qdw,
Linde:2018hmx,
Garcia-Saenz:2018ifx,
Garcia-Saenz:2018vqf,
Achucarro:2018vey,
Achucarro:2018ngj,
Achucarro:2019pux,
Bjorkmo:2019aev,
Grocholski:2019mot,
Fumagalli:2019noh,
Bjorkmo:2019fls,
Christodoulidis:2019mkj,
Christodoulidis:2019jsx,
Aragam:2019khr,
Mizuno:2019pcm,
Bravo:2019xdo,
Achucarro:2019mea,
Garcia-Saenz:2019njm,
Chakraborty:2019dfh,
Bjorkmo:2019qno,
Wang:2019gok,
Ferreira:2020qkf,
Braglia:2020fms,
Palma:2020ejf,
Fumagalli:2020adf,
Braglia:2020eai
}), in particular about geometrical aspects related to the curved field space described by the metric $G_{\indI \indJ}$, the possibility to inflate along trajectories characterised by a strongly non-geodesic motion in field space, and the corresponding distinct observational signatures.

\subsection*{Stochastic inflation}

Standard Perturbation Theory (SPT) during inflation treats perturbatively quantum fluctuations around supposedly homogeneous classical background fields. This distinct treatment is not only conceptually unsatisfactory, but it is also expected to break down in the presence of very light scalar fields whose large-scale evolutions are dominated, not by their classical dynamics, but instead by quantum diffusion effects.
The stochastic approach aims at dealing directly with the super-Hubble parts of the quantum fields driving inflation (see Refs.~\cite{STAROBINSKY1982175,Starobinsky:1986fx,NAMBU1988441,NAMBU1989240,Kandrup:1988sc,Nakao:1988yi,Nambu:1989uf,Mollerach:1990zf,Linde:1993xx,Starobinsky:1994bd} for the first papers on the subject). The corresponding theory, resulting from a \textit{coarse-graining} procedure, can be thought of as an EFT for long-wavelength modes during inflation.
More precisely, and concentrating for definiteness
on test scalar fields evolving in de Sitter space, the scalar fields are divided into 
infrared (IR) and UV parts delineated by a constant physical scale, the first one corresponding to the ``coarse grained'' super-Hubble parts of the quantum fields, with comoving momenta smaller than the time-dependent cutoff $k_\sigma(N)=\sigma a(N) H$, with a small positive parameter $\sigma \ll 1$ and where $N=\ln a$ is the number of \efolds.
The IR sector of the theory can be understood as an open system receiving a continuous flow of UV modes as they cross the growing coarse-graining scale $k_\sigma$.
Strikingly, the effect of this flow can be understood as classical random kicks added to the deterministic dynamics of the IR fields. More technically, the IR fields verify stochastic, so-called \textit{Langevin} equations, rather than the deterministic equations verified by the background fields in SPT.

An excellent agreement between the stochastic formalism and usual quantum field theory techniques has been found in a number of studies, mostly in the paradigmatic setup of the $\lambda \phi^4$ theory in de Sitter space, but also including backreaction in the single-field slow-roll regime~\cite{Tsamis:2005hd,Prokopec:2007ak,Finelli:2008zg,Finelli:2010sh,Garbrecht:2013coa,Garbrecht:2014dca,Onemli:2015pma,Cho:2015pwa}. This agreement is noteworthy because the computations of correlation functions are almost immediate in the stochastic theory, at least in the simplest contexts: it enables one to determine without effort what would be the results of intricate loop calculations in renormalised perturbative quantum field theory. Moreover, and importantly, the stochastic formalism enables one to resum the IR divergences of perturbative QFT, and derive fully non-perturbative results (such as equilibrium distributions in de Sitter space), a subject that has attracted a lot of attention and has been investigated using a variety of methods (see, e.g., Refs.~\cite{Seery:2007we,
Enqvist:2008kt,
2009JCAP...05..021S,
Burgess:2009bs,
Seery:2010kh,
Gautier:2013aoa,
Guilleux:2015pma,
Gautier:2015pca,
Markkanen:2017rvi,
LopezNacir:2019ord,
Gorbenko:2019rza,
Mirbabayi:2019qtx,
Adshead:2020ijf, 
Moreau:2020gib,
Cohen:2020php
}).

The stochastic formalism is not only useful for such formal investigations, as well as to tackle the issues related to eternal inflation~\cite{Linde:1986fc,Linde:1986fd,Goncharov:1987ir}, but it can also be used to compute observationally relevant quantities such as the power spectrum, higher $n$-point functions and other statistical properties of the adiabatic curvature perturbation $\zeta$ generated during inflation.
This is achieved with the help of the separate universe approach, which states that each region of the universe slightly larger than the Hubble radius evolves like a separate FLRW universe that is locally homogeneous and evolves independently from its neighbours~\cite{Wands:2000dp}.
Then, patching these regions enables one to 
deduce the curvature perturbation on even larger scales, identified as the fluctuation of the local number of $e$-folds of expansion $N(\bfx)$, a method known as the $\delta N$ formalism~\cite{Salopek:1990jq,Sasaki:1995aw,Sasaki:1998ug,Lyth:2004gb}. 
Its generalisation to stochastic inflation was called the \emph{stochastic-$\delta N$ formalism}~\cite{Fujita:2013cna,Fujita:2014tja,Vennin:2015hra,Kawasaki:2015ppx,Assadullahi:2016gkk,Vennin:2016wnk,Pinol:2018euk}, and it enables one to compute the statistical properties of $\zeta$ in a non-perturbative manner (see also Refs.~\cite{Rigopoulos:2003ak,Rigopoulos:2004gr,Rigopoulos:2004ba,Rigopoulos:2005xx,Rigopoulos:2005ae} for a related approach), reducing to SPT in a suitable classical limit, while being able to treat the regime where quantum diffusion effects dominate.
This has notably proved useful recently to compute the abundance of primordial black holes (PBH) resulting from the collapse of local overdensities generated during inflation~\cite{Kawasaki:2015ppx,Pattison:2017mbe,Ezquiaga:2018gbw,Biagetti:2018pjj,Ezquiaga:2019ftu,Panagopoulos:2019ail} (see, e.g., Refs.~\cite{Bullock:1996at,GarciaBellido:1996qt,Ivanov:1997ia,Yokoyama:1998pt} for early applications of the stochastic formalism in this context), a field that regained attention as PBHs are considered as candidates for LIGO/Virgo gravitational wave sources~\cite{Bird:2016dcv,Clesse:2016vqa,Sasaki:2016jop}, a possibly important component of dark matter (see, e.g., Refs.~\cite{Carr:2016drx,Carr:2020gox}), as well as possible explanations of the microlensing events found by OGLE~\cite{Niikura:2019kqi} and even of the hypothetical Planet 9~\cite{Scholtz:2019csj,Witten:2020ifl}.

Despite many achievements, and the fact that stochastic inflation with multiple fields has already been studied (see \cite{GarciaBellido:1993wn,GarciaBellido:1994vz,GarciaBellido:1995br} for first works at the early stage of stochastic inflation), we stress that it has never been formulated in a manner that is generally covariant under field redefinitions, nor derived from first principles in this context. This, together with the many recent developments concerning the geometrical aspects of nonlinear sigma models, constitute the main motivations of this work.

\subsection*{Path integrals and Hamiltonian action}

In the present paper, we begin by showing a ``heuristic" derivation of the phase-space Langevin equations of stochastic inflation in the general context of multifield inflation with curved field space, by working at the level of the classical equations of motion, but we also propose a rigorous path-integral derivation solving the ambiguities of this heuristic approach.
Path integrals are ubiquitous in physics, from statistical physics and quantum mechanics to field theories. In the context of stochastic inflation, they appear in a manner quite similar to the path-integral representation of the Brownian motion of a system linearly coupled to a thermal bath that is integrated out~\cite{Feynman:1963fq}, the role of the system and the bath being respectively replaced by the IR and UV sectors~\cite{Morikawa:1989xz,Calzetta:1999zr,Matarrese:2003ye,Liguori:2004fa,Levasseur:2013ffa,Levasseur:2013tja,Levasseur:2014ska,Moss:2016uix,Tokuda:2017fdh,Prokopec:2017vxx,Tokuda:2018eqs} (see also Refs.~\cite{Calzetta:1995ys,Calzetta:1996sy,Calzetta:1999xh,Parikh:2020nrd} for the use of similar tools in other gravitational contexts).

Path integrals are first constructed on a discrete time (and space for field theories that we shall focus on from now on) grid as the integral over all possible discrete \textit{jumps} from a field’s value to any other one, with fixed initial and final values. In the continuous limit, it corresponds to an integral over all the possible \textit{paths} to go from a fixed initial point to a fixed final one, thus justifying its name as ``integration over possible histories". Microscopically, the law governing the probability of a given jump between times $N_{j-1}$ and $N_j$ is dictated by the unitary operator $\hat{U}_j=\ee^{-i\hat{H}(\phi,\pi;N_j)(N_j-N_{j-1})}$, where $\hat{H}$ is the Hamiltonian operator of the system, and $\phi$ and $\pi$ denote the corresponding fields and momenta.
In this fundamental phase-space approach, the action entering in the final expression for the path integral over the values of the fields and momenta is called the \textit{Hamiltonian action} and reads $S=\int \mathrm{d}^4x \left[ \pi \dot{\phi} - \mathcal{H}(\phi,\pi) \right]$ where $\mathcal{H}$ is the Hamiltonian density associated with $H$. Note that when the Hamiltonian (density) is at most quadratic in momenta, it is possible to perform exactly the path integration over them, and express the theory as a path integral over fields only.
However one would recover the standard Lagrangian action only when the terms quadratic in momenta are field-independent~\cite{Weinberg:1995mt}, which is neither the case in general, nor in our situation of interest.

\subsection*{Partition function, ``in-in" formalism and doubling of the degrees of freedom}

In particle physics, transition amplitudes between asymptotic ``in" and ``out" states can be deduced from time-ordered correlation functions. The latter can themselves be derived from the generating functional $Z[J]$, i.e. the partition function with sources, which has a convenient  path-integral representation. In cosmology, one rather looks for the expectation values of operators in some ``in" state defined in the far past (typically the Bunch-Davies vacuum), as well as the corresponding causal equations of motion that they verify. However these can also be deduced from a generating functional expressed as a path integral, with the important peculiarity, for this ``in-in” partition function, that the path integral turns out to be performed on a Closed-Time-Path (CTP) of integration in the time domain, as represented in Fig.~\ref{fig: CTP} in the main body of this paper.
Working with this CTP amounts to considering a ``doubling of the histories": one along the forward branch, and one along the reverse one, and with doubled degrees of freedom, one version for each of the two paths. 
Naturally, there is no doubling of the genuine physical degrees of freedom in the theory, but only as dummy variables inside the path integral:
the two copies of the degrees of freedom are treated independently at any time \textit{but} the final one, at which the two branches of the CTP close, and boundary conditions must be imposed. Of course, the ``in-in" formalism was not intended for cosmology in the first place, but rather developed in the field of non-equilibrium statistical and quantum field theories, in which it is also known as the \textit{Schwinger-Keldysh formalism}~\cite{Schwinger:1960qe,Keldysh:1964ud}, proving extremely useful to describe quantum and thermal fluctuations, dissipation, decoherence and many other effects in various areas of physics (see, e.g., Refs.~\cite{Kamenev-book,Calzetta:2008iqa,altland_simons_2010}).

\subsection*{Coarse-graining}

Stochastic inflation corresponds to a low-energy effective version of the full theory that can be described by the ``in-in" path integral as explained above. To derive it, one must thus identify the relevant degrees of freedom (the super-Hubble modes in our case),
and integrate out of the theory the other ones (the sub-Hubble modes).  After splitting the full system into our subsystem of interest composed of IR fields, plus a bath of UV fluctuations, one can perturbatively integrate out explicitly the UV modes of the description.
However, remembering that ``integrating out is different from truncating", the UV fluctuations will leave an imprint on the IR dynamics, and this will be the source of the explicit noise and randomness in the equations of motion for the long-wavelength fields. This concept of \textit{coarse-grained effective action} is widely used in physics, from the study of Brownian processes in statistical physics, to the applications of renormalisation in field theories and decoherence in quantum mechanics, but was also applied to the cosmological context~\cite{Calzetta:1995ys,Calzetta:1999zr,Levasseur:2013ffa,Tokuda:2017fdh,Tokuda:2018eqs}. The coarse-graining procedure can also be understood at the level of the density matrix, which for a bipartite system (IR and UV sectors) can give the EFT for an open system (the IR modes) by tracing out the environment (the UV modes) and obtaining the \textit{reduced density matrix}. Be it at the level of the partition function or the density matrix, the coarse-graining approach within the in-in formalism is powerful because it enables one to control the approximations that are made and possibly derive next-order corrections~\cite{Burgess:2014eoa,Burgess:2015ajz,Collins:2017haz,Hollowood:2017bil}. 

\subsection*{Langevin equations, multiplicative noise and ambiguity of the discretisation scheme}

As we explain in the body of this paper, the effect of the UV modes on the IR dynamics is encapsulated in the \textit{influence action}. After careful investigation and introduction of auxiliary variables, it can be shown that this results in an explicit noise term in the equations of motion for the IR fields, with a covariance dictated by the (real part of the) power spectrum of the UV modes. The long-wavelength fields thus verify Langevin equations, with a deterministic drift coming from the ordinary background dynamics, but supplemented by a diffusion term due to the random kicks. Crucially, the effect of the small-scale, quantum fluctuations on the long-wavelength, classicalised IR fields, can be interpreted as a classical noise. Hence, the resulting theory describes genuinely quantum effects, albeit in a classically-looking stochastic manner.

Langevin equations have been studied for a long time in the context of Brownian processes, signal theory, etc. They constitute Stochastic Differential Equations (SDE) rather than Ordinary Differential Equations (ODE), and this difference is crucial.
Indeed, consider the simplest example of the Brownian motion of a particle, due to shocks with its environment at a given temperature; its position is a random quantity whose statistical 
properties may be determined.
However for a given realisation, the position of the particle, although being a continuous function of time, is not a differentiable function of time due to the properties of the white noise that affects its dynamics.
Thus, the mathematical understanding of trajectories and in particular time derivatives of the position of the particle, is intricate and leads to interesting subtleties. Of course, a discrete-time interpretation of the dynamics is always possible and may even be clearer, and complications arise when going to the continuous-time limit of the description.
A famous example (for statistical physicists) of possible difficulties is met when the noise is \textit{multiplicative}, that is when its amplitude (or covariance) is itself a function of the random variable that verifies the Langevin equations. Then, there is an ambiguity when going from the discrete-time representation to the continuous one: at which time exactly should the random variable that enters the noise amplitude be evaluated? When dealing with ODEs, we are used to forget about these subtleties because any choice of a discrete scheme leads to the same physical result.
However, this is not the case any more for SDEs with multiplicative noise, for which different scheme choices, usually parameterised by a number $\alpha$ between $0$ and $1$, lead to different values for physical quantities like statistical averages, Probability Density Functions (PDF), etc.
Amongst the infinite number of possible choices for $\alpha$, two have been particularly investigated for their interesting properties, the prepoint, $\alpha=0$ It\^o discretisation~\cite{ito1944109}, and the midpoint, $\alpha=1/2$ Stratonovich~\cite{stratonovich1966new} one. Indeed, while It\^o is widely used in applied and computational mathematics for its appealing mathematical properties (the covariance matrix can be arbitrarily reduced in any frame to identify independent noises, the noise at a given time step only depends on the values of the random variables at previous time steps, etc.), Stratonovich may be preferred in theoretical physics, where changes of variable are ubiquitous, because the standard chain rule for the derivative of composite functions is only verified in that case. In particular, this last property simplifies discussions about general covariance of the equations.
In this respect, it is important to highlight that, while a given SDE, interpreted with different schemes, defines different physical theories, it is always possible to describe the same physics by using different discretisation schemes. Indeed, one knows how to go from one continuous form of a SDE understood in a given discretisation scheme, to another form with a different scheme, while leaving the physics unaffected.

Keeping this in mind, whether the conventional form of the Langevin equations of stochastic inflation should be interpreted according to It\^o or Stratonovich schemes has already been discussed in the literature. On one hand, the Stratonovich scheme  has  been  advocated  by  the fact that white noises should be treated as the limit of colored noises when the smooth decomposition between short and long-wavelength modes becomes sharp~\cite{Mezhlumian:1991hw}. 
On the other hand, it has been suggested that only the It\^o scheme could be invariant under reparameterisation of the time variable~\cite{Vilenkin:1999kd}, and consistently reproduce one-loop QFT computations in the $\lambda \varphi^4$ theory~\cite{Tokuda:2017fdh}.
Eventually, it has  also been argued  that  the  choice  between  the  two prescriptions exceeds the accuracy of the stochastic approach~\cite{Vennin:2015hra}.
In our previous paper~\cite{Pinol:2018euk}, we tackled for the first time the issue of the discretisation ambiguity of the Langevin equations of stochastic inflation in the multifield context, and we discovered various conceptual issues with the stochastic description of IR fields during inflation, that we called ``inflationary stochastic anomalies". 

\subsection*{Inflationary stochastic anomalies}

In stochastic inflation, the covariance matrix of the noises entering the Langevin equation is proportional to the (real part of the) power spectra of the UV modes.
However the UV modes themselves evolve according to linear equations of motion
(at first order in perturbation theory for the UV modes) whose ``coefficients" are set by the values of the IR fields that constitute the random variables of interest.
Thus, the noise amplitude for the IR fields clearly depends on their own values, which corresponds to a multiplicative noise. Actually, the situation is even more intricate since rigorously the power spectra of UV modes (and thus the noise amplitude) cannot be simply expressed as functions of the IR fields at the current time, but rather are solutions of differential equations that involve them. This situation is called \textit{non-Markovian}, in contrast to Markov processes where the noise amplitude only depends
on the random variables at the time step of evaluation, and not at previous times.

However even letting aside the non-Markovian difficulty, the multiplicative noise results in the discretisation scheme ambiguity discussed above, and since the derivation of the Langevin equations does not \textit{a priori} come with any prescription regarding their discrete-time version, one should choose how to interpret them (i.e. prescribe a value for the parameter $\alpha$) based on physical criteria.
However, in our previous paper~\cite{Pinol:2018euk}, we found that no choice was satisfactory because of the following. The standard chain rule for the derivative of composite functions is only verified in the Stratonovich $\alpha=1/2$ case.
Thus, for any other choice, the Langevin equations as they are usually shown do not respect general covariance under field redefinitions.
However at that time we thought the Stratonovich choice was not satisfactory neither, even if for a different reason: only in the It\^o case is the frame of reduction of the noise matrix (necessary to identify independent Gaussian white noises and solve the Langevin equations numerically or proceed further analytically) irrelevant to the final result, 
as already known in statistical physics contexts (see e.g. Refs.~\cite{ryter1980properties,deker1980properties, dilemma,vKampen-manifold,GRAHAM1985209}).
So we were left with a dilemma: breaking of general covariance following the It\^o  interpretation or spurious frame-dependence in the Stratonovich  one?
It is important to note that, although more striking in the multifield context, this ambiguity is also present in single-field models of stochastic inflation. 
Although we showed that, for such a single scalar field in the overdamped limit, the difference between the two prescriptions is numerically small in the final correlation functions, the conceptual issue was still remaining. By including a tadpole diagram cancelling the frame dependence in the Stratonovich scheme, a covariant and frame-independent formulation was proposed in Ref.~\cite{Kitamoto:2018dek}, considering the overdamped limit (i.e. in field space and not in phase space) of test scalar fields in de Sitter space and in a Markovian approximation.
In this paper, we will show how inflationary stochastic anomalies are solved in full generality from first principles.

\subsection*{Structure of the paper}

The structure of the paper is as follows. We begin by introducing in Sec.~\ref{sec: heuristic} the definitions and the concepts behind stochastic inflation in phase space with several scalar fields and a general field-space metric, and developing an intuitive approach to derive ``heuristically" the Langevin equations for the coarse-grained fields and their momenta. We also highlight the conceptual issues behind these equations and their derivation using the classical equations of motion. Notably, we review in Sec.~\ref{sec: stochastic anomalies} why these equations suffer from ``inflationary stochastic anomalies", an issue that we solve by using the Stratonovich discretisation satisfying general covariance, and identifying that the quantum nature of the fluctuating fields entails the existence of a preferred noise frame. The corresponding covariant It\^o SDE, which can readily be used in numerical and analytical computations, are also derived as one of our main results.
In Sec.~\ref{sec: effective hamiltonian action}, we turn to the rigorous derivation of stochastic inflation using a path-integral approach.
This enables one to solve the other conceptual issues of the heuristic approach and to keep a better control over the approximations made throughout, paying a particular attention to the doubling of the degrees of freedom and the necessary boundary conditions imposed at the UV/IR transition by the Closed-Time-Path of integration.
We also show how the identification of covariant Vilkovisky-DeWitt variables in phase space, is crucial to maintain general covariance. We derive the influence action for the long-wavelength fields and momenta, resulting from integrating out the UV modes, and we show how the coarse-grained effective action can be interpreted to derive Langevin equations with manifestly real noises.
We finish in Sec.~\ref{sec:Markovian} by showing, in the Markovian limit, the phase-space, covariant Fokker-Planck equation corresponding to our multifield Langevin equations, as well as some analytical approximations for the noises' amplitudes. These results can be used in practical applications of our covariant multifield stochastic inflation framework. 
Sec.~\ref{sec: conclusions} is then devoted to conclusions and future prospects. Eventually, we gathered in appendices some technical details as well as a summary of our notations.
We adopt natural units, $c=\hbar=1$ throughout this paper.

\subsection*{Main results} 

We gather here in a few lines the main results of the paper:

\begin{itemize}
    \item ``Inflationary stochastic anomalies" are solved by the observation that the quantum nature of the fluctuating fields provides one with a natural frame for reducing the noise covariance matrix: the one of the independent creation and annihilation operators. This leads to a unique set of independent Gaussian white noises in the Langevin equations (up to a constant, irrelevant, orthogonal matrix), and highlights the genuine quantum origin of their stochasticity.

    \item 
    The Langevin equations as they are usually derived must be interpreted with the Stratonovich discretisation scheme and the preferred frame mentioned above, but they are easier to interpret and use after transforming them to their It\^o version. The corresponding noise-induced terms can then be used to define covariant time-derivatives compatible with It\^o calculus, $\ItoD_N$, see Eqs.~\eqref{eq: ItoD for X}--\eqref{eq: ItoD for V}.
    The resulting, It\^o-covariant, phase-space, Langevin equations for multifield inflation with curved field space and including back-reaction on the metric are eventually found to be: 
    \bae{
        \boxed{
        \ItoD_N\phiIR^\indI=\frac{\piIR^\indI}{H}+\xi^{Q\indI}, \qquad \ItoD_N\piIR_\indI=-3\piIR_\indI-\frac{V_\indI}{H}+\xi^\covP_\indI\,.      \label{Langevin-intro}
        }
    }
    Here,  $V_\indI$ denotes the gradient of the potential, $H$ is the local Hubble scale, given in terms of the infrared fields $\phiIR^\indI$ and momenta $\piIR_\indI$ by the Friedmann equation~\eqref{laspeIR}, and indices are raised with the inverse field-space metric.
    We also find the auto-correlation of the Gaussian white noises to be given by, for $\covX=(Q,\covP)$:
    \bae{
        \boxed{
        \braket{\xi^{\covX\indI}(N)\xi^{\covY\indJ}(N^\prime)}\equiv A^{\covX\covY\indI\indJ}(N) \delta(N-N^\prime)
        =\Re \calP^{\covX\covY\indI\indJ}(N ;\ksigma(N))
        \delta(N-N^\prime)\,,
        \label{noise-properties-intro}
        }
    }
    with $\calP^{\covX\covY\indI\indJ}$ the dimensionless power spectra of the UV modes $(Q^\indI,\covP_\indI)$ that follow the EoMs~\eqref{eq: UV EoM} deduced from the action~\eqref{eq: cov S2}, and evaluated at the scale $\ksigma(N)=\sigma a(N) H$ that joins the IR sector at the time $N$.

    \item When the dynamics can be approximated as Markovian, it is possible to derive the phase-space Fokker-Planck equation for the one-point scalar PDF $P(\phiIR^\indI,\piIR_\indJ;N)$ as
    \begin{empheq}[box=\fbox]{align}
	\partial_N P=&-\phasecovD_{\varphi^I}\left[\frac{G^{IJ}}{H}\varpi_J P\right]+\partial_{\varpi_I}\left[\left(3\piIR_\indI+\frac{V_\indI}{H}\right)P\right] \\
	&+\frac{1}{2}\phasecovD_{\varphi^I}\phasecovD_{\varphi^J}(A^{QQ\indI\indJ}P)+\phasecovD_{\varphi^I}\partial_{\varpi_J}(A^{Q\covP\indI}{}_\indJ P)
	+\frac{1}{2}\partial_{\varpi_I}\partial_{\varpi_J}(A^{\covP\covP}{}_{\indI\indJ}P), \nonumber
    \end{empheq}
    with $\phasecovD_{\phiIR^\indI}$ a phase-space covariant derivative defined by its action on 
    field-space vectors: $\phasecovD_{\phiIR^\indI}\genU^\indJ=\nabla_\indI\genU^\indJ+\Gamma_{\indI\indL}^\indK\piIR_\indK\partial_{\piIR_\indL}\genU^\indJ$, where $\nabla_\indI$ is the usual field-space covariant derivative. Under a slow-varying approximation, we further provide some analytical estimates for the noise properties in Eqs.~\eqref{eq: power spectra massive case-QQ}--\eqref{eq: power spectra massive case-PP}.
    
\end{itemize}

\section{Stochastic formalism: heuristic approach}
\label{sec: heuristic}

In this section we introduce the concepts and definitions used throughout the paper, by showing a heuristic derivation, made at the level of the classical equations of motion, of the Langevin equations in the general class of multifield models described by the action \eqref{S-intro}. Our analysis is valid beyond the test approximation, i.e. it takes into account the backreaction of the scalar fields on the spacetime metric. Moreover, we do so using a phase-space Hamiltonian language, without assuming any slow-roll regime (see, e.g., Refs.~\cite{Habib:1992ci,Tolley:2008na,Enqvist:2011pt,Kawasaki:2012bk,Rigopoulos:2016oko,Moss:2016uix,Grain:2017dqa,Tokuda:2017fdh,Prokopec:2017vxx,Ezquiaga:2018gbw,Tokuda:2018eqs,Cruces:2018cvq,Firouzjahi:2018vet,Pattison:2019hef,Fumagalli:2019ohr,Prokopec:2019srf,Ballesteros:2020sre} for previous works on the subject, albeit not in this general multifield context, and sometimes with different results and approaches). Eventually, we highlight the limitations of this heuristic approach, and stress the non-Markovian character of the IR dynamics.

\subsection{Generalities and ADM formalism}

The general action of several scalar fields minimally coupled to gravity that we consider is given by
\bae{\label{eq: general S}
	S=\int\dd^4x\sqrt{-g}\left[\frac{1}{2}\Mpl^2\Ricci-\frac{1}{2}g^{\mu\nu}G_{\indI\indJ}(\phifull)\partial_\mu\phifull^\indI\partial_\nu\phifull^\indJ-V(\phifull)\right].
}
Here $\Ricci$ is the Ricci scalar associated with the spacetime metric $g_{\mu\nu}$, $G_{\indI\indJ}$ denotes the metric of the field space, curved in general, spanned by the scalar fields $\phi^\indI$, and $V(\phi)$ denotes the scalar potential. 
In the ADM formalism~\cite{Arnowitt:1962hi,Salopek:1990jq}, the spacetime metric is written in the form
\bae{\label{eq: ADM form}
    \dd s^2=-\lapse^2\dd t^2+\spmet_{ij}(\dd x^i+\shift^i\dd t)(\dd x^j+\shift^j\dd t)\,,
}
where $\lapse$ is the lapse function, $\shift^i$ is the shift vector, and $\spmet_{ij}$ is the spatial metric.
The action~\eqref{eq: general S} then reads $S=\int \dd t \dd^3 x \Lag$ with the Lagrangian density
\bae{\label{eq: ADM action}
    \Lag=\lapse\sqrt{\spmet}\left[\frac{\Mpl^2}{2}\left(\Ricci^{(3)}+\excurv_{ij}\excurv^{ij}-\excurv^2\right)
    +\frac{1}{2\lapse^2}G_{\indI\indJ}\momv^\indI\momv^\indJ-\frac{1}{2}G_{\indI\indJ}\spmet^{ij}\partial_i\phifull^\indI\partial_j\phifull^\indJ-V\right]\,,
}
where $\spmet={\rm det}(\gamma_{ij})$ and $\Ricci^{(3)}$ is the Ricci curvature of the spatial hypersurfaces. Here, spatial indices are lowered and raised with $\spmet_{ij}$ and its inverse $\spmet^{ij}$,
\bae{
    \excurv_{ij}=\frac{1}{2\lapse}\left(2\shift_{(i | j)}-\dot{\gamma}_{ij}\right),
}
is the extrinsic curvature of spatial slices (where dots denote time derivatives, the symbol $|$ denotes the spatial covariant derivative associated with the spatial metric $\gamma_{ij}$, 
and parentheses signal symmetrisation), and one has
\bae{
   \momv^\indI=\dot{\phi}^\indI-\shift^i\partial_i\phifull^\indI.
}

The Lagrangian~\eqref{eq: ADM action} does not depend upon the time derivatives of $\lapse$ and $\shift^i$. This shows that the lapse function and the shift vector are not dynamical variables, and that the only dynamical variables are $\phi^I$ and $\gamma_{ij}$ whose canonically conjugate momenta are given by 
\begin{eqnarray}
 \pifull_\indI&=&\var{\Lag}{\dot{\phifull}^\indI}=\frac{\sqrt{\spmet}}{\lapse}G_{\indI\indJ}v^J\,, \\
   \pifull^{ij}&=&\var{\Lag}{\dot{\spmet}_{ij}}=\frac{\Mpl^2}{2}\sqrt{\spmet}(\excurv\spmet^{ij}-\excurv^{ij})\,.
\end{eqnarray}
The Hamiltonian density is given by the Legendre transform $\Ham=\pifull_\indI\dot{\phifull}^\indI+\pifull^{ij}\dot{\spmet}_{ij}-\Lag$, or equivalently the action can be written in a Hamiltonian form as (see e.g. Ref.~\cite{Langlois:1994ec} for the single-field case) 
\bae{\label{eq: Hamiltonian action}
    S=\int\dd^4x\left[\pifull_\indI\dot{\phifull}^\indI+\pifull^{ij}\dot{\spmet}_{ij}-\Ham\right],
}
where
\bae{\label{eq: Hamiltonian}
    \Ham= \sqrt{\gamma} \left( \lapse C+\shift^i C_i\right),
}
and the so-called constraints read
\bae{
    C& \equiv \frac{2}{\spmet\Mpl^2}\left[\pifull_{ij}\pifull^{ij}-\frac{1}{2}\left(\pifull^i_i\right)^2 \right]-\frac{\Mpl^2}{2}\Ricci^{(3)} +  \frac{1}{2\spmet} G^{\indI\indJ}\pifull_\indI\pifull_\indJ+G_{\indI\indJ}  \frac{\spmet^{ij}}{2} \partial_i\phifull^\indI\partial_j\phifull^\indJ+V, \\
    C_i& \equiv -2 \left( \frac{\pi^j_i}{\sqrt{\gamma}} \right)_{| j}+\frac{1}{{\sqrt{\gamma}}}\pifull_\indI\partial_i\phifull^\indI=\frac{1}{\sqrt{\gamma}} \left(-2 \partial_k \left( \gamma_{ij} \pifull^{jk}  \right)+\pifull^{jk} \partial_i \gamma_{jk} +\pifull_\indI\partial_i\phifull^\indI \right) \,.
}
The Hamilton equations $\dot \gamma_{ij}=\frac{\delta}{\delta \pi^{ij}} \left( \int \dd^3 x \Ham \right)$ and $\dot \pi^{ij}=-\frac{\delta}{\delta \gamma_{ij}} \left(\int \dd^3 x \Ham \right)$
give the dynamical parts of the Einstein equations, whose explicit form will not be needed in what follows, while the variation with respect to the lapse and shift enforce the energy and momentum constraints 
\begin{equation} \label{eq: constraints}
C=C_i=0\,.
\end{equation}
Eventually, Hamilton equations in the scalar sector can be written in the compact form 
\bae{
    \dot{\phifull}^\indI&=  \frac{\lapse}{\sqrt{\spmet}}G^{\indI\indJ}\pifull_\indJ+\shift^i\partial_i\phifull^\indI, \label{eq: rescaled phi EoM} \\
    \covD_t\pifull_\indI&=-\sqrt{\spmet}\lapse V_\indI+\covD_i\left(\sqrt{\spmet}\lapse G_{\indI\indJ}\spmet^{ij}\partial_j\phifull^\indJ\right)+\covD_i\left(\shift^i\pifull_\indI\right)\,. \label{eq: rescaled pi EoM}
}
Here $V_\indI=\partial V/\partial\phifull^\indI$, while $\covD_t$ and $\covD_i$ are field-space covariant spacetime derivatives, whose actions on field-space vectors $\genU^\indI$ and covectors $\genV_\indI$ read 
\bae{
    \covD_\mu\genU^\indI=\partial_\mu\genU^\indI+\Gamma^\indI_{\indJ\indK}\left(\partial_\mu\phifull^\indJ\right)\genU^\indK\,, \quad  \covD_\mu\genV_\indI=\partial_\mu\genV_\indI-\Gamma^\indK_{\indI\indJ}\left(\partial_\mu\phifull^\indJ\right)\genV_\indK\,, \quad  \text{with } \mu\in\{t,i\},
    \label{def-full-covariant-derivative}
}
and where $\Gamma^\indI_{\indJ\indK}$ are the Christoffel symbols associated with the field-space metric $G_{\indI\indJ}$.

\subsection{Gauge choice and smoothing procedure}
\label{gauge-smoothing}

In the stochastic framework, all fields (actual scalar fields as well as the spacetime metric) are divided into a classical IR component and a quantum UV component, which are the counterparts of respectively the background and the fluctuations in standard perturbation theory (SPT). An important difference between the two setups is that the fields' IR components have large-scale fluctuations, which is nothing else than what the stochastic theory aims at describing. Hence, gauge issues, which usually only concern the equivalent of the UV part, also apply to the IR sector.
As standard in stochastic inflation, we will deal with fluctuations of scalar type only, letting aside vector and tensor perturbations for our purpose here, something that is not restrictive as we elaborate on below.

A convenient gauge to study multifield inflation in SPT is the spatially flat gauge, in which all genuine (scalar type) fluctuating degrees of freedom are in the scalar fields. The same holds true in the stochastic context, and we will use the scalar gauge freedom so that spatial slices are homogeneous, with no fluctuation, neither on small nor on large scales. In what we can call the \emph{stochastic spatially flat gauge}, we thus have
\begin{equation}
    \spmet_{ij}(N,\mathbf{x})=a^2(N)\delta_{ij} \quad \textrm{and} 
    \quad a(N) \propto \ee^N\,,
\end{equation}
where $a$ is spatially constant, and we choose to work with the time variable $N$ such that $N= \ln (a)$ up to an arbitrary constant.
This choice is convenient and conceptually relevant (see Refs.~\cite{Sasaki:1995aw,Lyth:2004gb,Pattison:2019hef}). 
In the same manner as in SPT, in this gauge, the local number of \efolds of expansion computed in each (super)-Hubble patch is then identical~\cite{Salopek:1990jq,Lyth:2004gb}, and simply coincides with $N$. Saying it otherwise, neglecting any shear component that are suppressed on large scales in standard situations, the flat gauge coincides with the uniform-$\eN$ gauge. This way, the stochastic formalism enables one to determine how the inhomogeneities of the scalar fields evolve in different patches, with a local clock that is deterministic and shared by all patches. We will not do this in this paper, but this implies that we can, at least in principle, easily use our results in the framework of the stochastic-$\delta N$ formalism to deduce the properties of the large-scale curvature perturbation $\zeta$ (see, e.g., Refs.~\cite{Fujita:2013cna,Fujita:2014tja,Vennin:2015hra,Kawasaki:2015ppx,Assadullahi:2016gkk,Vennin:2016wnk,Pinol:2018euk}).\footnote{Anticipating somewhat on following elements of the paper, let us mention that this important fact holds even when taking into account the tensor and vector modes. First, at quadratic order in the action as considered in the paper, the UV parts of the tensor and vector modes are decoupled from the UV parts of the scalar ones. More importantly, the tensor degrees of freedom, properly defined non-perturbatively in a way that they do not affect the spatial volume element, are such that at leading-order in the gradient expansion, their IR parts are time-independent and locally homogeneous in each $\sigma$-Hubble patch (while the vector modes vanish). Hence, they can be transformed away by a choice of spatial coordinates. This does not affect neither the local Hubble parameter,  nor the proper time~\cite{Lyth:2004gb}, and thus our time variable $N$ is a local clock that is deterministic and shared by all patches, despite the existence of large-scale tensor fluctuations.}
In the stochastic spatially flat gauge, covariant spatial derivatives reduce to usual ones, the curvature of spatial slices $\Ricci^{(3)}$ vanishes, and $\sqrt{\spmet}=a^3$. To simplify equations, we also  \textit{rescale momenta}, $\pi_\indI \to a^3 \pi_\indI$, which we will use from now on and for the rest of this paper.\\

Let us now discuss the smoothing procedure splitting any quantity between its IR and UV components, first in the simpler context of quantum field theory in a fixed de Sitter background. For each quantity written in Fourier space as  $X(N,\mathbf{x})=\int\dk\ee^{i\kdx} X(N,\mathbf{k})$, its IR component is defined by coarse-graining as
\bae{
	X_\IR(N,\mathbf{x})=\int\dk\ee^{i\kdx}W\left(\frac{k}{\ksigma(N)}\right)X(N,\mathbf{k}),
	\label{def-XIR}
}
with some window function $W$ such that $W \simeq 1$ when its argument is small, and $W \simeq 0$ when its argument is large, i.e., smearing out short-wavelength modes $k>\ksigma(N) \equiv \sg a(N) H$, corresponding to a constant smoothing physical scale $\lambda_{{\rm s}}=(\sg H)^{-1}$. 
In this context, $\sg \ll 1$ is a small parameter ensuring that the smoothing scale is somewhat larger than the Hubble scale --- allowing for a gradient-, i.e., a $\sg$-expansion --- and therefore that the infrared component can be considered as classicalised. 
As usual in physics, the details of this coarse-graining procedure should not affect physical observables, i.e. in this context, the properties of fluctuations on physical scales $\lambda \gg \lambda_{{\rm s}}$.
Like in the context of the renormalisation group, a smooth window function seems physically motivated and desirable. 
However, this comes in general at the expense that the resulting description involves colored noises~\cite{Winitzki:1999ve,Matarrese:2003ye,Liguori:2004fa}, which are more difficult to handle analytically than white noises.
In this paper, we will conservatively use the simpler choice of a sharp window function $W(x)=\theta(1-x)$, which is largely used in the literature.
This has the advantage of being intuitive, and this will enable us to use the well-developed machinery of stochastic differential equations with white noises.
However, as we will see in section~\ref{sec: effective hamiltonian action}, in the path-integral approach in which short-wavelength fluctuations are integrated out, special attention has to be paid to the integration measure’s split into IR and UV sectors~\cite{Rigopoulos:2016oko,Moss:2016uix,Tokuda:2017fdh,Tokuda:2018eqs}.

For notational simplicity, we will also write $\ksigma(N)= \sg a(N) H$ in the case of scalar fields backreacting on spacetime, although the time-dependent Hubble scale $H$
is not defined \textit{a priori} in such a stochastic context, but should emerge as an IR quantity itself. One will find indeed that the quantity $1/\lapseIR$ plays the role of a ``local Hubble parameter'' (see Eq.~\eqref{laspeIR}), in agreement with the literature in SPT~\cite{Lyth:2004gb}.
One can thus imagine self-consistently defining the smoothing scale such that Eq.~\eqref{def-XIR} is verified for $X=\cal{N}$ with $\ksigma=\sg a /\lapseIR$.
We will not consider further this slight ambiguity, that is also present in single-field inflation while not being addressed in the literature to the best of our knowledge, and simply assume that the smoothing scale can be defined at least implicitly through a procedure similar to the one suggested above.

\subsection{Stochastic equations}
\label{subsec: stochastic equations}

We now decompose the scalar fields and the metric components into IR and UV parts as
\bae{\label{eq: IRUV decomposition}
    \bce{
        \dps
    	\phifull^\indI=\phiIR^\indI+Q^\indI, &
    	\dps
    	\pifull_\indI=\piIR_\indI+P_\indI=\piIR_\indI+\covP_\indI+\Gamma^\indK_{\indI\indJ}\piIR_\indK Q^\indJ, \\ 
    	\dps
    	\lapse=\lapseIR+\lapseUV, & 
    	\dps
    	\shift^i=a^{-2}\delta^{ij}\partial_j\shiftUV,
    }
}
where $\phiIR^\indI$, $\piIR_\indI$, and $\lapseIR$ are IR quantities, and one can fix $\shift^i_\IR=0$ as it is a pure gauge choice in the long-wavelength limit. 
The second term in the decomposition of $\pifull_\indI$, where $\Gamma_{IJ}^K$ is evaluated at the infrared values of fields $\phiIR^\indI$, may seem arbitrary, but it ensures that the UV quantity $\covP_\indI$ transforms at linear order in a covariant manner under field redefinitions, as we prove in Sec.~\ref{subsec: covariant perturbations}

In the heuristic stochastic approach, one simply substitutes the decomposition~(\ref{eq: IRUV decomposition}) into the original EoM~(\ref{eq: rescaled phi EoM}) and (\ref{eq: rescaled pi EoM}), keeping all nonlinearities in the IR sector --- albeit working at leading-order in the gradient expansion --- but keeping only linear terms in UV quantities.
One thus obtains
\bae{
	\phiIR^{\indI\prime}=\frac{1}{H} G^{\indI\indJ}\piIR_\indJ+\int\dk\ee^{i\kdx}\left[W^\prime\left(\frac{k}{\ksigma}\right)\phifull^I(N,\mathbf{k})
		+\left(1-W\left(\frac{k}{\ksigma}\right)\right)\UVE^{Q\indI}(N,\mathbf{k})\right],
    \label{first}
}
and
\bae{
    \covD_N\piIR_\indI&=-3\piIR_\indI- \frac{1}{H} V_\indI+\int\dk\ee^{i\kdx}\left[W^\prime\left(\frac{k}{\ksigma}\right)  \left( \pifull_\indI(N,\mathbf{k})-\Gamma_{\indI\indJ}^\indK\piIR_\indK\phifull^\indJ(N,\mathbf{k}) \right)\right. \nonumber \\
     &\qquad \qquad \qquad \qquad \qquad \qquad \qquad \left.+\left(1-W\left(\frac{k}{\ksigma}\right)\right)\UVE^\covP_\indI(N,\mathbf{k})\right], \label{second}
}
where a prime $^\prime$ denotes a simple derivative with respect to $N$ and we denote the covariant time derivative $\covD_N\piIR_\indI=\partial_N\piIR_\indI-\Gamma_{\indI\indJ}^\indK\phiIR^{\indJ\prime}\piIR_\indK$ --- covariant with respect to field redefinitions of the IR fields --- by the same symbol $\covD_N$ as in the fully nonlinear Eq.~\eqref{def-full-covariant-derivative} for simplicity.
Here, $\UVE^{Q\indI}$ and $\UVE^\covP_\indI$, whose expressions are given below in Eqs.~\eqref{eq: EQ} and \eqref{eq: EP}, stand for the linearised EoM in Fourier space, and the expression of $H \equiv 1/\lapseIR$ in terms of $\phiIR^\indI$ and $\piIR_\indI$ will be given in Eq.~\eqref{laspeIR}.

In SPT, which one formally recovers in the limit $\ksigma \to 0$, one assumes that the dynamics of fluctuations decouples from the one of the background, in which case one has $\UVE^{Q\indI}=\UVE^\covP_\indI=0$. The terms in $W^\prime$ vanish in this limit and thus each of the equations~\eqref{first} and \eqref{second} splits into two parts, one for the background and one for the fluctuations.
In the heuristic approach to stochastic inflation, one still assumes that UV fluctuations obey the same evolution equations $\UVE^{Q\indI}=\UVE^\covP_\indI=0$ as in SPT. However, due to the time-dependence of the coarse-graining scale $\ksigma(N)$, the IR dynamics is affected by the flow of UV modes joining the IR sector, an effect described by the terms involving the time-derivative of the window function, $W^\prime$. Thus writing
\bae{\label{eq: noises}
	\xi^{Q\indI}(x)=\int\dk\ee^{i\kdx}
    W^\prime\left(\frac{k}{\ksigma(N)}\right)	
    Q^\indI(N,\mathbf{k}), \quad
	\xi^{\covP}_\indI(x)=\int\dk\ee^{i\kdx}
	W^\prime\left(\frac{k}{\ksigma(N)}\right)
	\covP_\indI(N,\mathbf{k}),
}
one obtains the desired effective equations of motion for infrared fields and momenta,
\bae{\label{eq: Langevin in heuristic}
	\phiIR^{\indI\prime}=\frac{1}{H} G^{\indI\indJ}\piIR_\indJ+\xi^{Q\indI}, \quad\quad
	\covD_N\piIR_\indI=-3 \piIR_\indI-\frac{1}{H} V_\indI+\xi^{\covP}_\indI,
}
the so-called Langevin equations. In this description, one assumes that the UV quantities $Q^I(N,\mathbf{k})$ and $\covP_\indI(N,\mathbf{k})$, which in fact are quantum operators, can be described classically as they join the IR sector at the time $N_\sigma(k)$ such that $k=\ksigma(N_\sigma(k))$. 
Hence, the $\xi$'s can be interpreted as classical random noises, and when computing their statistical properties, one identifies ensemble averages with expectation values of the corresponding operators in the quantum vacuum state of the theory.
As we treat UV fluctuations at linear order, one can consider that the $\xi$'s obey a Gaussian statistics, with zero mean and fully characterised by their two-points correlations. In the case of the sharp window function $W(x)=\theta(1-x)$, the latter
can be easily computed and
are directly related to the power spectra of the UV fluctuations when they reach the coarse-graining scale: 
\bae{\label{eq: noise correlation}
    \braket{\xi^{\covX\indI}(x)\xi^{\covY\indJ}(x^\prime)}=
    \underbrace{\calP^{\covX\covY\indI\indJ}(N,\ksigma(N))\frac{\ksigma^\prime}{\ksigma}}_{\textstyle \begin{array}{c}
	A^{\covX\covY\indI\indJ}(N)\end{array}}
    \frac{\sin \ksigma r}{\ksigma r} \delta(N-N^\prime),
}
where $r=|\mathbf{x}-\mathbf{x}^\prime|$ is the comoving distance between the spacetime points $x$ and $x^\prime$, and with the dimensionless power spectra ${\cal P}$ such that
\bae{\label{eq: power spectra}
   \braket{Q^{\covX\indI}(N,\mathbf{k})Q^{\covY\indJ}(N,\mathbf{k}^\prime)}= (2\pi)^3\delta^{(3)}(\mathbf{k}+\mathbf{k}^\prime)\frac{2\pi^2}{k^3}\calP^{\covX\covY\indI\indJ}(N,k)\,.
}
Here, we used a condensed notation adapted to our phase-space description where $\covX=(Q,\covP)$ refers both to UV fields and covariant momenta, i.e. $\xi^{\covX\indI}=(\xi^{Q \indI },\,\xi^{\covP \indI}=G^{\indI \indJ}(\phiIR)	\xi^{\covP}_\indJ)$ and $Q^{\covX\indI}=(Q^{\indI},\,\covP^{\indI} = G^{\indI \indJ}(\phiIR) \covP_{\indJ})$.  
The auto-correlation of the noises $A^{\covX\covY\indI\indJ}$ is a contravariant rank-2 tensor since it inherits the transformation properties of the UV modes $(Q^{\indI},\,\covP^{\indI})$. Because of the presence of the delta function $\delta(N-N^\prime)$ coming from the time derivative of the step window function, the $\xi$'s can be regarded as white noises. This property would not hold true had we chosen a smooth window function.
Notice also that the noise correlations~\eqref{eq: noise correlation} are proportional to $\ksigma^\prime/\ksigma\times\sin \ksigma r/\ksigma r$. First, the ratio $\ksigma^\prime/\ksigma=1-\epsilon$ in Eq.~\eqref{eq: noise correlation}, with $\epsilon$ in Eq.~\eqref{epsilon}, may be approximated by unity.
Indeed, this slow-roll correction is likely to be too precise for the accuracy of the coarse-graining procedure, and other authors also proposed considering a slightly time-dependent parameter $\sigma$ such that $\sigma H$ is exactly constant~\cite{Starobinsky:1994bd}. Second, the precise form $\sin{\ksigma r}/\ksigma r$ of the apparent spatial correlation in the noises' two-point function depends on the choice of the window function $W$. However, since we neglected any spatial dependence of the IR fields, this oscillating and decaying term should only be understood as a step theta function $\theta(1-\ksigma r)$ taking values $1$ inside a ``$\sigma$-Hubble patch" and $0$ outside, in agreement with the separate universe approach.
The $\xi$'s can thus be understood as $\ksigma$-patch-independent Brownian noises, the evolution of each $\sg$-Hubble patch being determined only by the local physics.
In this paper we will discuss only one-point statistics (one ``$\sg$-Hubble patch" statistics to be accurate), the idea being that starting from one progenitor $\sg$-Hubble patch, the observable universe at the end of inflation is made of many $\sg$-Hubble patches that emerge from the same initial condition. Hence, by ergodicity, 
the ensemble average of the stochastic evolution of one $\sg$-Hubble patch can also be seen as the spatial average among these $\sg$-Hubble patches. Moreover, the study of the one-point statistics is not as restrictive as it may seem, and it is actually possible to extract detailed spatial information from it. Indeed, any two $\sigma$-Hubble patches initially share the same dynamics, until they become separated by the physical distance $(\sigma H)^{-1}$ and subsequently evolve independently. Using this time-scale correspondence, Starobinsky and Yokoyama have shown in Ref.~\cite{Starobinsky:1994bd} how, once the Fokker-Planck operator for the one-point probability density function (PDF) is known, one can determine the evolution equation for the two-point PDF, or even any $n$-point PDF (at different spatial and temporal locations), and thus, at least in principle, retrieve all the statistical information (see, e.g., \cite{Markkanen:2019kpv,Markkanen:2020bfc,Moreau:2019jpn} for recent applications).\footnote{Naturally, given the hard cutoff in the spatial correlations of the stochastic noises, spatial correlations can be reliably computed only when the relevant length scales are well above $H^{-1}$, but that is overwhelmingly the case for observationally relevant scales.} This logic is also put to good use in the stochastic-$\delta N$ approach,
with which one can compute Fourier space correlation functions of the observable large-scale curvature perturbation $\zeta$ (and not only statistics of the inflationary fields) (see, e.g., Refs.~\cite{Fujita:2013cna,Fujita:2014tja,Vennin:2015hra,Kawasaki:2015ppx,Assadullahi:2016gkk,Vennin:2016wnk,Pinol:2018euk}).

Eventually, note that first-principles methods to compute the power spectra will be reviewed in Sec.~\ref{subsec: classicalisation}, and that analytical estimates will be discussed in Sec.~\ref{sec:Markovian}.
Before explaining in Sec.~\ref{limitations} why this heuristic approach to stochastic inflation is not fully satisfactory, let us now fill in the gaps in the above description by 
characterising the dynamics of the UV fluctuations.

\subsection{Dynamics of UV fluctuations}

First, one needs to relate the perturbations of the non-dynamical parts of the metric, $\lapseUV$ and $\psi$, to the genuine degrees of freedom: the UV parts of the scalar fields and momenta $Q^\indI$ and $\covP_\indI$. For this, it is important to notice that the energy and momentum constraints~\eqref{eq: constraints} contain no time derivative. Hence, contrary to the Hamilton equations~\eqref{first} and \eqref{second}, no explicit noise enters in their IR/UV decomposition, and they can be straightforwardly split into independent equations on large and small scales.

The momentum constraint gives no information on large scales, in agreement with the fact that all choices of threadings are equivalent at leading-order in the gradient expansion, while the small scale component gives the expression of $\lapseUV$:
\bae{
	\lapseUV=\frac{\piIR_\indI Q^\indI}{2\Mpl^2H^2}.
	\label{lapseUV}
}
As for the energy constraint, its long-wavelength limit is non-trivial and is equivalent to the first Friedmann equation, while the small scale part relates $\psi$ to $\lapseUV$ and $Q^\indI,\covP_\indI$:
\bae{
    \frac{3\Mpl^2}{\lapseIR^2} &\equiv 3\Mpl^2H^2=\frac{1}{2} G^{\indI\indJ}\piIR_\indI\piIR_\indJ+V, \label{laspeIR} \\	
    2\Mpl^2H^2 \frac{k^2}{a^2}\shiftUV&=\piIR_\indI\covP^\indI+V_\indI Q^\indI+6 \Mpl^2 H^3 \lapseUV\,.	
    \label{eq: perturbation energy const}
}
The equation~\eqref{laspeIR} confirms that $1/\lapseIR$ plays the role of a local Hubble parameter, with the usual Friedmann constraint holding in each $\ksigma$-patch. 
In this respect, note that if one converts $\piIR_\indI$ to $\phiIR^{\indI\prime}$ with use of the IR EoM~(\ref{eq: Langevin in heuristic}), the Friedmann equation would include an explicit noise term. This demonstrates the conceptual advantage of the Hamiltonian language over the Lagrangian one in the stochastic formalism.

Equipped with the constraints~\eqref{lapseUV} and \eqref{eq: perturbation energy const}, one can express $\UVE^{Q\indI}$ and $\UVE^\covP_\indI$ in the condensed form: 
\bae{
    \UVE^{Q\indI}(N,\mathbf{k})=&-\covD_N Q^{\indI}(N,\mathbf{k})+\frac{\covP^\indI(N,\mathbf{k})}{H}+M^2_{\covP Q}{}^\indI{}_\indJ Q^\indJ(N,\mathbf{k}),
    \label{eq: EQ} \\
	\UVE^\covP_\indI(N,\mathbf{k})=&-\covD_N \covP_\indI(N,\mathbf{k})- 3\covP_\indI(N,\mathbf{k})-\frac{k^2}{a^2 H}
		Q_\indI(N,\mathbf{k})	-\frac{1}{H}	M^2_{QQ\indI\indJ} Q^J(N,\mathbf{k})-M^2_{Q\covP\indI}{}^\indJ\covP_\indJ,
	\label{eq: EP}
}
where indices are lowered and raised with the IR metric in field space $G_{\indI \indJ}(\phiIR)$ and its inverse, and
\bae{
		M^2_{QQ\indI\indJ}&=V_{;\indI\indJ}-R_\indI{}^{\indK\indL}{}_\indJ\piIR_\indK\piIR_\indL
		+\frac{1}{2\Mpl^2H}(V_\indI\piIR_\indJ+\piIR_\indI V_\indJ)+\frac{3\piIR_\indI\piIR_\indJ}{2\Mpl^2}, \label{M2QQ}
		\\
		M^2_{Q\covP\indI\indJ}&=M^2_{\covP Q\indI\indJ}=\frac{\piIR_\indI\piIR_\indJ}{2\Mpl^2H^2}\,, \label{M2QP}
}
with $V_{;\indI\indJ} \equiv \nabla_\indJ V_\indI=V_{,\indI \indJ}-\Gamma_{\indI \indJ}^\indK V_{\indK}$ 
the covariant Hessian of the potential, and 
$R^\indS{}_{\indI\indJ\indK}\equiv\Gamma^\indS_{\indI\indK,\indJ}-\Gamma^\indS_{\indI\indJ,\indK}
    +\Gamma^\indR_{\indI\indK}\Gamma^\indS_{\indJ\indR}-\Gamma^\indR_{\indI\indJ}\Gamma^\indS_{\indK\indR}$ the Riemann tensor of the field space.
To obtain the expressions~\eqref{eq: EQ}--\eqref{eq: EP}, and in accordance with treating the UV modes linearly, we have simplified the infrared ``coefficients'' by neglecting the noise terms in Eq.~\eqref{eq: Langevin in heuristic}, and similarly, we used $\UVE^{Q\indI}=0$ in Eq.~\eqref{eq: EP}. As expected, 
the equations
$\UVE^{Q\indI}=\UVE^\covP_\indI=0$
are equivalent to
the EoM for linear perturbations in SPT~\cite{Sasaki:1995aw}, with background fields replaced by IR ones, i.e. their combination gives
\bae{\label{eq: UV eom mode}
    \covD_N^2 Q^\indI+\left(3-\epsilon\right)\covD_N Q^\indI+\left(\frac{k^2}{a^2H^2}\delta^{\indI}_{\indJ} + \frac{M^2{}^\indI{}_\indJ}{H^2}\right)Q^\indJ=0,
}
where (consistently neglecting noise terms in the second equality)
\bae{
	\epsilon \equiv-\frac{\dotH}{H} = \frac{\piIR_\indI\piIR^\indI}{2\Mpl^2H^2},
	\label{epsilon}
}
and the mass matrix reads
\bae{
    M^2{}^\indI{}_\indJ=V^\indI{}_{;\indJ}-R^{\indI\indK\indL}{}_\indJ\piIR_\indK \piIR_\indL-\frac{H}{a^3\Mpl^2}\covD_N\left(\frac{a^3}{H}\piIR^\indI\piIR_\indJ\right). 
}

\subsection{Limitations of the heuristic approach}
\label{limitations}

Although qualitatively satisfying, the above heuristic approach to stochastic inflation  suffers from a number of technical and conceptual issues. 

\begin{itemize}
    \item  When going from Eqs.~\eqref{first}--\eqref{second} to \eqref{eq: noises}, we attributed $\phifull^\indI(N,\ksigma(N))$,
the Fourier component of the full field at the transition time, to the UV part $Q^I$ (and similarly for momenta), in a rather arbitrary manner.

\item We assumed
that the UV modes obey $\UVE^{Q\indI}=\UVE^\covP_\indI=0$, i.e. the same equations as in standard perturbation theory with background fields replaced by IR ones.

\item Despite the fact that $\phiIR^\indI$ and $\piIR_\indI$ are real, the noise correlation 
$\braket{\xi^{Q\indI}\xi^{\covP\indJ}}$ has an imaginary component, owing to the fact that the quantum operators $Q^\indI(N,\ksigma(N))$ and $\covP^\indI(N,\ksigma(N))$ do not commute. To interpret Eq.~(\ref{eq: Langevin in heuristic}) as proper real stochastic equations, one has to replace by hand $\braket{\xi^{Q\indI}\xi^{\covP\indJ}} \to \frac12 \left( \braket{\xi^{Q\indI}\xi^{\covP\indJ}}+\braket{\xi^{\covP\indJ}\xi^{Q\indI}}\right)=\Re \left[ \text{$\braket{\xi^{Q\indI}\xi^{\covP\indJ}}$} \right]$, i.e. to take the (real) vev of hermitian operators only (see e.g. Ref.~\cite{Grain:2017dqa}).\footnote{This problem is not present for the $\xi^Q \xi^Q$ and $\xi^\covP \xi^\covP$ correlations which are real, as the (real space) $Q
^I$ (and the $\covP^I$ separately) are hermitian operators that commute with one another at equal times, see also Eqs.~\eqref{Q-quantisation} and \eqref{two-point-vacuum}.}

\item In addition to these difficulties, there remains an ambiguity in the treatment of stochastic differential equations of the type~\eqref{eq: Langevin in heuristic} as the continuous limit of discrete processes. In a previous letter~\cite{Pinol:2018euk}, 
we emphasised the role of such discretisations and unveiled the presence of \emph{inflationary stochastic anomalies}, potentially inducing spurious frame dependences or breaking the covariance of the theory.

\end{itemize}

All these difficulties are related and motivates a careful treatment of quantum aspects of the problem.
First, in Sec.~\ref{sec: stochastic anomalies}, we will discuss and solve the issue of the aforementioned stochastic anomalies. Critical to this resolution
is the identification of \textit{independent} Gaussian white noises --- as required from a proper mathematical treatment of stochastic differential equations --- in one-to-one correspondence with the \textit{independent} quantum creation and annihilation operators necessary for the quantisation of the UV modes.
Second, in Sec.~\ref{sec: effective hamiltonian action}, we solve the other difficulties related to IR/UV interactions by working at the level of the action and integrating out the quantum UV modes in the closed-time-path formalism. We do so paying a particular attention to issues of covariance and, following Refs.~\cite{Tokuda:2017fdh,Tokuda:2018eqs},
to the integration measure’s split into IR and UV sectors. Notably, the fact that UV modes become IR, but not the reverse, entails the existence of fluctuations without dissipation, in contrast to ordinary open systems.

\subsection{To be or not to be Markovian}
\label{subsec:Markovian?}

Here we would like to stress an ever-present subtlety, be it in the heuristic approach or in a proper quantum field theory treatment. It lies in the fact that the effective dynamics of the coarse-grained scalar fields is stricly speaking non-Markovian (see, e.g., Refs.~\cite{Boyanovsky:1994me,Rau:1995ea,Boyanovsky:1998pg,Xu:1999aq,Berera:2007qm,Farias:2009wwx,Farias:2009eyj,Gautier:2012vh,Buldgen:2019dus} for related discussions in 
various areas of physics). 
For the equations~\eqref{eq: Langevin in heuristic} to describe a Markov process, characterised by the absence of memory, one would need the statistical properties of the noises to be a function of the infrared variables $(\phiIR^\indI,\piIR_\indI)$ at current time $N$.
However, the power spectra of the UV modes~\eqref{eq: power spectra}, or in a related manner their mode functions, are not even functions of the IR variables. They are simply solutions of the differential equations~\eqref{eq: EQ} and \eqref{eq: EP}, whose ``coefficients" depend on the IR variables, and that are evaluated at time $N$ for the mode with wave number $k_\sigma(N)$. Moreover, this effective ``background" for the UV dynamics is described by coarse-grained fields whose values were affected by previous realisations of the noises. The dynamics described by such equations is thus very rich and complex. 

In this respect, we would like to stress that the bulk of this paper as well as its main result~\eqref{Langevin-intro}--\eqref{noise-properties-intro} do not involve any Markovian approximation, as our emphasis is on the first-principle derivation of these manifestly covariant equations. This means that our Langevin-type equations can be in principle solved numerically together with the dynamics of the UV modes dictating the noise properties.
They can also serve as a basis for future analytical works, and in this context, it can be convenient to resort to the Markovian approximation, at the expense for instance of assuming some slow-varying regime. We discuss such analytical estimates in Sec.~\ref{sec:Markovian}. One of the advantage of the Markovian approximation is that one can then write a Fokker-Planck (FP) equation for the (one-point) probability density function of the IR fields and momenta, with the result~\eqref{eq: phase space fokker-planck}.
Such an equation is easier to handle numerically or analytically than the Langevin-type equations, and indeed covariance is even more manifest with such a formulation. However, we stress again that our main results hold more generally.

\section{Stochastic anomalies and their solution}
\label{sec: stochastic anomalies}

A generic difficulty in the description of stochastic processes is that stochastic equations like the one~\eqref{eq: Langevin in heuristic} are not mathematically defined unless specifying 
their discretisation schemes (see e.g. Refs.~\cite{risken1989fpe,vankampen2007spp}).
In particular, in our context, different choices of discretisations (among which It\^o and Stratonovich are the most famous ones) can lead to a violation of the EoM's covariance, and/or an unphysical noise-frame dependence as we pointed out previously~\cite{Pinol:2018euk}.
We first review such \emph{stochastic anomalies} in Sec.~\ref{discretisation}. To make the physics easier to grasp, we sometimes restrict ourselves to the particular case of a Markov process. Indeed, this enables one to write the so-called Fokker-Planck equation, corresponding to the Langevin equations with a no-memory noise, that dictates the deterministic evolution of the probability density function for the IR fields and their momenta.
We then explain in Sec.~\ref{subsec: classicalisation} why the particular framework of stochastic inflation,
where the classical stochastic noise emerges from a quantum field theory description, provides us with a preferred frame for the reduction of the noise auto-correlation matrix:
the one of independent creation and annihilation quantum operators. 
Stochastic anomalies are thus solved when interpreting the Langevin equations~\eqref{eq: Langevin in heuristic} with this particular choice of frame and in the Stratonovich scheme.
However, this resolution is rather formal and in order to make it more explicit, we introduce \emph{stochastically-parallel-transported vielbeins} in Sec.~\ref{vielbeins-Ito}. Strikingly, with the use of such vielbeins, the Langevin equations~\eqref{eq: Langevin in heuristic} interpreted in the Stratonovich scheme can be recast in the form~\eqref{eq: covariant Langevin Ito} interpreted in the It\^o scheme, featuring covariant derivatives adapted to It\^o calculus. These equations constitute one of the main results of this paper: they are manifestly covariant, readily adapted to numerical implementations, and when supplemented with a Markovian approximation, they lead to the phase-space Fokker-Planck equation~\eqref{eq: phase space fokker-planck}.

\subsection{Ambiguity of the discretisation scheme}
\label{discretisation}

A stochastic differential equation (SDE), or equivalently its solution as a stochastic integral, is mathematically defined as the infinitesimal-step, continuous limit of a finite-step, discrete summation, in the same way that the Riemann integral is defined.
From step $i$ to step $i+1$, the integrand must be evaluated at some time between times $N_i$ and $N_{i+1}$,
expressed as $(1-\discAlpha)N_i+\discAlpha N_{i+1}$ with the parameter $0\le\discAlpha\le1$. The Riemann integral of differential functions is independent of this discretisation choice of $\discAlpha$ in the continuous limit. However, due to the non-differentiability of the stochastic noise, the stochastic integral does depend on $\discAlpha$. 
Conveniently for our purpose, let us explain these subtleties in a situation where the stochastic variables $\genX^\indI$ are coordinates on a manifold, endowed with a metric whose components in these coordinates are $G_{\indI \indJ}$ (see
Appendix~\ref{appendix: stochastic calculus} for generic mathematical properties independently of this specific context). Let us further assume that the stochastic process under study is described by a deterministic drift $h^\indI$ as well as noises' amplitudes $g^{\genX\indI}_A$ that all transform as vectors under redefinitions of the coordinates $\genX^\indI$, and such that the corresponding set of Langevin equations reads 
\bae{\label{eq: dXdt}
    \dif{\genX^\indI}{N}=h^\indI+g^{\genX\indI}_\indA \circ_\alpha \xi^\indA, \quad \braket{\xi^\indA(N)\xi^\indB(N^\prime)}=\delta^{\indA\indB}\delta(N-N^\prime), \quad
    \delta^{AB} g_A^{\genX\indI} g_B^{\genX\indJ}=A^{\genX\genX IJ},
}
together with a specified discretisation scheme $\discAlpha$ represented by the symbol $\circ_\alpha$.
Here $A^{\genX\genX IJ}$ stands for the auto-correlation of the effective noises $\xi^\indI=g^{\genX\indI}_A\xi^\indA$, i.e. $\langle \xi^\indI(N) \xi^\indJ(N) \rangle=A^{\genX\genX IJ}\delta(N-N^\prime)$, and transforms as a rank-2 tensor under coordinate transformations.\footnote{Notice that in this more general mathematical context, we label the noises' amplitudes $g^{\genX\indI}_A$ and their auto-correlation $A^{\genX\genX \indI\indJ}$ by the stochastic variables $\genX^\indI$ that receive the stochastic kicks via the Langevin equation. This notation is slightly different from the one used in the specific multifield stochastic inflation context, where the noises' amplitudes $g^{\covX\indI}_A$ and their auto-correlation $A^{\covX\covX \indI\indJ}$ are labeled by the UV modes $(Q^\indI, \covP^\indI)$ that are responsible for the stochastic kicks received by the IR fields $(\phiIR^\indI,\piIR^\indI)$.}
It is important to understand that prescribing this auto-correlation is not sufficient to define the corresponding SDE, but that one needs to specify the full set of $g^{\genX\indI}_A$'s, i.e. the decomposition of the $\xi^\indI$'s onto a set of independent noises $\xi^\indA$, what we may call in what follows an orthonormal frame for the noises.
The setup~\eqref{eq: dXdt} encapsulates the specific case of stochastic inflation when momenta are neglected (although a decomposition into independent noises $\xi^A$ has not been identified yet), i.e., only the first of the Langevin equations~\eqref{eq: Langevin in heuristic} 
is considered here. We do so to simplify the presentation, but the discussion will be extended next to the more general setup of Langevin equations in phase space.

The simplest situation for such kind of SDE is when the noise amplitude $g^{\genX\indI}_\indA$ is only a (deterministic) function of time, in which case the noise is called \emph{additive}, and all discretisations have the same continuous limit. In the generic case however, the noise also depends on the stochastic variables $\genX^\indI$ at time $N$, in which case it is called \emph{multiplicative} and choices of discretisations matter. As we stressed in Sec.~\ref{subsec:Markovian?}, the stochastic equations for the coarse-grained scalar fields are even more complicated as, contrary to standard SDEs, the dependence of the noise on the stochastic variables is only indirect. This is the reason why we sometimes refer to them as Langevin-type equations.
Despite this, let us begin by explaining the covariance issue in the simplest context of a Markovian description in which the $g^{\genX\indI}_\indA$'s explicitly depend on $\genX^\indI(N)$.
In that case, there is a one-to-one correspondence between the Langevin equations~\eqref{eq: dXdt} with a given discretisation scheme $\discAlpha$ and a Fokker-Planck
partial differential equation for 
the transition probability from the initial state $\genX^\indI_\mathrm{ini}(N_\mathrm{ini})$ to $\genX^\indI(N)$, sometimes simply referred to as 
the probability density function (PDF) of the stochastic variables, $P\left(N;\genX^\indI\right)$: 
\bae{\label{eq: covariant fokker-planck}
    \partial_N\Ps=-\N_\indI(h^\indI\Ps)+\alpha\N_\indI\left[g^{\genX\indI}_\indA \N_\indJ \left(g^{\genX\indJ}_\indA\Ps\right)\right]-\left(\alpha-\frac{1}{2}\right)\frac{1}{\sqrt{G}}\partial_\indI\partial_\indJ\left(\sqrt{G} A^{\genX\genX IJ}\Ps\right)\,.
}
Here, $\N_\indI$ 
is the usual field-space covariant derivative and we defined a rescaled PDF $\Ps=P/\sqrt{G}$, with $G=\textrm{det}(G_{\indI \indJ})$, where the subscript $s$ indicates that it is a scalar under redefinition of the coordinates.
From this expression, it is possible to identify two particular
values of $\alpha$.
It is indeed possible to set to zero
the second term in Eq.~\eqref{eq: covariant fokker-planck} with the choice of a \emph{prepoint}, $\discAlpha=0$ discretisation, called the It\^o scheme. Another interesting option is to keep this second term but to set to zero the third one by preferring a \emph{midpoint}, $\discAlpha=1/2$ discretisation, called the Stratonovich scheme. In the rest of this section, we will review the pros and cons of each of these two choices, keeping in mind that our derivation of the Langevin equations in the previous section did not come with any prescription for $\discAlpha$, so at this stage one should discriminate between the possibilities of discretisation based on physical arguments.

\subsubsection{It\^o scheme}
\label{Ito-scheme}

The It\^o scheme, corresponding to $\discAlpha=0$, is widely used in applied and computational mathematics 
because it has the advantage of expressing explicitly the stochastic variables at time $N_{i+1}$ in terms of known values at $N_i$. Not only is it conceptually clear, but it is also easy to implement numerically, which explains its widespread use in various areas of science.
However, this description seems to suffer from a fundamental issue in our context: in the Fokker-Planck (FP) equation~\eqref{eq: covariant fokker-planck}, where the third term survives for $\discAlpha=0$, only partial derivatives $\partial_I$ appear, rather than covariant derivatives $\N_I$. 
Thus, the FP equation as it is breaks covariance. More precisely, the problem is not that this equation is formulated in terms of non-covariant objects, i.e., that is not manifestly covariant, it is that it is not consistent with $\Ps$ being a scalar quantity.

Actually, this fundamental flaw can already be seen at the level of the Langevin-type equations, even when the process is not assumed to be Markovian. Indeed, the standard chain rule for the derivation of composite functions of the stochastic variables $\genX^\indI$ does not hold in the It\^o prescription, but
gets corrected by the auto-correlation of the noise.
This so-called It\^o's lemma states that, under a change of variables $\genX^\indI\to\Xt^{\indIt}=\Xt^{\indIt}(\genX^\indI)$, the infinitesimal variations read~\cite{ito1944109}:
\bae{
	\dd\Xt^{\indIt}=\pdif{\Xt^{\indIt}}{\genX^\indI}\dd\genX^\indI+\frac{1}{2}\frac{\partial^2\Xt^{\indIt}}{\partial\genX^\indI\partial\genX^\indJ}A^{\genX\genX\indI\indJ}\dd N.
	\label{eq: Ito-lemna}
}
We prove such kinds of exotic properties of stochastic calculus in Appendix~\ref{appendix: stochastic calculus} and refer the interested reader to it. However the form of this lemma can be easily understood:
a white noise is not a differentiable function because its infinitesimal variation $\dd \xi$ is proportional to $\sqrt{\dd N}$ rather than $\dd N$, thus $\dd \xi/\dd N$ diverges when $\dd N \rightarrow 0$. Therefore at order $\dd N$ even the second derivative of $\genX^I$ matters in the Taylor expansion~\eqref{eq: Ito-lemna}. The conclusion is that the standard infinitesimal variation $\dd\genX^\indI$ does not transform as a vector, contrary to the expectation for the infinitesimal variation of a coordinate on a manifold.
Thus, although equations~\eqref{eq: dXdt} and \eqref{eq: Langevin in heuristic} are covariant under the standard chain rule, they are actually not if they are interpreted in the It\^o sense, precisely because the standard chain rule is not verified.
This fact forbids us to interpret the Langevin equations derived in the heuristic approach
with the It\^o scheme. However, covariance and It\^o together are not doomed to fail, and it is actually possible to define covariant derivatives compatible with It\^o calculus that compensate for the breaking of the standard chain rule. 
A possible such derivative for the coordinates $\genX^\indI$ is given by
\bae{\label{eq: ItoD for X}
	\ItoD\genX^\indI=\dd\genX^\indI+\frac{1}{2}\Gamma^\indI_{\indJ\indK}A^{\genX\genX\indJ\indK}\dd N,
}
which we show to transform as a vector in It\^o calculus in Appendix~\ref{appendix: stochastic calculus}.\footnote{Related notions of It\^o covariant derivatives have been discussed in the literature independently of the context of inflation in Ref.~\cite{GRAHAM1985209}.}
There we also derive It\^o-covariant derivatives for vectors $\genU^I$ and covectors $\genV_I$ when they are subject to Langevin equations with noises $g^{\genU\indI}$ and $g^{\genV}_{\indI}$:
\bae{
	&\quad \ItoD\genU^\indI=\covD\genU^\indI+\frac{1}{2}\left(\Gamma^\indI_{\indJ\indS,\indK}-\Gamma^\indM_{\indJ\indS}\Gamma^\indI_{\indM\indK}\right)\genU^\indS A^{\genX\genX\indJ\indK}\dd N+\Gamma^\indI_{\indJ\indK}A^{\genX\covU\indJ\indK}\dd N, \\
	&\quad 	\ItoD\genV_\indI=\covD\genV_\indI-\frac{1}{2}\left(\Gamma^\indS_{\indI\indJ,\indK}+\Gamma^\indM_{\indI\indJ}\Gamma^\indS_{\indK\indM}\right)\genV_\indS A^{\genX\genX\indJ\indK}\dd N
	-\Gamma^\indK_{\indI\indJ}A^{\genX\covV\indJ}{}_\indK\dd N, \label{eq: ItoD for V}
}
where the quantities $A^{\genX\covU\indI\indJ}=g^{\genX\indI}_\indA g^{\covU\indJ}_\indA$ and $A^{\genX\covV\indI}{}_\indJ=g^{\genX\indI}_\indA g^\covU_{\indJ\indA}$ are the cross-correlations between the coordinate noise $g^{\genX\indI}_\indA$ and the covariant combinations of (co)vector noise:
\bae{
    g^{\covU\indI}_\indA=g^{\genU\indI}_\indA+\Gamma^\indI_{\indJ\indK}\genU^\indJ g^{\genX\indK}_\indA, \qquad g^\covV_{\indI\indA}=g^\genV_{\indI\indA}-\Gamma_{\indI\indJ}^\indK\genV_\indK g^{\genX\indI}_\indA.
}
Note also that the difference between these It\^o-covariant derivatives and usual covariant derivatives for vectors and covectors, $\covD\genU^\indI=\dd\genU^\indI+\Gamma^\indI_{\indJ\indK}\genU^\indJ \dd \genX^\indK$ and $\covD\genV_\indI=\dd\genV_\indI-\Gamma^\indJ_{\indI\indK}\genV_\indJ \dd \genX^\indK$, only contains terms proportional to noise amplitudes squared.

Had we obtained Langevin equations of the type~\eqref{eq: Langevin in heuristic} but with $\dd$ and $\covD$ replaced by $\ItoD$, then they would be covariant under field redefinitions (and induced redefinitions of momenta) if and only if, interpreted in It\^o. 
Actually, we will see 
in section~\ref{vielbeins-Ito} that exactly these It\^o-covariant derivatives emerge when interpreting our equations in the Stratonovich scheme, and reformulating them in the  It\^o-language. 
However for the moment, one should abandon the It\^o scheme together with equations~\eqref{eq: Langevin in heuristic}, as covariance would then be lost. Let us now discuss the second most popular discretisation.

\subsubsection{Stratonovich scheme} 
\label{Strato}

The midpoint, or Stratonovich, discretisation corresponds to $\discAlpha=1/2$. Physicists like it because it is intuitive to use in analytical calculations: as proved in Appendix~\ref{appendix: stochastic calculus}, the standard chain rule applies, hence it is easy to check the covariance of a given equation and straightforward to perform changes of variables. Saying it more trivially: when physicists make ``naive'' computations by applying standard rules in a stochastic context, like what we did in Sec.~\ref{sec: heuristic}, they are implicitly using the Stratonovich scheme.
As we can clearly see from the FP equation~\eqref{eq: covariant fokker-planck}, it is the only choice that respects covariance.
Again, this can be understood already at the level of Langevin equations, since because the standard chain rule applies, infinitesimal variations $\dd \phiIR^\indI$ and $\covD \piIR_\indI$ 
are well vectors and covectors of the field space. 
Nonetheless, although general covariance is respected, this description is not yet satisfactory.
Indeed, when the noise is multiplicative (which it is in most interesting scenarios), the second term in Eq.~\eqref{eq: covariant fokker-planck} depends explicitly on the identification of an orthonormal frame for the noises, through the appearance of $g^{\genX\indI}_\indA$.
However, the only outcome of our derivation for stochastic inflation so far has been the auto-correlation of the effective noises, for instance $\braket{\xi^{Q\indI}(N)\xi^{Q\indJ}(N^\prime)}=\frac{\dotksigma}{\ksigma}\calP^{QQ\indI\indJ}(\ksigma)\delta(N-N^\prime)$ in a given $\sigma$-Hubble patch. 
Of course, it is always possible to reduce the auto-correlation matrix in a frame where it is diagonal,
i.e. to find a ``square-root matrix'' $g_A^{\genX\indI}$ verifying $\delta^{AB} g_A^{\genX\indI} g_B^{\genX\indJ}=A^{\genX\genX IJ}$.
However, such a frame is not unique, and an ambiguity remains: the physics described by the FP equation~\eqref{eq: covariant fokker-planck} depends on the choice of this frame. This is easily seen if, after a choice $g^{\genX\indI}_\indA$, one performs a rotation to another orthonormal frame in which the noise is diagonal again, with an orthonormal matrix 
$\Rot^{\Bt}_{\phantom{\Bt}\indA}$ such that the ``square-root matrix'' of the noise correlations changes without affecting its auto-correlation: $g_\indA^{\genX\indI}=\Rot^{\Bt}_{\phantom{\Bt}\indA}\gt_{\Bt}^{\genX\indI}$. Then, the second term in the FP equation transforms as
$\N_\indI\left[g_\indA^{\genX\indI} \N_\indJ(g_\indA^{\genX\indJ} P_s)\right]=\N_\indI\left[\Rot^{\Bt}_{\phantom{\Bt}\indA} \Rot^{\Ct}_{\phantom{\Bt}\indA}\gt_{\Bt}^{\genX\indI} \N_\indJ(\gt_{\Ct}^{\genX\indJ} P_s)\right]+ \N_\indI\left[\Rot^{\Bt}_{\phantom{\Bt}\indA}\left(\N_\indJ \Rot^{\Ct}_{\phantom{\Bt}\indA} \right)\gt_{\Bt}^{\genX\indI}\gt_{\Ct}^{\genX\indJ}  P_s \right]$.
Because the matrix $\Rot$ is orthonormal, $\Rot^{\Bt}_{\phantom{\Bt}\indA} \Rot^{\Ct}_{\phantom{\Bt}\indA}=\delta^{\Bt\Ct}$, the result would be frame-independent if there was only the first of these two terms.
However since $\Rot$ can depend on the position in field space, its (covariant) derivative is not zero and there is no reason in general for $\Rot^{\Bt}_{\phantom{\Bt}\indA}\left(\N_\indJ \Rot^{\Ct}_{\phantom{\Bt}\indA} \right)$
to vanish, hence the frame-dependence of the result.
Actually this difficulty holds for all discretisation schemes except when $\discAlpha=0$, the It\^o case where this second term in Eq.~\eqref{eq: covariant fokker-planck} is killed. That also explains why the It\^o scheme is often preferred in numerical implementations: 
it is possible to use an algorithm to reduce the noise correlation matrix in an orthonormal frame, and the result does not depend on the choice of such frame.
However this apparently unsolvable ambiguity, breaking of covariance in It\^o or spurious frame-dependence in Stratonovich, is solved by the understanding that in stochastic inflation there is actually a preferred frame in which the noise is diagonal, and that it is given by the basis of independent creation and annihilation operators of the quantum UV modes. This result has close links to the classicalisation of light scalar fields on super-Hubble scales as we shall see now.

\subsection{Classicalisation and frame of independent creation and annihilation operators} 
\label{subsec: classicalisation}

As is well known in the context of multifield inflation (see e.g. Refs.~\cite{Salopek:1988qh,GrootNibbelink:2001qt,Tsujikawa:2002qx,Weinberg:2008zzc}), the quantum operators $\hat{Q}^\indI$ and $\hat{\covP}_\indI$ should be decomposed on a $N_\mathrm{fields}$-dimensional set of independent creation and annihilation operators (labed by the index $\indA$) as:
\bae{
    \bce{
        \dps
        \hat{Q}^\indI(N,\mathbf{k})=Q^\indI_\indA(N,k)\hat{a}^\indA_\mathbf{k}+
        \Bigl(Q^{\indI}_\indA(N,k)
        \Bigr)^*\hat{a}^{\indA \dagger}_{-\mathbf{k}}, \\[10pt]
        \dps
        \hat{\covP}_\indI(N,\mathbf{k})=\covP_{\indI\indA}(N,k)\hat{a}^\indA_\mathbf{k}+\left(\covP_{\indI\indA}(N,k)\right)^*\hat{a}^{\indA \dagger}_{-\mathbf{k}},
    }
    \quad \text{with } \left[\hat{a}^\indA_\mathbf{k},\hat{a}^{\indB \dagger}_\mathbf{k^\prime}\right]=(2\pi)^3 \delta^{\indA\indB}\delta^{(3)}(\mathbf{k}-\mathbf{k^\prime}),
    \label{quantisation}
}
where note that indices $\indA, \indB \ldots$ can be raised and lowered with the symbol $\delta_{\indA\indB}$, so that the up or down position has no particular meaning. One should therefore follow the evolution of the $2N_\mathrm{fields}^2$ complex mode functions $Q^\indI_\indA(N,k)$ and $\covP_{\indI\indA}(N,k)$ verifying the first order differential equations $E^{Q\indI}=0=E^{\covP}_{\indI}$~\eqref{eq: EQ}--\eqref{eq: EP}, or equivalently solve $N_\mathrm{fields}$ times (corresponding to the label $\indA$) the coupled $N_\mathrm{fields}$-dimensional system of second-order differential equations~\eqref{eq: UV eom mode} verified by each set of $Q^\indI_\indA(N,k)$, each time with different initial conditions. 
This stems from the fact that in order to define a vacuum state $\ket{0}$ and to quantise the fluctuations when all relevant momenta are sub-Hubble, one should identify $N_\mathrm{fields}$ independent fields, each coming with its own creation and annihilation operators.
Note that in a generic system of coordinates or/and curved field space, the field fluctuations $\hat{Q}^\indI$ do not verify the above property, as they are kinetically coupled. However, their projections on a set of vielbeins, or even clearer, on a set of parallel-transported vielbeins (see Sec.~\ref{vielbeins-Ito}), naturally provide independent degrees of freedom inside the Hubble radius.

Relatedly, let us remark that even with a fixed vacuum state annihilated by the $\hat{a}^\indA_\bfk$'s, 
there is no unique choice of independent operators verifying the commutation relations in Eq.~\eqref{quantisation}. Indeed, once such a set is given, any other one related by a unitary transformation $\U$ provides another suitable set, i.e. 
the equations~\eqref{quantisation} take the same form in terms of the barred quantities such that
\bae{
    &\hat{\bar{a}}^{\At}_\bfk=\U^{\At}{}_\indB\hat{a}^{\indB}_\bfk \qquad \text{and} \label{rotation-a-operators} \\
    &\barQ^\indI_\At=Q^\indI_\indB(\U^\dagger)^\indB{}_\At, \quad \bar{\covP}_{\indI\At}=\covP_{\indI\indB}(\U^\dagger)^\indB{}_\At\,, \label{rotation-basis}
}
without changing neither the operators $\hat{Q}^\indI$ and $\hat{\covP}_\indI$ nor the vacuum state $\ket{0}$.
This arbitrariness is of course equivalently visible at the level of the quantisation conditions.
Indeed, the commutation relations 
\bae{\label{eq: commutation relations}
    &[\hat{Q}^\indI(N,\mathbf{x}),\hat{\covP}_\indJ(N,\mathbf{x}^\prime)]=\frac{i \delta^I_J}{a^3(N)}  \delta^{(3)}(\mathbf{x}-\mathbf{x}^\prime) \,,  \\
    &[\hat{Q}^\indI(N,\mathbf{x}),\hat{Q}^\indJ(N,\mathbf{x}^\prime)]=[\hat{\covP}_\indI(N,\mathbf{x}),\hat{\covP}_\indJ(N,\mathbf{x}^\prime)]=0,
}
impose the following relations on the mode functions:
\bae{
    &Q^\indI_\indA(N,k) \covP^*_{\indJ\indA}(N,k)-\textrm{c.c.}=\frac{i \delta^I _J}{a^3(N)}, \label{i-part}\\
    &Q^\indI_\indA(N,k)Q^{*\indJ}_\indA(N,k)-\textrm{c.c.}=\covP_{\indI\indA}(N,k) \covP^*_{\indJ\indA}(N,k)-\textrm{c.c.}=0, \label{Q-quantisation}
}
where, as before, the sum over $\indA$ is implicit.\footnote{As usual, these relations, once verified at some initial time, hold at all time by virtue of the equations of motion verified by the mode functions.} It is then apparent that two sets of mode functions related by a time-independent unitary matrix like in Eq.~\eqref{rotation-basis} are equally valid and describe the same physics.

Once this quantisation is in place, the two-point vacuum expectation value of the quantum UV operators $\hat{Q}^{\covX\indI}=\left(\hat{Q}^\indI,\hat{\covP}^\indI\right)$ at a given time $N$ are given by 
\bae{
    \braket{ 0| \hat{Q}^{\covX\indI}(N,\mathbf{k})\hat{Q}^{\covY\indJ}(N,\mathbf{k^\prime}) | 0}=(2\pi)^3 \delta^{(3)}(\mathbf{k}+\mathbf{k^\prime}) \delta^{AB} Q^{\covX\indI}_\indA(N,k)\left(Q^{\covY\indJ}_\indB(N,k)\right)^*,
    \label{two-point-vacuum}
}
thus providing the power spectra entering into the properties of the noises~\eqref{eq: noise correlation}.
However, to describe their statistics at time $N$, only the mode $\ksigma(N)$ matters.
Crucially, this mode is far outside the Hubble radius for $\sg \ll 1$, which we indeed considered from the start to ensure that the gradient-expansion is valid and that the infrared fields can be considered as classical (see Sec.~\ref{gauge-smoothing}).
Relatedly, in this regime, 
the complex mode functions $Q^\indI_\indA(N,k)$ and $\covP_{\indI\indA}(N,k)$ become real to a very good accuracy
(up to an irrelevant constant unitary matrix), corresponding to fluctuations being in a highly squeezed state~\cite{PhysRevD.42.3413,Albrecht_1994,Polarski:1995jg,KIEFER_1998}. This property is well known for a single light scalar field, and we will see in Sec.~\ref{sec:Markovian} that it also holds for multiple scalar fields in the massless approximation. More interestingly, we also show there that it is actually valid in the slow-varying approximation for light scalars of masses $m_i<3/2 H$ (see Sec.~\ref{sec:generic}), the situation of interest for the stochastic formalism.\footnote{In this framework, the presence of heavy degrees of freedom with masses $m_i>3/2 H$, for which Eq.~\eqref{eq: squeezing} is not applicable (because the mode functions of heavy fields have genuinely time-dependent phases on super-Hubble scales),
is not problematic, as their mode functions are anyway $\sigma$-suppressed at coarse-graining scale crossing.}
Using this, it is thus possible to forget about the complex conjugates ${}^*$ and to 
consider:
\bae{\label{eq: squeezing}
    \hat{Q}^\indI(N,\mathbf{k})\underset{k\ll aH}{\simeq}Q^\indI_\indA(N,k)\left(\hat{a}^\indA_\mathbf{k}+\hat{a}^{\indA \dagger}_{-\mathbf{k}}\right), \qquad
    \hat{\covP}_\indI(N,\mathbf{k})\underset{k\ll aH}{\simeq}\covP_{\indI\indA}(N,k)\left(\hat{a}^\indA_\mathbf{k}+\hat{a}^{\indA \dagger}_{-\mathbf{k}}\right).
}
It is then natural to define the variables
$\bb_\mathbf{k}^A=\hat{a}^\indA_\mathbf{k}+\hat{a}^{\indA \dagger}_{-\mathbf{k}}$ where we forgot the hat on purpose. 
Indeed, these are the only ``quantum" operators that we are left with on super-Hubble scales and they all commute with one another, i.e. 
$ \left[\bb_\mathbf{k}^A,\bb^{\indB}_\mathbf{k^\prime}\right]=0$, hence the fluctuations can be understood as classical.\footnote{Of course, the canonical commutation relations for fields and momenta still hold, and whether cosmological perturbations completely lost their quantum nature or not is a field of research that has its own dedicated literature, see e.g.  Refs.~\cite{SUDARSKY_2011,Martin_2016,martin2019cosmic,ashtekar2020emergence} for recent references. In this paper, we shall be conservative and consider that super-Hubble fluctuations can well be treated classically.}
It can easily be checked that this definition of the $\bb_\mathbf{k}^A$'s endows them with
Gaussian statistics with
$\braket{\bb^A_\mathbf{k}}=0$ and $\braket{\bb^A_\mathbf{k}\bb^B_\mathbf{k^\prime}}=(2\pi)^3\delta^{AB}\delta^{(3)}(\mathbf{k}+\mathbf{k^\prime})$, where the brackets of quantum vacuum expectation values can now be understood as statistical ensemble averages for the stochastic fields $\bb_\mathbf{k}^A$.
The noises~\eqref{eq: noises} can now be expressed as
\bae{
    &\xi^{Q\indI}(x)=\fsigma Q^\indI_\indA(N,\ksigma(N)) \xi^\indA(x), \quad
    \xi_{\covP}^{\indI}(x)= \fsigma \covP_{\indI\indA}(N,\ksigma(N)) \xi^\indA(x), \text{ with }  \fsigma=\sqrt{\frac{\ksigma^3}{2\pi^2}\frac{\dotksigma}{\ksigma}},
}
where, again, the ratio $\ksigma^\prime/\ksigma$ may be approximated by
unity. We also defined
\bae{ 
    \xi^\indA(x)=\fsigma^{-1}\int \dk \ee^{i \mathbf{k} \cdot \mathbf{x}}\dif{\theta(k-\ksigma(N))}{N}\bb^A_\mathbf{k},
}
that are independent Gaussian white noises normalised to almost unity in a given $\sg$-Hubble patch:
\bae{
    \braket{\xi^\indA(x) \xi^\indB(x^\prime)} = 
    \frac{\sin{\ksigma r}}{\ksigma r} \delta^{AB} \delta(N-N^\prime), \text{ with } r=|\mathbf{x}-\mathbf{x^\prime}|.
}
Recalling that the spatial correlation $\sin{\ksigma r}/\ksigma r$ should be approximated by the theta function $\theta(1-\ksigma r)$ taking values $1$ inside a $\sigma$-Hubble patch and $0$ outside, one eventually finds
\bae{\label{eq: normalized-independent-noises}
    \braket{\xi^\indA(x) \xi^\indB(x^\prime)} = 
    \begin{cases}
    \delta^{AB} \delta(N-N^\prime),& \text{if } r=|\mathbf{x}-\mathbf{x^\prime}| \leq \left(\sigma a H \right)^{-1}, \\
    0,              & \text{otherwise.}
\end{cases}
}
Strikingly, these are the same noises
$\xi^\indA$ that appear both in $\xi^{Q\indI}$ and $\xi^\indI_{\covP}$. This means that we were able  to ``decorrelate" the $2N_\mathrm{fields}$ correlated noises by expressing them in terms of $N_\mathrm{fields}$ uncorrelated ones. In a mathematical language, one would say that the noise amplitude
$A^{\covX\covY\indI\indJ}$, with $(\covX,\covY) \in (Q,\covP)$, can be understood as a bilinear form whose matrix is of dimension $2N_\mathrm{fields} \times 2N_\mathrm{fields}$, but of rank $N_\mathrm{fields}$ only, and can thus be reduced. 
The Langevin equations~\eqref{eq: Langevin in heuristic} are thus rewritten as
\bae{\label{eq: Langevin independent noise}
	\phiIR^{\indI\prime}=\frac{1}{H}G^{\indI\indJ}\piIR_\indJ+ \fsigma Q^\indI_\indA \circ \xi^A, \quad\quad
	\covD_N\piIR_\indI=-3\piIR_\indI-\frac{V_\indI}{H}+\fsigma {\covP}_{\indI\indA} \circ \xi^A,
}
where now the Stratonovich interpretation, indicated by the simple symbol $\circ \equiv\circ_{1/2}$, is non-ambiguous as independent noises $\xi^A$ have been identified, and where covariance is respected as the standard chain rule applies.
To be precise, there is strictly speaking
a family of possible independent noises $\xi^\indA$, but, taking into account that we used real $(Q^\indI_\indA,\covP_{\indI \indA})$ variables in Eq.~\eqref{eq: Langevin independent noise}, these noises are simply related by a constant rotation $U$ (when restricting Eq.~\eqref{rotation-a-operators} to real orthogonal matrices). Like in Eq.~\eqref{rotation-basis}, this induces a constant rotation $U^T$
of the noises' amplitudes $(Q^I_A,\covP_{\indI\indA})$, which, as we have seen in Sec.~\ref{Strato}, does not lead to any ambiguity. Modulo this irrelevant rotation, we can hence talk about \textit{the} frame of independent noises used in the Langevin-type equations~\eqref{eq: Langevin independent noise}.
Eventually, one has to be careful there with the covariant time derivative $\covD_N$, as it contains stochastic noises through the time derivative $\phiIR^{\indI\prime}$. It should hence also be discretised in the
Stratonovich scheme: 
\bae{\label{eq: Strato D_N}
    \covD_N\genU^\indI=\genU^{\indI\prime}+\Gamma^\indI_{\indJ\indK}\genU^\indJ\circ\phiIR^{\indK\prime}, \qquad \covD_N\genV_\indI=\genV_\indI{}^\prime-\Gamma_{\indI\indJ}^\indK\genV_\indK\circ\phiIR^{\indJ\prime}\,,
}
where the symbol $\circ$ indicates that when discretised, the term on its left should be evaluated at the midpoint, i.e. one has for instance:
\bae{\Gamma_{\indI\indJ}^\indK\genV_\indK\circ\phiIR^{\indJ\prime} \equiv \frac{G^{\indJ \indL}}{H}\Gamma_{\indI\indJ}^\indK\genV_\indK \piIR_\indL + \left(\fsigma \Gamma_{\indI\indJ}^\indK\genV_\indK  Q^\indI_\indA \right) \circ \xi^\indA\,.
}

Now that we understood that the frame of independent creation-annihilation operators provides the right frame in which to formulate the Langevin equations with a Stratonovich discretisation, all issues of covariance and frame-dependences are solved, but this resolution is still somewhat formal. In order to make this resolution more readily apparent, we will go one step further and derive equivalent Langevin-type equations in the It\^o scheme, which are easier to deal with numerically.

\subsection{It\^o-covariant Langevin equations}
\label{vielbeins-Ito}

In order to go from the Stratonovich Langevin equations~\eqref{eq: Langevin independent noise} to It\^o ones with the same physical content, one will introduce vielbeins defining a local orthonormal frame along the IR trajectory. This additional structure will eventually disappear from the final It\^o Langevin equations, while generating It\^o-covariant derivatives, but for this, one has to be careful about their definitions.
Let us first consider a given point in field space, say the initial condition for $\phiIR^I$
in a given realisation of the stochastic processes. At this point, it is possible to reduce the metric 2-form $G_{\indI\indJ}$ to identity $\delta_{\alpha\beta}$ by using projectors $e^\indI_\alpha$ from one basis to the other. Then they verify the following relations \emph{at this point}:
\bae{
    G_{\indI\indJ}e^\indI_\alpha e^\indJ_\beta=\delta_{\alpha \beta}, \quad  \text{and} \quad \delta^{\alpha\beta}e^\indI_\alpha e^\indJ_\beta=G^{\indI\indJ}.
}
For these variables to constitute a set of vielbeins, these relations should hold along the whole IR trajectory. We thus ask this property to be conserved along the trajectory, $\covD_N\left(G_{\indI\indJ}e^\indI_\alpha e^\indJ_\beta\right)=0$.
Because $\covD_N G_{\indI\indJ}=0$ by definition, if we write
$\covD_N e^\indI_\alpha = \Omega_{\alpha}^{\phantom{\alpha}\beta} e^\indI_\beta$, we find that the matrix $\Omega$ must be anti-symmetric, parameterizing the local rotation of the orthonormal frame. Then, which anti-symmetric matrix to choose is a matter of convenience. For example, a popular choice is the decomposition in the so-called adiabatic/entropic basis~\cite{Gordon:2000hv,GrootNibbelink:2001qt}, defined by a Gram-Schmidt orthogonalisation process applied to the successive covariant derivatives of $\phiIR^{\indI\prime}$, in which case the entries of the anti-symmetric matrix
correspond to covariant turn rates of the background trajectory in field space.
An even simpler choice in some sense is to use \emph{parallel-transported} vielbeins which verify $\covD_N e^\indI_\alpha=0$, i.e. to chose $\Omega=0$. These or other choices of vielbeins may have their own advantages for the analytical understanding of the behaviour of UV fluctuations (see sections~\ref{sec:light} and \ref{sec:generic}).
In the following, we make the choice $\Omega=0$ but we stress that this is merely for convenience, and that the resulting It\^o-covariant Langevin equations do not depend on this choice, as any set of vielbeins disappear altogether from the final result.

More important is to note that again, the covariant time derivative $\covD_N$ is a stochastic derivative, with an underlying discretisation corresponding to the Stratonovich scheme, as defined in Eq.~\eqref{eq: Strato D_N}. Therefore the parallel transport of vielbeins must be realised in the following stochastic way:
\bae{
    e^{\indI}_\alpha{}^\prime=-\Gamma^\indI_{\indJ\indK}e^\indJ_\alpha\circ\phiIR^{\indK\prime}.
}
We call these vielbeins \emph{stochastically-parallel-tranported} vielbeins, as this equation defining them is nothing but a Langevin equation. The vielbeins thus really become new stochastic variables, i.e., the collection of stochastic processes reads $\genS^\indn=\left(\phiIR^\indI,\piIR_\indI,e^\indI_\alpha\right)$: coordinates on the field space, covectors and vectors. Notice that with these definitions the indices $\alpha$ and $\beta$ can be raised and lowered with the metric $\delta_{\alpha\beta}$, i.e. the up or down position makes no difference. We define the projections of the UV modes along those vielbeins, $Q^\alpha_\indA$ and $\covP^\alpha_\indA$ as 
\bae{
    Q^\alpha_\indA=G_{\indI\indJ}e^{\indI\alpha}Q^\indJ_\indA, \qquad \covP^\alpha_\indA=e^{\indI\alpha}\covP_{\indI\indA}.
    \label{def-projected-perturbations}
}
These variables are scalars in field space and one deduces from Eq.~\eqref{eq: UV eom mode} and by virtue of our choice $\Omega=0$ that they verify the simple second-order differential equation:
\bae{\label{eq: UV eom mode vielbein}
    Q^{\alpha}_\indA{}^{\prime\prime}+\left(3-\epsilon\right)Q^{\alpha}_\indA{}^\prime+\left(\frac{k^2}{a^2H^2}\delta^\alpha_\beta + \frac{M^2{}^\alpha{}_\beta}{H^2} \right) Q^\beta_\indA=0,  \quad \text{with } M^2{}^\alpha{}_\beta =e_\indI^\alpha e^\indJ_\beta M^2{}^\indI{}_\indJ\,.
}
An advantage of the parallel-transported vielbeins is thus that the perturbations $Q^\alpha_\indA$ in such a basis are not kinetically coupled but only mix via the projection of the mass matrix, $M^2{}^\alpha{}_\beta$, which we will use for analytical estimates in Sec.~\ref{sec:Markovian}. This is of course equivalent to our statement in Sec.~\ref{subsec: classicalisation} that these projected fields are independent deep inside the Hubble radius, making easier the quantisation process.

Independently of this, let us now reformulate our system of Langevin equations~\eqref{eq: Langevin independent noise} with the new set of stochastic variables augmented by the vielbeins:
\bae{\label{eq: Langevin independent noise vielbein}
    \bce{
        \dps
	    \phiIR^{\indI\prime} & \dps \hspace{-5pt}
	    =\frac{1}{H}G^{\indI\indJ}\piIR_\indJ+\fsigma e^\indI_\alpha Q^\alpha_\indA\circ\xi^A, \\[5pt]
	    \dps
    	\piIR_\indI{}^\prime & \dps \hspace{-5pt}
    	=-3\piIR_\indI-\frac{V_\indI}{H}+\Gamma_{\indI\indJ}^\indK\piIR_\indK\circ\phiIR^{\indJ\prime}+\fsigma G_{\indI\indJ}e^\indJ_\alpha\covP^\alpha_\indA\circ\xi^\indA, \\[5pt]
    	e^{\indI}_\alpha{}^\prime & \dps \hspace{-5pt}
    	=-\Gamma^\indI_{\indJ\indK}e^\indJ_\alpha\circ\phiIR^{\indK\prime}.
    }
}
Although it is not manifest, we know these equations respect general covariance, and as
is proved in Appendix~\ref{appendix: stochastic calculus}, it is always possible to move from a given discretisation scheme to another one in the continuous description, by adding a noise-induced deterministic drift of the form (going from Stratonovich to It\^o), $\frac{1}{2}\left(\partial g^{\indn}_\indA/\partial \genS^\indm\right)g^{\indm}_\indA$
to the equation of motion for the process $\genS^\indn$. Let us then find the equivalent It\^o description of equations~\eqref{eq: Langevin independent noise vielbein}, 
keeping in mind that the $Q^\alpha_\indA,\covP_{\alpha\indA}$ are not given functions of the IR stochastic variables, but rather solutions of differential equations 
that involve them. Using Eq.~\eqref{eq: scheme conversion} with $\discAlpha=1/2$, we thus find:
\bae{
    \phiIR^{\indI\prime}&=\frac{1}{H}G^{\indI\indJ}\piIR_\indJ+\fsigma e^\indI_\alpha Q^\alpha_\indA\xi^\indA \nonumber \\
    &\qquad +\frac{\fsigma^2}{2}\left(\pdif{e^\indI_\alpha}{e^\indJ_\beta}Q^\alpha_\indA\right)\times\left(-\Gamma^\indJ_{\indK\indL}e^\indK_\beta e^\indL_\gamma Q^\gamma_\indA\right),
    &&\hspace{-10pt}\text{[noise of $e^\indJ_\beta$]} \label{Ito-first} \displaybreak[0] \\
    \piIR_\indI{}^\prime&=-3\pi_\indI-\frac{V_\indI}{H}+\frac{G^{\indJ \indL}}{H}\Gamma_{\indI\indJ}^\indK\piIR_\indK  \piIR_\indL 
    +\fsigma\left(G_{\indI\indJ}e^\indJ_\alpha\covP^\alpha_\indA+\Gamma_{\indI\indJ}^\indK\piIR_\indK e^\indJ_\alpha Q^\alpha_\indA   \right)\xi^\indA \nonumber \\
    &\qquad +\frac{\fsigma^2}{2}\left(\pdif{\Gamma_{\indI\indJ}^\indK}{\phiIR^\indL}\piIR_\indK e^\indJ_\alpha Q^\alpha_\indA+\pdif{G_{\indI\indJ}}{\phiIR^\indL}e^\indJ_\alpha\covP^\alpha_\indA\right)\times\left(e^\indL_\beta Q^\beta_\indA\right)
    &&\hspace{-10pt}\text{[noise of $\phiIR^\indL$]} \nonumber \\
    &\qquad +\frac{\fsigma^2}{2}\left(\Gamma_{\indI\indJ}^\indK\pdif{\piIR_\indK}{\piIR_\indL}e^\indJ_\alpha Q^\alpha_\indA\right)\times\left(G_{\indL\indS}e^\indS_\beta\covP^\beta_\indA+\Gamma_{\indL\indS}^\indR\piIR_\indR e^\indS_\beta Q^\beta_\indA\right) 
    &&\hspace{-10pt}\text{[noise of $\piIR_\indL$]} \nonumber \\
    &\qquad +\frac{\fsigma^2}{2} \pdif{e^\indJ_\alpha}{e^\indK_\beta}\left(G_{\indI\indJ}\covP^\alpha_\indA+\Gamma_{\indI\indJ}^\indK\piIR_\indK  Q^\alpha_\indA \right)\times\left(-\Gamma^\indK_{\indL\indM}e^\indL_\beta e^\indM_\gamma Q^\gamma_\indA\right),
    &&\hspace{-10pt}\text{[noise of $e^\indK_\beta$]} \displaybreak[0] \\
    e^{\indI}_\alpha{}^\prime&=-\frac{G^{\indK \indL}}{H}\Gamma^\indI_{\indJ\indK}e^\indJ_\alpha \piIR_\indL - \fsigma \Gamma^\indI_{\indJ\indK} e^\indJ_\alpha e^\indK_\beta Q^\beta_\indA \xi^A
    \nonumber \\
    &\qquad +\frac{\fsigma^2}{2}\left(-\pdif{\Gamma^\indI_{\indJ\indK}}{\phiIR^\indL}e^\indJ_\alpha e^\indK_\beta Q^\beta_\indA\right)\times\left(e^\indL_\gamma Q^\gamma_\indA\right) 
    &&\hspace{-10pt}\text{[noise of $\phiIR^\indL$]} \nonumber \\
    &\qquad +\frac{\fsigma^2}{2}\left(-\Gamma^\indI_{\indJ\indK}\pdif{(e^\indJ_\alpha e^\indK_\beta)}{e^\indL_\gamma} Q^\beta_\indA\right)\times\left(-\Gamma^\indL_{\indR\indS}e^\indR_\gamma e^\indS_\delta Q^\delta_\indA\right), 
    &&\hspace{-10pt}\text{[noise of $e^\indL_\gamma$]}
}
where the absence of a symbol $\circ_\alpha$ means that the underlying discretisation is the It\^o scheme. We recognise the appearance of the corrective terms $\propto A^{QQ \indI\indJ}$ and $A^{Q\covP \indI}{}_\indJ$ that are needed to make It\^o-Langevin equations covariant.
For instance, the last term in Eq.~\eqref{Ito-first}, coming from the Stratonovich to It\^o conversion, reads 
$-\frac12\fsigma^2 \Gamma^{\indI}_{\indK \indL} e^\indK_\alpha Q^\alpha_\indA e^\indL_\gamma Q^\gamma_\indA=-\frac12 \Gamma^{\indI}_{\indK \indL} A^{QQ \indK \indL}$
where
\bae{
    A^{QQ \indK \indL}=\fsigma^2 Q^\indK_\indA(N,\ksigma(N)) Q^\indL_\indA(N,\ksigma(N))=\frac{\ksigma^\prime}{\ksigma} {\cal P}^{QQ\indK\indL}(N,\ksigma(N)),
}
is intrinsically defined independently of the vielbeins (see Eqs.~\eqref{eq: noise correlation} and \eqref{eq: power spectra}).\footnote{Since we have taken into account the classicalisation of the perturbations on super-Hubble scales and considered the mode functions to be real, notice that here the cross-correlation $A^{Q\covP \indI}{}_{\indJ}=\fsigma^2 Q^\indI_\indA P_{\indJ\indA}$ is automatically real. 
Anyway, the reality of the auto-correlation of the noises will be rigorously proven from the path-integral approach in Sec.~\ref{sec: effective hamiltonian action}.}
With similar manipulations, one can 
rewrite the equations in It\^o with use of the covariant derivatives previously defined in Eqs.~(\ref{eq: ItoD for X})--(\ref{eq: ItoD for V}), as
\bae{\label{eq: covariant Langevin Ito}
    \bce{
        \dps
        \ItoD_N\phiIR^\indI & \dps \hspace{-5pt}
        =\frac{\piIR^\indI}{H}+\fsigma e^\indI_\alpha Q^\alpha_\indA \xi^{\indA}, \\
        \dps
        \ItoD_N\piIR_\indI & \dps \hspace{-5pt}
        =-3\piIR_\indI-\frac{V_\indI}{H}+ \fsigma e_\indI^\alpha\covP_{\alpha,\indA}\xi^{\indA}, \\
        \dps
        \ItoD_N e^\indI_\alpha & \dps \hspace{-5pt}
        =0.
    }
}
This self-consistency of the Langevin equations is quite remarkable given the degree of complexity of these stochastic differential equations.
In particular, as announced, the vielbeins disappear completely from any physical quantity computed from the first two equations in Eq.~\eqref{eq: covariant Langevin Ito}, as they do not appear in the It\^o-covariant derivatives, and as it is only the auto-correlation of the effective noises $(\xi^{Q\indI}=\fsigma e^\indI_\alpha Q^\alpha_\indA \xi^{\indA}= f_\sigma
Q^\indI_\indA \xi^{\indA},\,\xi^{\covP}_{\indI}=
\fsigma e_\indI^\alpha\covP_{\alpha,\indA}\xi^{\indA}=\fsigma \covP_{\indI,\indA}\xi^{\indA})$ that matter in It\^o. Moreover, the covariance of Eqs.~\eqref{eq: covariant Langevin Ito} is manifest.

Now that the question of ``stochastic anomalies" is solved, let us present a more rigorous derivation of the Langevin equations in phase space, with use of the coarse-grained effective hamiltonian action in a path-integral approach. As already mentioned, this will enable one to correctly treat the IR-UV interactions at the time of crossing the coarse-graining scale,  
so that all noises are manifestly real and that UV modes dictating their properties obey the same EoM as in SPT.

\section{Coarse-grained effective Hamiltonian action}\label{sec: effective hamiltonian action}

In this section, we derive the covariant Langevin-type equations of multifield stochastic inflation in a midpoint discretisation scheme, based on functional methods borrowed from non-equilibrium quantum field theory. We will begin by reviewing some of these notions and explaining the roadmap and principles of the computation, before turning to the computation itself. In particular, we will have to deal with several difficulties. First, as usual in cosmology we want to describe the fields dynamics, or their equal-time ``in-in" correlation functions. This is different from QFT in Minkoswki spacetime where we are interested in scaterring ``in-out" amplitudes. Thus, the time integration contour should follow a closed-time-path (CTP). 
Second, we want to compute an effective action for the long wavelength IR modes coupled to the bath of short wavelength UV modes, with interactions that should be specified upon physical arguments. In particular, we will solely consider the IR/UV couplings coming from the flow of UV modes joining the IR sector, i.e. the couplings specific to the time-dependent split between the two sectors.
Last but not least, we want to pay particular attention to the covariance of the theory. 
On the UV side, it is known that the perturbations $\dphi^\indI$ do not transform beyond linear order as genuine vectors under field redefinitions. In SPT, this subtlety is only relevant when computing the action at cubic order or higher in perturbations. However, in a stochastic context, we have to take this into account even when considering the action at quadratic order in fluctuations, as the part of the action that is linear in $\dphi^\indI$ does not vanish, but rather plays a crucial role in determining the UV-IR transition. 
This also applies to momentum perturbations $\dpi_\indI$ that do not even transform as covectors at linear order, and for which quadratic corrections are also needed.
On the IR side, based on the arguments developed in the previous section, we will also interpret our path integral as the continuous limit of discrete integrations with an explicit scheme corresponding to the midpoint (Stratonovich-like) discretisation, to ensure that the standard chain rule for changes of variables 
is appplicable.

\subsection{Roadmap and principles of the computation}

\begin{figure}
    \centering
    \includegraphics[width=0.9\hsize]{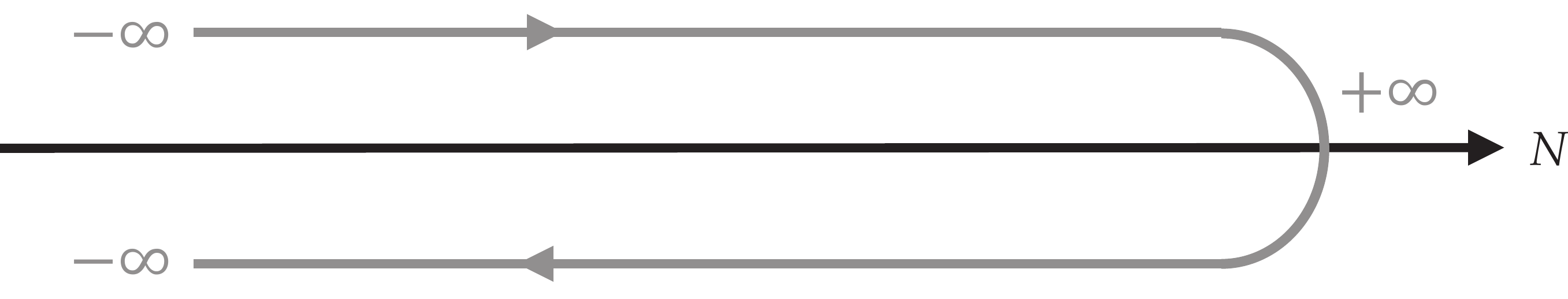}
    \caption{ Closed-time-path $C$ of integration used in the ``in-in" formalism.} 
   \label{fig: CTP}
\end{figure}

In particle physics, one wishes to compute transition amplitudes between asymptotic ``in" states (in the far past) and ``out" states (in the far future), which
are defined long before interactions are switched on, and long after they are switched off. The situation is crucially different in cosmology, where one is rather interested in vacuum expectation values of quantum operators. Actually, the notion of a future, asymptotic, ``out" state is more intricate in a cosmological context as particles keep interacting at least gravitationally in curved spacetimes; boundary conditions can only be imposed in the far past, when the wavelengths of relevant fields are much smaller than the Hubble radius. Expectation values in such time-dependent contexts can be deduced from the ``in-in" partition function (rather than the ``in-out" one for particle physics) in the presence of external currents, $Z\left[J_{\indX\indI}\right]$, which is the generating functional of all correlation functions defined as expectation values in the initial 
(vacuum) state. This ``in-in" generating functional, 
which can be thought of as summing over all possible ``out" states, can be computed using a closed-time-path contour of integration~\cite{Schwinger:1960qe,Keldysh:1964ud,Jordan:1986ug,Calzetta:1986ey,2009AdPhy..58..197K,
Weinberg:2005vy,Calzetta:2008iqa} shown in Fig.~\ref{fig: CTP}, and according to:
\bae{
	Z\left[\JUV_{\indX\indI}\right]=
	\int_C\mD \phi^{\indX\indI}\exp\left(i\SHam\left[\phi^{\indX\indI}\right]+i\int\dd^4x\JUV_{\indX\indI}\phifull^{\indX\indI}\right)\,,
	\label{ZJ}
}
where the subscript $C$ precises the contour of integration. Here, in accordance with first principles in the path-integral approach, $\SHam$ denotes the classical action in the Hamiltonian form~\eqref{eq: Hamiltonian action}, and notations are similar as before: the index $X$ denotes position or momentum in phase space, i.e. the Hamiltonian action depends on the scalar fields
$\phi^{Q\indI}=\phi^I$ and on their (contravariant) momenta $\phi^{P\indI}=G^{IJ}(\phi^\indK)\pi_\indJ$. 
Formally, it will prove useful for computations to lower and raise the index $X$ with the appropriate metric $\frac{1}{i}\sigma_{2XY}=\big(\begin{smallmatrix}0&-1\\1&0  \end{smallmatrix}\big)$ and its inverse. 
Note that contrary to ``in-out" partition functions in which $Z[0]$ must be computed as the sum over all vacuum bubbles to enforce a correct normalisation, for ``in-in" partition functions, it is trivial to see that $Z[0]=1$ as the norm of the initial vacuum state should be.
Eventually, we note that the condensed notation $\mD\phi^{XI}$ that we will use throughout should really be understood as the canonical phase-space measure $\prod_{I,J} \mD\phi^I\mD\pi_J$.

To avoid the formal path integral along the closed contour $C$, one can divide it in two path integrals over the forward ($+$) and backwards ($-$) parts of the time contour $C=C^+\cup C^-$.
This boils down to doubling the number of degrees of freedom in the path integral,
with $\phifull^{\indX\indI\pm}$ living respectively on $C^+$ and $C^-$. The $\pm$ fields and momenta should be considered independent, except
at future infinity where the time path closes (truly, at any time later than the ones of interest), and where we use the usual boundary conditions that $\pm$ fields coincide, i.e. $\phifull^{\indI+}(+\infty)=\phifull^{\indI-}(+\infty)$, but that momenta $\pifull_\indI^\pm$ are left unconstrained.
The time flow being reversed on the $C^-$ branch, the path integral to perform can be rewritten along a forward contour
only as
\bae{
	Z\left[\JUV_{\indX\indI}^\pm\right]=\int_{C^+}\mD\phifull^{\indX\indI\pm}\exp\left(i\SHam\left[\phifull^{\indX\indI+}\right]-i\SHam\left[\phifull^{\indX\indI-}\right]+i\int\dd^4x\JUV_{\indX\indI}^+\phifull^{\indX\indI+}-i\int\dd^4x\JUV_{\indX\indI}^-\phifull^{\indX\indI-}\right).
}
In practice, we will make use of the so-called Keldysh basis
(letting aside the various indices here):
\bae{
    \bpme{
       \phifull^\KelC \\
        \phifull^\KelD
    }=\bpme{
        1/2 & 1/2 \\
        1 & -1
    }\bpme{
        \phifull^+ \\
        \phifull^-
    } \quad \Leftrightarrow \quad
    \bpme{
        \phifull^+ \\
        \phifull^-
    }=  \underbrace{  \bpme{
        1 & 1/2 \\
        1 & -1/2
    }}_{\textstyle K}\bpme{
        \phifull^\KelC \\
        \phifull^\KelD
    },
    \label{classical-quantum-def}
}
where $\phifull^\KelC$ and $\phifull^\KelD$ are respectively referred to as the classical and quantum components of the fields, and $K$ is the matrix of change of basis. The rationale for this denomination is that among the solutions of the saddle point equations for the Keldysh action $\SHam\left[\phifull^{\indX\indI+}\right]-\SHam\left[\phifull^{\indX\indI-}\right]$, there is always one with vanishing quantum component and classical component obeying the classical equation of motion $\delta S/\delta \phifull^{\indX\indI}=0$.
Although we are using natural units with $\hbar=1$, one can intuitively think of the quantum component as
$\hbar$-suppressed, and indeed the stochastic equations we will derive, with classical equations of motion corrected by noises of quantum-mechanical origin, can be seen as semi-classical equations of motion.

Introducing covariant notations, latin indices $\inda$, $\indb$,  $\cdots$ for the $\pm$ fields, and Fraktur indices $\Kela$, $\Kelb$, $\cdots$ for the Keldysh label $\KelC/\KelD$, the corresponding change of basis can be summarised as $\phi^\inda=K^\inda{}_\Kela \phi^\Kela$.
To keep compact expressions, we use
the convention of summation over repeated indices, and we will use well-chosen metrics to raise and lower them:
the metric $\sigma_{3}{}_{\inda\indb}=\diag(1,-1)_{\inda\indb}$ in the $\pm$ basis; and the corresponding one $\sigma_1{}_{\Kela\Kelb}=\left(\begin{smallmatrix}0&1\\1&0\end{smallmatrix}\right)_{\Kela\Kelb}$ in the Keldysh basis. Note that in the latter, the matching condition of the CTP branches at future infinity reads $\phifull^{\indI\KelD}(+\infty)=0$, again with no constraint on momenta.
The generating functional  in this basis reads (note that the Jacobian $|K|$ is unity)
\bae{
    Z\left[\JUV_{\indX\indI}^\Kela\right]&=\int\mD\phifull^{\indX\indI\Kela}\exp\left(i\SHam\left[\phifull^{\indX\indI\Kela}\right]+i\int\dd^4x\JUV_{\indX\indI}^\KelD\phifull^{\indX\indI\KelC}+i\int\dd^4x\JUV_{\indX\indI}^\KelC\phifull^{\indX\indI\KelD}\right),
}
where $\SHam\left[\phifull^{\indX\indI\Kela}\right]=\SHam\left[\phifull^{\indX\indI\KelC}+\phifull^{\indX\indI\KelD}/2\right]-\SHam\left[\phifull^{\indX\indI\KelC}-\phifull^{\indX\indI\KelD}/2\right]$.

\bfe{width=0.9\hsize}{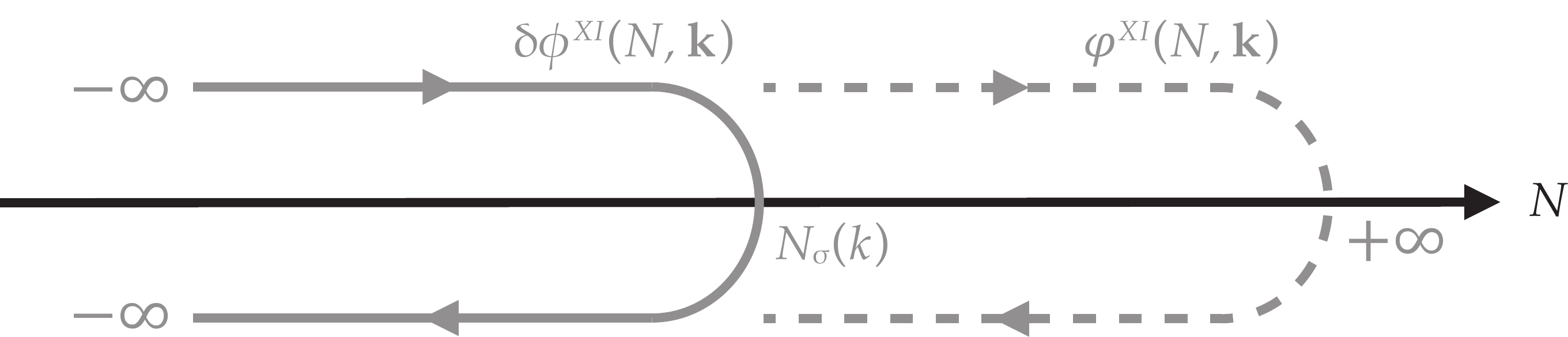}{
 In the stochastic approach, the original path integral along the closed contour $C$ shown in Fig.~\ref{fig: CTP} is divided for each wavenumber $\bfk$
into a path integral over UV fields $\dphi^{\indX\indI}(N,\bfk)$, and into one over IR fields $\phiIR^{\indX\indI}(N,\bfk)$. Hence, the corresponding UV path is closed at the transition time $N_\sg(k)$ with the boundary condition $\dphi^{\indI\KelD}(N_\sg(k),\bfk)=0$~\eqref{eq: boundary conditions}, while the IR path starts there with the other boundary condition $\phiIR^{\indI\KelC}(N_\sg(k),\bfk)=0$, and is closed at future infinity with the boundary condition $\phiIR^{\indI\KelD}(+\infty)=0$.}{fig: CTP_UVIR}

Formally, the ``in-in" vacuum expectation value of any operator at time $t_\star$ can be computed by introducing this operator at this particular time on the closed-time-path of integration and with vanishing currents. Equivalently, $n$-point functions can be derived by calculating the $n^{\mathrm{th}}$ functional derivatives of the above generating functional $Z\left[\JUV_{\indX\indI}^\Kela\right]$ with respect to the external currents $\JUV_{\indX\indI}^\Kela$ and evaluating them at $\JUV_{\indX\indI}^\Kela=0$. 
As for the equations of motion verified by the expectation values of $\phi^{\indX\indI}$, they can be determined by extremising the quantum effective action, defined as the Legendre transformation of $W\left[\JUV_{\indX\indI}^\Kela\right]=-i \ln Z\left[\JUV_{\indX\indI}^\Kela\right]$.
However in the following we choose to follow another route: based on the physical distinction between quantum short-wavelength modes and classical long-wavelength ones, we want to derive
a classical (albeit stochastic) effective theory for the latter only, and then compute expectation values within this new theory.
This amounts in our setup to derive what can be called the ``coarse-grained effective Hamiltonian action" that governs the dynamics of the coarse-grained scalar fields in a Hamiltonian language (see e.g. Refs.~\cite{Calzetta:1999zr,Calzetta:2008iqa,Vacca:2012vt} for related concepts).
Indeed, based on the scale separation provided by the physical Hubble radius $H^{-1}$, the original fields can be written in real space as IR+UV:
$\phi^{\indX\indI\Kela}(x)=\varphi^{\indX\indI\Kela}(x)+\delta\phi^{\indX\indI\Kela}(x)$, where we have in mind the Fourier cutoff $\ksigma(N)$ discussed in Sec.~\ref{sec: heuristic}.\footnote{Note that the cutoff $\ksigma(N)=\sigma a(N)H$ is not deterministic because the Hubble parameter depends on the stochastic realisations of the fields. Hence, the cutoff scale has the same status in the path-integral and in the heuristic approaches, i.e. it is understood to be defined self-consistently as mentioned at the end of Sec.~\ref{gauge-smoothing}. The following discussion is independent of this subtlety.} The Fourier components of the fields and momenta thus verify:
\bae{\label{eq: decompositon Fourier}
\phifull^{\indX\indI\Kela}\left(N,\bfk\right)=\left\{
    \begin{array}{ll}
        \delta \phi^{\indX\indI\Kela}\left(N,\bfk\right),  & \text{ if } N<\tsigma(k),  \\
        \phiIR^{\indX\indI\Kela}\left(N,\bfk\right),  & \text{ if } N>\tsigma(k),
    \end{array}
\right.
}
where $\tsigma(k)$ represents the time at which the modes of modulus $k$ cross the UV/IR cutoff, and hence at which boundary conditions need to be specified for the fields. 
Because the UV modes $\delta \phi^\Kela(N,\bfk)$ stop being defined at the time $\tsigma(k)$, their time path actually closes at this particular time, which enforces the boundary condition 
$\delta \phi^{\indI\KelD}\left(\tsigma(k),\bfk\right)=0$, 
like for the full fields whose time path closes at future infinity.
Inversely, the time path of IR modes $\phiIR^\Kela(N,\bfk)$ 
begins at that time, with vanishing initial conditions for the classical component of the 
fields, $\phiIR^{\indI\KelC}\left(\tsigma(k),\bfk\right)=0$ (see Fig.~\ref{fig: CTP_UVIR}).\footnote{As we will see, the stochastic equations derived heuristically in section~\ref{sec: heuristic} actually concern the classical component of the fields, so that the boundary condition $\phiIR^{\indI\KelC}\left(\tsigma(k),\bfk\right)=0$ agrees with the fact that in the stochastic approach, IR fields with wavevectors $\bfk$ do not exist before the time $\tsigma(k)$.}
Note that again, neither IR nor UV momenta are constrained at the time $\tsigma(k)$.

Because these conditions will be crucial to specify interactions between IR and UV modes, we rewrite them together:
\bae{\label{eq: boundary conditions}
\phifull^{\indI\Kela}\left(\tsigma(k),\bfk\right)=\left\{
    \begin{array}{ll}
         \delta \phi^{\indI\KelC}\left(\tsigma(k),\bfk\right), & \text{ if } \Kela=\KelC,  \\
        \phiIR^{\indI\KelD}\left(\tsigma(k),\bfk\right),  & \text{ if } \Kela=\KelD,
    \end{array}
\right.
}
assigning the Fourier component at the transition time,  either fully to the UV part for the classical component, or fully to the IR part for the quantum component.
It will become clear when discussing IR-UV interactions in the discretised version of the path integral 
that, indeed, no boundary condition is required for the momenta,
because in the path integral they can be evaluated at intermediate time steps and we can avoid to specify their values at the exact time $N_\sigma(k)$. 
It was actually shown in the context of a single test scalar field in de Sitter that these boundary conditions at $N_\sigma(k)$ enables the to-be-found stochastic description to correctly 
reproduce the propagators of the corresponding free QFT~\cite{Tokuda:2017fdh,Tokuda:2018eqs}.
Now that this decomposition into IR and UV fields is fully specified, one may rewrite 
the generating functional (at vanishing currents for simplicity) as
\bae{
    Z&=\int\mD\phiIR^{\indX\indI\Kela}\exp\left(i\Seff\left[\phiIR^{\indX\indI\Kela}\right]\right), \qquad \text{with} \label{Z} \\ \exp\left(i\Seff\left[\phiIR^{\indX\indI\Kela}\right]\right)&=\int\mD\dphi^{\indX\indI\Kela}\exp\left(i\SHam\left[\phiIR^{\indX\indI\Kela}+\dphi^{\indX\indI\Kela}\right]\right), \label{Seff}
}
where the path integral over the UV modes has to be performed explicitly.
Then, we will see that upon the introduction of auxiliary stochastic variables describing possible deviations from the classical EoM, one needs not perform the path integral over the IR fields, but simply observe which IR trajectories have non-zero weights in the remaining path integral, and hence obtain the desired Langevin equations.
However before that, let us  note that Eq.~\eqref{Seff} provides only a ``naive" expression, and that one has to be careful  about the fact that field perturbations themselves do not transform covariantly under field redefinitions beyond linear order~\cite{Vilkovisky:1984st,Gong:2011uw}. 
To ensure that the resulting effective theory respects general covariance, the path integral should be expressed in terms of covariant objects, and we now turn to the identifications of suitable ones in our Hamiltonian formulation.

\subsection{Covariant perturbations in the Hamiltonian language}
\label{subsec: covariant perturbations}

It is well known that field perturbations are not covariant objects beyond linear order. This subtlety is usually irrelevant if one is only interested in the Gaussian properties of the inflationary fluctuations, because SPT is defined around homogeneous fields $\phibg^{\indX \indI}(N)$ that 
solve the classical equations of motion, $\frac{\delta \SHam}{\delta \phi^{\indX \indI}}\big\rvert_{\phibg^{\indX\indI}}=0$, and any non-covariant contribution coming from the linear action in terms of $\delta \phi^{\indX\indI}$ thus vanishes.
However, the aim of stochastic inflation is precisely to take into account the difference between the effective equations of motion verified by the coarse-grained scalar fields and the classical equations of motion verified by $\phibg^{\indX \indI}(N)$ in SPT. In this context, the part of the action that is linear in $\delta \phi^{\indX\indI}$ not only does not vanish but actually plays a crucial role. Thus, in stochastic inflation, perturbations should be covariant objects at least up to
quadratic order, even to describe only Gaussian statistics and contrary to SPT.

In anticipation of our later setup, we define the perturbations at some spacetime point $x$ by the displacements of 
the full inflaton fields $\phifull^\indI(x)$ and their conjugate momenta $\pifull_\indI(x)$ from
their coarse-grained values $\phiIR^\indI(x)$ and $\piIR_\indI(x)$ (the homogeneous background $\phibg$ and $\pibg$ 
would instead be used as reference fields in SPT): 
\bae{\label{eq: def of dphi}
    \dphi^\indI(x)=\phifull^\indI(x)-\phiIR^\indI(x), \qquad \dpi_\indI(x)=\pifull_\indI(x)-\piIR_\indI(x).
}
These finite displacements~(\ref{eq: def of dphi}) do not transform covariantly under field redefinitions beyond the linear approximation, and therefore one needs to relate them to contravariant/covariant infinitesimal perturbations.
The $\dphi$'s expansion has been already discussed in Ref.~\cite{Gong:2011uw}. 
The two neighbouring points in field space $\phifull(x)$ and $\phiIR(x)$ can be connected by a unique field-space geodesic, which we parameterise by the affine parameter $\lambda$ such that $\phifull(\lambda=0)=\phiIR(x)$ and $\phifull(\lambda=1)=\phifull(x)$ (see Fig.~\ref{fig: transport method}). 
We then define the Vilkovisky-DeWitt-type variable $Q^\indI$ by the ``initial velocity"
\bae{
    \left.\dif{\phifull^\indI}{\lambda}\right|_{\lambda=0}=Q^\indI.
}
This geometrical definition ensures that $Q^I$ lies in the tangent space of the point $\phiIR(x)$, i.e. that it behaves as desired as a vector under field redefinitions. Using the fact that, by definition, $\phifull^\indI(\lambda)$ verifies the geodesic equation 
\bae{
    \covD_\lambda^2\phifull^\indI(\lambda)=\dif{^2\phifull^\indI}{\lambda^2}+\Gamma^\indI_{\indJ\indK}\dif{\phifull^\indJ}{\lambda}\dif{\phifull^\indK}{\lambda}=0\,,
}
where $\covD_\lambda$ represents the covariant derivative with respect to the affine parameter $\lambda$, 
one can express 
$\phifull^\indI(\lambda)$ 
in terms of $Q^\indI$ by using the following expansion around $\lambda=0$:
\bae{
    \phifull^\indI(\lambda)&=\phifull^\indI(\lambda=0)+\left.\dif{\phifull^\indI}{\lambda}\right|_{\lambda=0}\lambda+\frac{1}{2}\left.\dif{^2\phifull^\indI}{\lambda^2}\right|_{\lambda=0}\lambda^2
    +\cdots \nonumber \\
    &=\phiIR^\indI+Q^\indI\lambda-\frac{1}{2}\Gamma^\indI_{\indJ\indK}Q^\indJ Q^\indK\lambda^2+\cdots,
}
thus obtaining the field perturbations 
\bae{\label{eq: dphi expansion}
    \dphi^\indI=\phifull^\indI(\lambda=1)-\phifull^\indI(\lambda=0)=Q^\indI-\frac{1}{2}\Gamma^\indI_{\indJ\indK}Q^\indJ Q^\indK+\cdots.
}
The non-tensorial feature of the Christoffel symbols explicitly shows the non-covariance of the finite perturbations $\dphi^\indI$ beyond the linear approximation.

The displacement $\dpi_\indI$ can also be expressed
in terms of a truly covariant tensor in a similar way. 
For that, let us consider a family $\pifull_\indI(\lambda)$ of covectors at each point along the geodesic $\phifull^\indI(\lambda)$, and such that $\pifull_\indI(\lambda=0)=\varpi_\indI(x)$ and $\pifull_\indI(\lambda=1)=\pifull_\indI(x)$.
It is then natural to define a second Vilkovisky-DeWitt-type variable $\covP_\indI$ by the ``initial momentum velocity" along the geodesic as
\bae{\label{eq: covP def}
    \covP_\indI=\covD_\lambda\pifull_\indI|_{\lambda=0}=\left( \dif{\pifull_\indI}{\lambda}-\Gamma_{\indI\indJ}^\indK\pifull_\indK\dif{\phifull^\indJ}{\lambda} \right)
    |_{\lambda=0}
=P_\indI-\Gamma^\indK_{\indI\indJ}\piIR_\indK Q^\indJ,
}
where on the right-hand side, the naive $P_\indI$ such that $ \dpi_\indI=\pifull_\indI(\lambda=1)-\pifull_\indI(\lambda=0)=P_\indI \lambda+{\cal O}(\lambda^2)$ does not even transform covariantly at linear order, contrary to $\covP_\indI$, whose intrinsic geometrical definition ensures that it transforms as a covector at all order in perturbation theory.
If one now imposes that the covectors $\covD_\lambda\pifull_\indI$ are parallel-transported along the geodesic:
\bae{
    0=\covD_\lambda^2\pifull_\indI=\dif{^2\pifull_\indI}{\lambda^2}-2\Gamma^\indK_{\indI\indJ}\dif{\phifull^\indJ}{\lambda}\dif{\pifull_\indK}{\lambda}-(\Gamma^\indS_{\indI\indJ,\indK}-\Gamma^\indS_{\indI\indR}\Gamma^\indR_{\indJ\indK}-\Gamma^\indR_{\indI\indJ}\Gamma^\indS_{\indR\indK})\pifull_\indS\dif{\phifull^\indJ}{\lambda}\dif{\phifull^\indK}{\lambda},
}
it is possible to express $\dpi_\indI$ in terms of $\covP_I$.
However, we note that imposing $\covD_\lambda^2\pifull_\indI=0$ is \emph{one} simple possible choice, but that others are possible, corresponding to a freedom in the identification of a suitable covariant momentum perturbation.
We refer the interested reader to Appendix~\ref{appendix: covP} for more details on this point, to which we will come back, and here just quote the relation between $\dpi_\indI$ and $\covP_I$ for this particular choice:
\bae{\label{eq: covariant momentum UV perturbation}
    \dpi_\indI=\covP_\indI+\Gamma^\indK_{\indI\indJ}\piIR_\indK Q^\indJ +\Gamma^\indK_{\indI\indJ}Q^\indJ \covP_\indK+\frac{1}{2}(\Gamma^\indS_{\indI\indJ,\indK}-\Gamma^\indS_{\indI\indR}\Gamma^\indR_{\indJ\indK}
    +\Gamma^\indR_{\indI\indJ}\Gamma^\indS_{\indR\indK})\piIR_\indS Q^\indJ Q^\indK+\cdots.
}
Equipped with these geometrically defined objects, we are now ready to compute the covariant effective action 
up to second order in the UV fields and momenta as
\bae{\label{eq: covariant effective action}
    \exp\left(i\Seff[\varphi^{\indX\indI\Kela}]\right)=\int\mD Q^{\covX\indI\Kela }\exp\left(i\So[\varphi^{\indX\indI\Kela}]+\Sl[\varphi^{\indX\indI\Kela},Q^{\covX\indI\Kela}]+\Sz[\varphi^{\indX\indI\Kela},Q^{\covX\indI\Kela}] \right),
}
with $Q^{\covX\indI\Kela}=(Q^{\indI\Kela}, \covP^{\indI\Kela})$ used as a short-hand notation, and like for Eq.~\eqref{ZJ}, $\mD Q^{\covX\indI\Kela}$ truly refers to the canonical phase-space measure $\prod_{\indI,\Kela,\indJ,\Kelb}\mD Q^{\indI\Kela} \mD \covP_{\indJ}^{\Kelb}$.

\subsection{Covariant CTP action and IR-UV interactions}

To investigate the effect of linear UV perturbations on the IR dynamics, we must first covariantly expand the action up to second order in the perturbations.
Starting from the Hamiltonian action~\eqref{eq: Hamiltonian action}, and expanding it up to second order in the fields' covariant UV perturbations $Q^{\covX\indI}$ as well as metric 
UV perturbations $\lapseUV$ and $\psi$,\footnote{Let us stress again that the stochastic approach is perturbative in the UV parts of the fields, and we only treat them up to quadratic order in this work (i.e. at the level of linearised perturbation theory). However no expansion is used at this stage for the IR parts of the fields, for which all nonlinearities are kept, at leading order in the gradient expansion.} one finds
\bae{
	\So&=\int\dd^4x\,a^3\left[\piIR_\indI \phiIR^{\indI\prime}-\frac{1}{H}\left(\frac{1}{2}G^{\indI\indJ}\piIR_\indI\piIR_\indJ+V+3\Mpl^2H^2\right)\right], \label{eq: S0} \\
	\Sl&=\int\dd^4x\,a^3\left[\covP_\indI
	\left(\phiIR^{\indI\prime}-\frac{\piIR^\indI}{H} \right) -Q^\indI\left(\covD_N\piIR_\indI+3\piIR_\indI+\frac{V_\indI}{H}\right)
	-\lapseUV\left(\frac{1}{2}\piIR_\indI\piIR^\indI+V-3\Mpl^2H^2\right)\right], \label{eq: cov S1} \\
	\Sz&=\int\dd^4x\,a^3\left[-3\Mpl^2H^3\lapseUV^2
	-\lapseUV\left(\piIR_\indI\covP^\indI+V_\indI Q^\indI+2\Mpl^2H^2\frac{\Laplacian}{a^2}\shiftUV\right)
	+\piIR_\indI Q^\indI\frac{\Laplacian}{a^2}\shiftUV 
	\right.\nonumber \\
	&\qquad\left.-\frac{1}{H}\left(\frac{1}{2}\covP_\indI\covP^\indI-\frac{1}{2}Q_I\frac{\Laplacian}{a^2}Q^\indI+\frac{1}{2}V_{;\indI\indJ}Q^\indI Q^\indJ
	-\frac{1}{2}R_\indI{}^{\indJ\indK}{}_\indL\piIR_\indJ\piIR_\indK Q^\indI Q^\indL\right)+\covP_\indI\covD_N Q^\indI\right. \nonumber \\
	&\qquad\left. + \left(\phiIR^{\indI\prime}-\frac{\piIR^\indI}{H} \right) \frac{1}{2}R_{\indI\indJ\indK}{}^\indL \piIR_L Q^\indJ Q^\indK  \right] \,.\label{eq: cov S2}
}
In usual perturbation theory, where the coarse-grained fields and momenta are replaced by their homogeneous values that are solutions of the classical equations of motion dictated by $\So$, 
the linear action $\Sl$ vanishes. 
It is thus sufficient to use covariant variables at linear order only, and one needs not bother about quadratic terms in 
$(Q^\indI, \covP_\indI)$ in Eqs.~\eqref{eq: dphi expansion} and \eqref{eq: covariant momentum UV perturbation}.
Here on the contrary,  $S^{(1)}$ does not vanish because the time derivatives of the coarse-grained fields and momenta differ slightly from their classical values (by a quantity one can interpret as a classical random noise as we shall find soon).
Relatedly, one can check that the manifest covariance of the result~\eqref{eq: S0}--\eqref{eq: cov S2} would not have hold if one had not expanded $\dphi^{\indX\indI}$ to quadratic order in covariant perturbations. Let us now examine its three contributions.

\paragraph*{Pure IR sector}

$\So$ governs the propagation and self-interactions of the IR fields, without consideration of the UV modes at all.
More generally, it should be interpreted as dictating the deterministic drift for the IR fields. Notice that in general, for generic potential $V$ and field-space metric $G_{\indI\indJ}$, this classical drift action can be non-linear in the IR modes. Thus, we will not bother writing explicitly a rather complex expression for $\So\left[\phiIR^{\indX\indI\Kela}\right]=\So[\phiIR^{\indX\indI\KelC}+\phiIR^{\indX\indI\KelD}/2]-\So[\phiIR^{\indX\indI\KelC}-\phiIR^{\indX\indI\KelD}/2]$, given that we will only be interested in its variation evaluated at classical IR solutions, which simply reads
\bae{\label{eq: var-S0}
\left.\var{\So\left[\phiIR^{\indX\indI\Kela}\right]}{\phiIR^{\indY\indJ\KelD}}\right|_{\phiIR^{\indY\indJ\KelD}=0}= \left.\var{\So\left[\phiIR^{\indX\indI}\right]}{\phiIR^{\indY\indJ}}\right|_{\phiIR^{\indX\indI}=\phiIR^{\indX\indI\KelC}}.
}
In a related manner,
one can see that because of its structure, $\So\left[\phiIR^{\indX\indI\Kela}\right]$ only contains odd powers of $\phiIR^{\indX \indI \KelD}$, the quantum component of the fields.
Thus, its expansion in quantum fields is trivial up to quadratic order, and one can actually write:
\bae{
\label{eq: S0 linear in varphi-q}
    \So\left[\phiIR^{\indX\indI\Kela}\right]= \left.
    \int\dd^4x
    \var{\So\left[\phiIR^{\indX\indI}\right]}{\phiIR^{\indY\indJ}(x)}\right|_{\phiIR^{\indX\indI}=\phiIR^{\indX\indI\KelC}}
    \phiIR^{\indY\indJ\KelD}(x)
    +\mathcal{O}\left(\phiIR^\KelD\right)^3\,.
}

\paragraph*{IR-UV interactions}

As already stated, we focus in this derivation on the IR-UV interactions stemming from the continuous flow of quantum UV modes to the open system of classical IR ones, which results from the time-dependent Fourier cutoff $\ksigma(N)$. Interestingly, these interactions are encoded in $\Sl$ in the time derivatives of the IR fields,
as seen in the heuristic approach (Sec.~\ref{sec: heuristic}).
We thus focus on those terms and neglect other ones in $\Sl$, which amounts to considering the time derivatives acting as $\delta(N-\tsigma(k))$ in Fourier space as we shall see now. 

However first, because we will write the discretised version of the linear action $\Sl$, it will prove useful to take a step back. The form of $\Sl$ that we displayed in Eq.~\eqref{eq: cov S1} is physically appealing because it makes explicitly appear 
the background EoM for the IR fields, times the UV perturbations, in an explicitly covariant form. However to find it we had first to integrate by parts the term $\int \dd^4 x a^3 Q^{\indI \prime} \piIR_\indI=-\int \dd^4 x  a^3 Q^\indI (\piIR_\indI^\prime + 3 \piIR_\indI)$ and then combine it with the change from $P_\indI$ to $\covP_\indI=P_\indI-\Gamma^\indK_{\indI\indJ}\piIR_\indK Q^J$ to form the covariant time derivative $\covD_N \piIR_\indI$.
Thus, going back to this previous version, the relevant terms in $\Sl$ can be more simply expressed (i.e. with less time derivatives) as  $\Sl \supset \int \dd^4x\,a^3 \left(\covP_\indI
\phiIR^{\indI\prime}-Q^\indI(\covD_N\piIR_\indI + 3 \piIR_\indI)\right)=\int \dd^4x \,a^3\left(P_I\phiIR^{\indI\prime}+Q^{\indI\prime}\piIR_I\right)$, which we will now take as our starting point to compute IR-UV interactions in the Keldysh basis. Note that each of these two terms is not covariant when taken separately, but that their sum is indeed covariant. Understanding why these terms with time derivatives are peculiar is easier in Fourier space,
and requires the investigation of the action in terms of Keldysh fields:
\bae{\label{eq: S1 Keldysh}
    &\Sl\left[\phiIR^{\indX\indI\Kela},Q^{\indX\indI\Kela} \right] \nonumber \\
    &\quad\supset\int \dd N  \dk\,a^3(N) \left[P_I^\KelC\left(N,\mathbf{k}\right)\phiIR^{\indI\KelD\prime}\left(N,\mathbf{k}\right)+Q^{\indI\KelC\prime}\left(N,\mathbf{k}\right)\piIR_I^{\KelD}\left(N,\mathbf{k}\right)\right] + \left( \KelC \leftrightarrow \KelD \right).
}
Now, remember that the same Fourier component of UV and IR modes can never be defined at the same time, 
except at the transition time $\tsigma(k)$. Of course it means that the support of these terms is of measure zero, and this is actually why terms without derivatives were neglected in $\Sl$.\footnote{Strictly speaking this discussion applies only to terms in $\Sl$ that are bilinear in IR and UV fields. Terms that are higher-order in IR quantities contain nonlinear IR-UV mode mixings, but the stochastic formalism does not aim at taking into account 
these couplings that are also present in Minkowski spacetime.} However the terms with derivatives that we kept play a special role. Let us look for example at the first term, going back to the discrete description of the phase-space path integral for the mode $\mathbf{k}$ around the time $\tsigma(k)$:
\bae{
    &\int_{\tsigma(k)-\Delta N}^{\tsigma(k)+\Delta N} \dd N\,a^3(N)
    P_I^\KelC\left(N,\mathbf{k}\right)\phiIR^{\indI\KelD\prime}\left(N,\mathbf{k}\right) \nonumber \\
    &\qquad=\Delta N\left[a^3(\tsigma(k)+\widetilde{\Delta N})P_I^\KelC\left(\tsigma(k)+\widetilde{\Delta N},\mathbf{k}\right)\frac{\phiIR^{\indI\KelD}\left(\tsigma(k)+\Delta N,\mathbf{k}\right)-\phiIR^{\indI\KelD}\left(\tsigma(k),\mathbf{k}\right)}{\Delta N} \right. \nonumber \\
    &\qquad\qquad\left. + a^3(\tsigma(k)-\Delta N+\widetilde{\Delta N})P_I^\KelC\left(\tsigma(k)-\Delta N+\widetilde{\Delta N},\mathbf{k}\right)\frac{\phiIR^{\indI\KelD}\left(\tsigma(k),\mathbf{k}\right)-\phiIR^{\indI\KelD}\left(\tsigma(k)-\Delta N,\mathbf{k}\right)}{\Delta N} \right] \nonumber  \\
    &\qquad =a^3(\tsigma(k)-\Delta N+\widetilde{\Delta N})\, P_I^\KelC\left(\tsigma(k)-\Delta N+\widetilde{\Delta N},\mathbf{k}\right)\phiIR^{\indI\KelD}\left(\tsigma(k),\mathbf{k}\right),
}
with $0<\widetilde{\Delta N}<\Delta N$ and  where we used the conditions~\eqref{eq: decompositon Fourier} to get the second equality. Note that while the fields' values
$\phiIR^\indI$ and $Q^\indI$ are evaluated on the discrete time grid at $N_\indj=j \Delta N$, the momenta $\piIR_\indI$ and $P_\indI$ can be evaluated at intermediate time steps $\widetilde{N_\indj}=j \Delta N + \widetilde{\Delta N}$, which enables us to compute IR-UV interactions without specifying boundary conditions at the exact time $\tsigma(k)$ for them.
In the same way, we find that the second term contributes $-a^3(\tsigma(k)+\widetilde{\Delta N})\,Q^{\indI\KelC} \left(\tsigma(k),\mathbf{k}\right)\piIR_{\indI}^{\KelD}\left(\tsigma(k)+\widetilde{\Delta N},\mathbf{k}\right)$. 
As for the third and the fourth term from the $(\KelC$-$\KelD)$ permutation in Eq.~\eqref{eq: S1 Keldysh}, they vanish by virtue of the boundary conditions~\eqref{eq: boundary conditions}. Thus, the interaction action that we consider can be rewritten in the continuous limit as
\bae{\label{eq: Sint}
	\Sint\left[ \varphi^{\indX\indI\Kela},Q^{\indX\indI\Kela}\right]&=\int\dd^4x\int\dk\delta(N-\tsigma(k))\ee^{i\kdx}a^3\left[P_\indI(x)^\KelC{\phiIR}^{\indI\KelD}(N,\mathbf{k})-Q^{\indI\KelC}(x) \piIR_\indI^\KelD(N,\mathbf{k})\right] \nonumber \\
    &=\int\dd^4x\,a^3Q^{\indX\indI\Kela}(x)\phiIR^\prime_{\indX\indI\Kela}(x),
}
where we introduced the pseudo ``time derivatives" $\phiIR^\prime_{\indX\indI\Kela}(x)$ with lower indices that are defined as
\bae{\label{eq: time-derivatives}
	\bce{
		\dps
    	\phiIR^\prime_{Q\indI\Kela}(x):=-\delta^\KelC_\Kela\int\dk\delta(N-\tsigma(k))\ee^{i\kdx}\piIR_\indI{}^\KelD(N,\mathbf{k}), \\[10pt] 
		\dps
		\phiIR^\prime_{P I \Kela}(x):=\delta^\KelC_\Kela\int\dk\delta(N-\tsigma(k))\ee^{i\kdx}\phiIR_\indI{}^{\KelD}(N,\mathbf{k}),
	}
}
thus restricting the couplings to be of the form $\mathrm{UV}^\KelC\times\mathrm{IR}^\KelD$. Those interaction terms will be the ones responsible for the modification of the classical equations of motion for the IR fields, once the UV perturbations are integrated out. Let us now investigate the dynamics of the latter.

\paragraph*{UV dynamics}
As a first comment, note that the last line of $\Sz$ seemingly
goes beyond the approximation of treating UV modes up to quadratic order in the action, as it consists of a quadratic term in the UV perturbations, multiplied by the ``background-like'' equation of motion
for the IR fields, i.e. by a quantity of order of the to-be-found noise. The careful reader will also have noticed that such a term is exactly of the kind that can ambiguously appear depending on the exact definition of a covariant momentum UV perturbation $\covP_\indI$, as we explain in Appendix~\ref{appendix: covP}. Because this arbitrariness can not affect the physics, we are free to make the choice ($\kappa= 1/2$) such that this term proportional to the Riemann tensor of the field space vanishes. This procedure also fixes the form of $\SE$ as is shown in Appendix~\ref{appendix: covP}, thus we conclude that this term does not affect the Gaussian properties of the theory and we will discard it in what follows, leaving for future work the investigation of this subtlety and of potentially interesting non-Gaussian features related to the geometry of the field space.

Extremising the action~\eqref{eq: S0}--\eqref{eq: cov S2} with respect to the non-dynamical fields $\lapseUV$ and $\psi$ that appear without any time-derivative, one recovers the local Friedmann equation~\eqref{laspeIR} as well as the expressions~\eqref{lapseUV} and \eqref{eq: perturbation energy const} for $\lapseUV$ and $\psi$ in terms of $Q^\indI$ and $\covP_\indI$. Plugging them back into the second-order action, one can write the latter in the condensed form
\bae{\label{eq: S2 with indices}
	\Sz=\frac{1}{2}\int\dd^4x\dd^4x^\prime Q^{\indX\indI}(x)\Lambda_{\indX\indY\indI\indJ}(x,x^\prime)Q^{\indY\indJ}(x^\prime),
}
where we used the non-covariant variables $Q^{\indX\indI}$ that naturally appear in $\Sint$
instead of the covariant perturbations $Q^{\covX\indI}$.
As a result, some of the following intermediate steps will not be manifestly covariant. However, one is perfectly allowed to use such non-covariant objects to make calculations, and then to switch back to covariant ones using the relation~(\ref{eq: covP def}) $\covP_\indI=P_\indI-\Gamma_{\indI\indJ}^\indK\piIR_\indK Q^\indJ$.
Instead of quoting the kernel $\Lambda_{\indX\indY\indI\indJ}$ corresponding to the non-covariant UV modes that only appear in intermediate steps, we rather show its covariant counterpart $\Lambda_{\covX\covY\indI\indJ}$, which is given by 
\bae{
	\bpme{
		\Lambda_{QQ\indI\indJ}, & \Lambda_{Q\covP\indI\indJ} \\
		\Lambda_{\covP Q\indI\indJ}, & \Lambda_{\covP\covP\indI\indJ}
	}=\delta^{(4)}(x-x^\prime)a^3\bpme{
		\frac{1}{H}\left(G_{\indI\indJ}\frac{\Laplacian}{a^2}-M^2_{QQ\indI\indJ}\right), & -G_{\indI\indJ}(\covD_N+3
		)-M^2_{Q\covP\indI\indJ} \\
		G_{\indI\indJ}\covD_N-M^2_{\covP Q\indI\indJ}, & -G_{IJ}/H
	},
}
where the differential operators act on the $x^\prime$ coordinates, and with $M^2_{QQ\indI\indJ}$ and $M^2_{Q\covP\indI\indJ}=M^2_{\covP Q\indI\indJ}$ already given in Eqs.~\eqref{M2QQ}--\eqref{M2QP}.

Extremising $\Sz$ 
with
respect to the covariant UV perturbations yields the following classical EoM for the UV fields: 
\bae{\label{eq: UV EoM}
	\int\dd^4x^\prime\Lambda_{\covX\covY\indI\indJ}(x,x^\prime)Q^{\covY\indJ}(x^\prime)=0, \quad \Leftrightarrow \quad
	\bce{
		\dps
		\covD_N Q^\indI=\frac{\covP^\indI}{H}+M^2_{\covP Q}{}^\indI{}_\indJ Q^\indJ, \\
		\dps
		\covD_N\covP_\indI=-3\covP_\indI+\frac{1}{H}\left(G_{\indI\indJ}\frac{\Laplacian}{a^2}-M^2_{QQ\indI\indJ}\right)Q^\indJ-M^2_{Q\covP\indI}{}^\indJ\covP_\indJ\,,
	}
}
which are nothing else than equations~(\ref{eq: EQ}) and (\ref{eq: EP}) $\UVE^{Q\indI}=0=\UVE^{\covP}_\indI$ found in the heuristic approach (Sec.~\ref{sec: heuristic}).
Strikingly, as we shall see in the next subsection, it is sufficient to know the (inverse of the) kernel operator $\Lambda$ to compute the corrections to the IR dynamics, due to their interactions with UV fluctuations as dictated by $\Sint$. Thus, the UV modes dictating these corrections can be understood as evolving according to $\Sz$ only, hence they verify the EoM~\eqref{eq: UV EoM}, similar to the one of SPT but with background fields replaced by their infrared counterparts.
This is an interesting improvement from the path-integral approach compared to the heuristic one, where we had to \emph{assume} that the dynamics of UV modes was decoupled from the one of IR ones, see the paragraph before Eq.~\eqref{eq: noises} and the one after Eq.~\eqref{M2QP}. \\

Eventually, in the path integral~\eqref{eq: covariant effective action}
with doubled degrees of freedom, the quadratic action written in terms of the fields in the Keldysh basis reads:
\bae{
	\Sz\left[ \varphi^{\indX\indI\Kela},Q^{\indX\indI\Kela}\right]=\frac{1}{2}\int\dd^4x\dd^4x^\prime Q^{\indX\indI\Kela}(x)\Lambda_{\indX\indY\indI\indJ\Kela\Kelb}(x,x^\prime)Q^{\indY\indJ\Kelb}(x^\prime),
}
where $\Lambda_{\indX\indY\indI\indJ\Kela\Kelb}$ is given by the basis transformation $\Lambda_{\indX\indY\indI\indJ\Kela\Kelb}=(K^T)_\Kela{}^a\Lambda_{\indX\indY\indI\indJ\inda\indb}K^\indb{}_\Kelb$, with $K$ given in Eq.~\eqref{classical-quantum-def}, and
with the $\pm$ basis operator
\bae{\label{eq: Lambda_ab}
    \Lambda_{\indX\indY\indI\indJ\inda\indb}=\diag\left(\Lambda_{\indX\indY\indI\indJ}(\phiIR^+),-\Lambda_{\indX\indY\indI\indJ}(\phiIR^-)\right).
}
Note that, as the differential operator $\Lambda_{\indX\indY\indI\indJ}$ depends on the IR fields $\phiIR^{\indX\indI}$, one has in principle to distinguish
between its evaluations on $+$ IR fields and on $-$ IR fields: $\Lambda_{\indX\indY\indI\indJ}(\phiIR^+)$ and $\Lambda_{\indX\indY\indI\indJ}(\phiIR^-)$.
However, as we already noticed, one can think of the expansion in the quantum components as an expansion in $\hbar$. In this respect, in order to derive the leading-order quantum effects,
it sufficient to use the expression of $\Lambda$ at zeroth-order:
\bae{\label{eq: Lambda_ab classical}
    \Lambda_{\indX\indY\indI\indJ\inda\indb}=\Lambda_{\indX\indY\indI\indJ}(\phiIR^\KelC)\sigma_{3\inda\indb}+\mathcal{O}\left(\varphi^\KelD\right)\delta_{\inda\indb}\,.
}
Although it may seem a crude approximation, we will check the consistency of this expansion in the next subsection, and explain why higher-order corrections are indeed 
not needed for our computation.

\subsection{Covariant 
coarse-grained effective Hamiltonian action and Langevin equations}
\label{subsec:effective Hamiltonian action-Langevin equations}

Now we have to gather the three contributions to the covariant coarse-grained Hamiltonian  effective action and perform the following path integral: 
\bae{
\exp\left(i\Seff[\varphi^{\indX\indI\Kela}]\right)=&\exp\left(i\So[\varphi^{\indX\indI\Kela}]\right)  \nonumber \\
	& \times
	\int \mD
	Q^{\indX\indI\Kela}\exp\left[i\int\dd^4x\,a^3\phiIR^\prime_{\indX\indI\Kela}Q^{\indX\indI\Kela}
	+\frac{i}{2}\int\dd^4x\dd^4x^\prime Q^{\indX\indI\Kela}\Lambda_{\indX\indY\indI\indJ\Kela\Kelb}Q^{\indY\indJ\Kelb}\right].
}
Note that we safely replaced the measure 
$\mD Q^{\covX\indI\Kela}$ by $\mD Q^{\indX\indI\Kela}$ in the path integral~\eqref{eq: covariant effective action} as the Jacobian for the transformation  
$Q^{\covX\indI\Kela}\to Q^{\indX\indI\Kela}$ is exactly one.
The Gaussian integral over the UV modes can be performed exactly to give:
\bae{
	\Seff[\varphi^{\indX\indI\Kela}]=&\, \So[\varphi^{\indX\indI\Kela}]\underbrace{+\frac{i}{2}\ln\left[\Det\left(\Lambda\right)\right]}_{\textstyle \begin{array}{c}
	\Sren[\varphi^{\indX\indI\KelC}]\end{array}} \underbrace{-\frac{1}{2}\int\dd^4x\dd^4x^\prime \left(a^3\phiIR_{\indX\indI\Kela}^\prime\right)_x(\Lambda^{-1})^{\indX\indY\indI\indJ\Kela\Kelb}{}_{xx^\prime}\left(a^3\phiIR^\prime_{\indY\indJ\Kelb}\right)_{x^\prime}}_{\textstyle \begin{array}{c}
	\SIA[\varphi^{\indX\indI\Kela}]\end{array}}. \label{def-influence}
}
The first term dictates the classical, background dynamics of the IR fields, and contains no new information compared to SPT. The second term is nothing but the usual QFT one-loop correction which can be computed in principle, and then reabsorbed by renormalisations of the bare parameters in the classical action $\SHam$.
We will thus omit this contribution in the following, although the renormalisation procedure is of course highly non-trivial to perform explicitly. More important for us is the third contribution $\SIA$ called the \emph{influence action}, describing the influence on the coarse-grained fields of the small-scale UV fluctuations that were integrated out
(or more generally, the influence of an environment on the system of interest~\cite{Feynman:1963fq}). In the rest of this subsection, we compute this influence action, discuss its physical interpretation and derive the resulting stochastic equations for the coarse-grained fields.\\

As $\Lambda^{-1}$ is nothing but the closed-time-path-ordered two point correlation function of UV modes, it is easier to express  
it first in the $\pm$ basis with latin indices $\inda$, $\indb$, $\cdots$, and then translate it into the Keldysh basis with use of the matrix of change of basis $K^\inda{}_\Kela$. So we first focus on 
\bae{\label{eq: two point function}
    i(\Lambda^{-1}&)^{\indX\indY\indI\indJ\inda\indb}(x,x^\prime) \nonumber \\
    & =\int \mD Q^{\indX\indI\inda} \exp{\left[\frac{i}{2}\int \dd^4 x \dd^4 x^\prime Q^{\indX\indI\inda} \Lambda_{\indX\indY\indI\indJ\inda\indb} Q^{\indY\indJ\indb}\right]} Q^{\indX\indI\inda}(x) Q^{\indY\indJ\indb}(x^\prime)  \\
    &=
    \bce{
        \dps
        \theta(N-N^\prime)\braket{\hat{Q}^{\indX\indI}(x)\hat{Q}^{\indY\indJ}(x^\prime)}+\theta(N^\prime-N)\braket{\hat{Q}^{\indY\indJ}(x^\prime)\hat{Q}^{\indX\indI}(x)}, & (\inda=+,\,\indb=+), \\
        \dps
        \braket{\hat{Q}^{\indY\indJ}(x^\prime)\hat{Q}^{\indX\indI}(x)}, & (\inda=+,\,\indb=-), \\
        \dps
        \braket{\hat{Q}^{\indX\indI}(x)\hat{Q}^{\indY\indJ}(x^\prime)}, & (\inda=-,\,\indb=+), \\
        \dps
        \theta(N^\prime-N)\braket{\hat{Q}^{\indX\indI}(x)\hat{Q}^{\indY\indJ}(x^\prime)}+\theta(N-N^\prime)\braket{\hat{Q}^{\indY\indJ}(x^\prime) \hat{Q}^{\indX\indI}(x)}, & (\inda=-,\,\indb=-), \nonumber
    } 
}
where the ordering of the quantum operators is determined by the chronological order along the closed-time path $C$, and the brackets $\braket{\cdots}$ denote usual vacuum expectation values of UV operators under the $\phiIR^{\indX\indI}$-dependent measure $\mD Q^{\indX\indI} \exp{\left(  i \Sz\left[\phiIR^{\indX\indI}, Q^{\indX\indI} \right] \right)}$. Note that these expectation values are unambiguously defined in the same way on both branches of the CTP, as we recall that for our computation, it is sufficient to evaluate the kernel 
$\Lambda$ with vanishing quantum components of the IR fields, i.e. $\Lambda(\varphi^\KelC)$, see Eq.~\eqref{eq: Lambda_ab classical}.
Thus, for each component of the $\Lambda^{-1}$ matrix we are able to forget the $\pm$ indices and we can compute them as in usual perturbation theory, but with background fields replaced by their IR counterparts.

With use of the dimensionless
unequal time two-point functions
\bae{
    (2\pi)^3\delta^{(3)}(\mathbf{k}+\mathbf{k}^\prime)\frac{2\pi^2}{k^3}\calP^{\indX\indY\indI\indJ}(N,N^\prime;
    k)=\Braket{\hat{Q}^{\indX\indI}(N,\mathbf{k})\hat{Q}^{\indY\indJ}(N^\prime
    ,\mathbf{k}^\prime)},
}
$\Lambda^{-1}$ can be expressed more explicitly as 
\bae{
    &(\Lambda^{-1})^{\indX\indY\indI\indJ\inda\indb}(x,x^\prime) \nonumber \\
    &=-i\int\dk\ee^{i\mathbf{k}\cdot(\mathbf{x}-\mathbf{x}^\prime)}\frac{2\pi^2}{k^3}
    \bpme{
        \begin{array}{c}
            \theta(N-N^\prime)\calP^{\indX\indY\indI\indJ}(N,N^\prime;k) \\
            +\theta(N^\prime-N){\calP^{\indX\indY\indI\indJ}}^*(N,N^\prime;k)
        \end{array} &
        {\calP^{\indX\indY\indI\indJ}}^*(N,N^\prime;k) \\
        \calP^{\indX\indY\indI\indJ}(N,N^\prime;k) &
        \begin{array}{c}
            \theta(N^\prime-N)\calP^{\indX\indY\indI\indJ}(N,N^\prime;k) \\
            +\theta(N-N^\prime){\calP^{\indX\indY\indI\indJ}}^*(N,N^\prime;k)
        \end{array}
    }^{\inda\indb},
}
where we used that $\calP^{\indY\indX\indJ\indI}(N^\prime,N;k)={\calP^{\indX\indY\indI\indJ}}^*(N,N^\prime;k)$ 
as a consequence of $\hat{Q}^{\indX \indI}(x)$ being hermitian operators, and hence $\hat{Q}^{\indX \indI \dagger}(N,\mathbf{k}) =\hat{Q}^{\indX \indI}(N,-\mathbf{k})$.
We now express $\Lambda^{-1}$ in the Keldysh basis as
\bae{\label{eq: Lambda in Keldysh}
    &(\Lambda^{-1})^{\indX\indY\indI\indJ}{}_{\Kela\Kelb}(x,x^\prime)=
(K^T)_\Kela{}^a \sigma_{3\inda\indb}\, (\Lambda^{-1})^{\indX\indY\indI\indJ\indb\indc}(x,x^\prime)\, \sigma_{3\indc\indd}
K^\indd{}_\Kelb  
    \nonumber \\
    &=-i\int\dk\ee^{i\mathbf{k}\cdot(\mathbf{x}-\mathbf{x}^\prime)} \frac{2\pi^2}{k^3}
    \bpme{
        0 & -2i\theta(N^\prime-N)\Im\calP^{\indX\indY\indI\indJ}(N,N^\prime;k)  
        \\ 
        2i\theta(N-N^\prime)\Im\calP^{\indX\indY\indI\indJ}(N,N^\prime;k) 
        &        \Re\calP^{\indX\indY\indI\indJ}(N,N^\prime;k)
    }_{\Kela\Kelb}\,,
}
where we used $\theta(N^\prime-N)+\theta(N-N^\prime)=1$.
The influence action $\SIA$ in Eq.~\eqref{def-influence}
can then be explicitly obtained after contracting twice with  $\phiIR^\prime_{\indX\indI}{}^\Kela(x) \propto 
\delta^\Kela_\KelD\delta(N-\tsigma(k))$ (note that the position of the $\Kela$ index has been flipped compared to Eq.~\eqref{eq: time-derivatives} with use of the $\sigma_1^{\Kela\Kelb}$ metric), retaining only the $\KelD$-$\KelD$ component of Eq.~\eqref{eq: Lambda in Keldysh} evaluated at equal times $N=N^\prime=\tsigma(k)$:
\bae{\label{eq:IA}
    \SIA=&\,\frac{i}{2}\int\dd^4x\dd^4x^\prime \left(a^3\phiIR_{\indX\indI}{}^{\KelD}\right)_x(\Re\Pi^{\indX\indY\indI\indJ})_{xx\prime}\left(a^3\phiIR_{\indY\indJ}{}^{\KelD}\right)_{x^\prime},
}
where  
\bae{\label{eq: Pi tensor}
    \Pi^{\indX\indY\indI\indJ}(x,x^\prime)&=\int\dk\ee^{i\mathbf{k}\cdot(\mathbf{x}-\mathbf{x}^\prime)} \frac{2\pi^2}{k^3} \delta(N-\tsigma)\delta(N^\prime-\tsigma)
    \calP^{\indX\indY\indI\indJ}(N,N^\prime;k) \nonumber \\
    &=\frac{k_\sg{}^\prime}{\ksigma}\frac{\sin\left(\ksigma|\mathbf{x}-\mathbf{x}^\prime|\right)}{\ksigma|\mathbf{x}-\mathbf{x}^\prime|}\calP^{\indX\indY\indI\indJ}(N,\ksigma)\delta(N-N^\prime),
}
with $\calP$ on the second line being simply the usual equal-time dimensionless two-point correlation function. Note that any higher-order correction in the quantum components of the IR fields, coming from evaluating $\Lambda$ beyond the leading order result~\eqref{eq: Lambda_ab classical}, would generate terms of order $\mathcal{O}(\phiIR^\KelD)^3$ in the influence action. Put it otherwise, our computation is exact up to quadratic order in the quantum components.

Although we have just seen that the contractions with the ``time derivatives" $\phiIR^\prime_{\indX\indI}{}^\Kela$ only kept
the information about the $\KelD$-$\KelD$ component of $\Lambda^{-1}{}_{\Kela\Kelb}$, it is still interesting to notice the ``causality structure"~\cite{2009AdPhy..58..197K,Calzetta:2008iqa} of this operator.

\paragraph*{Classical-Classical component.}
First, the ``cl-cl" component in Eq.~\eqref{eq:IA} is zero. It is also easy to check that there is no term independent of $\varphi^\KelD$ in the fully nonlinear, purely IR Keldysh action $\So[\varphi^{\indX\indI\KelC},\varphi^{\indX\indI\KelD}]$, as can be seen for example from its expansion~\eqref{eq: S0 linear in varphi-q} in the quantum components of the IR fields.
This means that for vanishing quantum components
$\varphi^{\indX\indI\KelD}=0$, the effective action $S_\mathrm{eff}=\So+S_\mathrm{IA}$ is zero: $\Seff[\varphi^{\indX\indI\KelC},0]=0$. This was expected because for $\phiIR^\KelD=0$, the fields coincide on the forward and backward parts of the closed-time contour and thus the two contributions cancel each other. A last interpretation is that the quantum components do not propagate alone and must mix to classical ones.
In this respect, note that although our derivation was done at lowest non-trivial order in the quantum components of the fields and momenta, with $\Lambda \to \Lambda(\varphi^{\KelC})$
in the path integral over UV modes, this property actually holds non-perturbatively. Indeed, any correction to the current computation would be proportional to powers of $\varphi^\KelD$, 
and thus would still be vanishing when evaluated on configurations with purely classical components.

\paragraph*{Classical-Quantum component.}
This component is interesting because, although non-zero in Eq.~\eqref{eq: Lambda in Keldysh}, it results in a vanishing contribution to the influence action
after contracting with the ``time-derivatives" 
$\phiIR^\prime_{\indX\indI}{}^{\Kela} \propto \delta^\Kela_\KelD$,
a property inherited from the boundary conditions~\eqref{eq: boundary conditions}.
If $\phiIR^{\indI\KelC}$ and $Q^{\indI\KelD}$ were not vanishing at $N_\sigma(k)$, $\SIA$ would be augmented by a cross-term of the form $\left[\varphi^\KelC \left(\Lambda^{-1}\right)_{\KelC,\KelD} \varphi^\KelD +\left( \KelC \leftrightarrow \KelD \right)    \right]$ and proportional to the imaginary part of the power spectrum.
The mixed ``q-cl"/``cl-q" components in the influence action are more generally known for describing the dissipation of the system (the IR modes) by backreacting on the environment (the UV modes), and being responsible for the famous fluctuation-dissipation theorem. 
Indeed, if no boundary condition was imposed at the time $\tsigma(k)$, the classical field configurations $\phiIR_{\indX\indI}{}^\KelC(x)$ would get an extra friction term in their equations of motion of the form 
$-2\int\dd^4x^\prime\,\Im\Pi^{\indX\indY\indI\indJ}(x,x^\prime)a^3(N^\prime)\phiIR_{\indY\indJ}{}^\KelC(x^\prime)$.
However in our setup of stochastic inflation, the continuous flow from UV to IR modes via the time-dependent cutoff $\ksigma(N)$ is unidirectional and we expect no such backreaction, and thus no dissipation.\footnote{The corresponding mass and friction terms entailed by this classical-quantum component were neglected by hand in Refs.~\cite{Morikawa:1989xz,Matarrese:2003ye,Levasseur:2013ffa}. Moss and Rigopoulos cast doubt on the naive way to perform the IR-UV decomposition by a time-dependent window function~\cite{Rigopoulos:2016oko,Moss:2016uix}, and indeed, in Refs.~\cite{Tokuda:2017fdh,Tokuda:2018eqs}, Tokuda and Tanaka carefully showed that the stochastic theory enables one to recover the free propagators only when choosing the appropriate boundary conditions in the Keldysh basis, with the consequence of prohibiting the classical-quantum component.}

\paragraph*{Quantum-Quantum component.} 
The ``q-q" component is the only one that survives in the influence action after contracting with the pseudo ``time-derivatives", and because it is quadratic in the quantum parts of the fields and momenta, it constitutes a non-trivial quantum correction to the classical dynamics, again describing the effects of the integrated out short-scale fluctuations on the IR sector. Let us now discuss its physical implications.\\

Strikingly, the influence action~\eqref{eq:IA} is purely imaginary. This implies that in the path integral~\eqref{Z} over the IR components, the weights of configurations with non-zero quantum components are exponentially suppressed.
This important fact warrants that our expansion in the quantum components of the fields (and momenta) is well justified.
In a related manner, even though we do not use the formalism of density matrices in our paper, 
it can be shown quite generally
that such imaginary ``$\KelD$-$\KelD$" component in the influence action acts to suppress the off-diagonal terms of the reduced density matrix $\rho_\mathrm{r}(\phiIR^+,\phiIR^-)$ obtained after tracing out the environment (the UV modes here), a process that can be understood as decoherence (see, e.g. Refs.~\cite{Feynman:1963fq,Calzetta:2008iqa}).
The exponential suppression of the quantum components of the fields in the weigth $\ee^{iS_\mathrm{IA}}$ of the path integral is of course reminiscent of statistical field theory. 
Following the seminal paper of Feynman and Vernon~\cite{Feynman:1963fq}, this insight is put to good use 
by performing what is sometimes called a Hubbard-Stratonovich transformation~\cite{1957SPhD....2..416S,1959PhRvL...3...77H}: introducing auxiliary fields $\xi^{\indX\indI}$, the exponential of the influence action can be rewritten as
\bae{
    \ee^{iS_\mathrm{IA}}=\int\mD\xi^{\indX\indI}P\left[\xi^{\indX\indI}; \phiIR^{\indX\indI\KelC}\right]\ee^{i\int\dd^4x\,a^3\xi^{\indX\indI}\phiIR_{\indX\indI}{}^\KelD},
    \label{HS}
}
where $P\left[\xi^{\indX\indI} ; \varphi^{\indX\indI\KelC}\right]$ denotes the Gaussian weight
\bae{
    P[\xi^{\indX\indI}; \phiIR^{\indX\indI\KelC}]=\sqrt{\Det( 2 \pi \Re\Pi)}^{-1}\exp\left[-\frac{1}{2}\int\dd^4x\dd^4x^\prime\xi^{\indX\indI}\left(\Re\Pi^{-1}_{\indX\indY\indI\indJ}\right)_{\phiIR^{\indX\indI\KelC}}\xi^{\indY\indJ}\right]\,,
    \label{Gaussian-weight}
}
and where the subscript $\phiIR^{\indX\indI\KelC}$
recalls that $\Pi$, as essentially the Green's function of $\Lambda(\phiIR^{\indX\indI\KelC})$, can thus be seen as a (complicated) functional of the IR classical components $\phiIR^{\indX\indI\KelC}$. The manipulation~\eqref{HS}--\eqref{Gaussian-weight} is a simple mathematical identity, in essence the inverse of a Gaussian integration. Yet, it offers a very useful physical insight. Indeed, the partition function~\eqref{Z} can now be rewritten as
\bae{
    Z&=\int \mD \phiIR^{\indX\indI\Kela}  \exp\left(i\Seff[\varphi^{\indX\indI\Kela}] \right)\, \label{Z-new} \\
   &=\int \mD \varphi^{\indX\indI\KelC } \int\mD\xi^{\indX\indI} P\left[\xi^{\indX\indI};\phiIR^{\indX\indI\KelC}\right]
   \int \mD \varphi^{\indX\indI\KelD} \,\mathrm{exp}(
     \underbrace{ i \So\left[\varphi^{\indX\indI\Kela}\right]+i\int\dd^4x\,a^3\xi^{\indX\indI}\phiIR_{\indX\indI}{}^\KelD}_{\textstyle \begin{array}{c}
	i \Steff[\varphi^{\indX\indI\Kela},\xi^{\indX\indI}]\end{array}}
   )\,,  \nonumber
}
with
$\int\mD\xi^{\indX\indI}P[\xi^{\indX\indI};\phiIR^{\indX\indI\KelC}]=1$ for any realisation of $\phiIR^{\indX\indI\KelC}$. Upon the introduction of the Hubbard-Stratonovich fields $\xi^{\indX\indI}$,  the imaginary quadratic interactions of the quantum components in $\Seff[\varphi^{\indX\indI\Kela}]$ have been turned into a real linear coupling between the quantum components and the auxiliary fields in the new real effective action $\Steff[\varphi^{\indX\indI\Kela},\xi^{\indX\indI}]$.
Of course, the physical interpretation behind Eq.~\eqref{Z-new} is that $P\left[\xi^{\indX\indI};\phiIR^{\indX\indI\KelC}\right]$ endows the $\xi$'s with Gaussian statistics with
\bae{\label{eq: noise stat.}
    \bce{
        \dps
        \braket{\xi^{\indX\indI}(x)} \equiv \int\mD\xi^{\indX\indI}\xi^{\indX\indI}(x)P\left[\xi^{\indX\indI};\phiIR^{\indX\indI\KelC}\right]=0, \\
        \dps
        \braket{\xi^{\indX\indI}(x)\xi^{\indY\indJ}(x^\prime)}\equiv \int\mD\xi^{\indX\indI}\xi^{\indX\indI}(x)\xi^{\indY\indJ}(x^\prime)P\left[\xi^{\indX\indI};\phiIR^{\indX\indI\KelC}\right]=
        \left.\Re\,\Pi^{\indX\indY\indI\indJ}(x,x^\prime)\right|_{\phiIR^{\indX\indI\KelC}}.
    }
}
The delta function $\delta(N-N^\prime)$ in $\Re\Pi$, see Eq.~\eqref{eq: Pi tensor}, indicates that $\xi$'s can be interpreted as Gaussian white noises, like in the heuristic approach, with amplitudes determined by the power spectra of the UV modes on the ``background" of the IR classical components. Additionally, it is interesting to notice that the reality of the noise is guaranteed in this first-principle derivation, contrary to the heuristic approach where this feature has to be added by hand (see Sec.~\ref{limitations}).
The partition function~\eqref{Z-new} together with equations~\eqref{eq: Pi tensor} and \eqref{eq: noise stat.} represent one of the main results of this paper.\\

It is now relatively straightforward to take into account the effect of the quantum components on the classical ones. Indeed, recall that our computation of $\Seff[\varphi^{\indX\indI\Kela}]$ was made up to quadratic order in the quantum components. 
Consistently neglecting cubic terms in the expression~\eqref{eq: S0 linear in varphi-q} for $\So$, the quantum components therefore enter only linearly in $\Steff[\varphi^{\indX\indI\Kela},\xi^{\indX\indI}]$, and the path integral over them can hence be performed explicitly, yielding the delta functional $\delta\left( \left.\var{\So\left[\phiIR^{\indX\indI}\right]}{\phiIR_{\indY\indJ}}\right|_{\phiIR^{\indX\indI}=\phiIR^{\indX\indI\KelC}}+a^3 \xi^{\indY\indJ}\right)$  
in the remaining path integral over the classical components of the IR fields, $\phiIR^\KelC$, and the auxiliary variables $\xi$.
Thus, the only trajectories with non-zero weights in the path integral are the ones that verify the following equations of motion:
\bae{\label{eq: raw Langevin}
    \phiIR^{\indI\KelC}{}^\prime=\frac{\piIR^{\indI\KelC}}{H}+\xi^{Q\indI}, \qquad  
    \piIR_\indI^{\KelC}{}^\prime=-3\piIR_\indI^\KelC-\frac{V_\indI\left(\phiIR^{\indI\KelC}\right)}{H}+\frac{1}{H}\Gamma^{\indK}_{\indI\indJ}\left(\phiIR^{\indI\KelC}\right)\piIR_\indK^\KelC\piIR^{\indJ\KelC} +\xi^{P}_\indI,
}
where we will omit to write explicitly ``cl" for simplicity in what follows.
While the first equation is already in a manifestly covariant form, the second one is not. However this is not surprising as neither $\piIR_\indI^\prime$ nor $P_\indI$ (that was integrated out), are covariant quantities themselves. However this equation does respect general covariance, as is seen by using $\xi^{P}_\indI=\xi^{\covP}_\indI+\Gamma^\indK_{\indI\indJ}\piIR_\indK\xi^{Q\indJ}$, as well as $\piIR_I^\prime=\covD_N\piIR_I+\Gamma^\indK_{\indI\indJ}\piIR_\indK\phiIR^{\indJ\prime}$ and replacing $\phiIR^{\indJ\prime}$ by its value according to the first Langevin equation. Eventually, the stochastic EoM~(\ref{eq: raw Langevin}) can be summarised in an explicitly covariant way as (again, removing the ``cl" exponent for conciseness) 
\bae{\label{eq: covariant Langevin}
    \phiIR^{\indI\prime}=\frac{\piIR^\indI}{H}+\xi^{Q\indI}, \qquad
    \covD_N\piIR_\indI=-3\piIR_\indI-\frac{V_I}{H}+\xi^{\covP}_\indI.
}
As we discussed at length in Sec.~\ref{sec: stochastic anomalies}, these Langevin equations should be understood as the continuous limit of a discrete process with a Stratonovich scheme. Moreover, the identification of independent quantum fields in the Bunch-Davies regime provides one, upon classicalisation, with an essentially  unique set of independent white noises with which to formulate these Stratonovich Langevin-type equations. As we also explained there, It\^o's discretisation also has a number of advantages, and one can convert the latter equations into the corresponding It\^o's ones as: 
\bae{\label{eq: Ito-Langevin-final}
    \boxed{
    \ItoD_N\phiIR^\indI=\frac{\piIR^\indI}{H}+\xi^{Q\indI}, \qquad \ItoD_N\piIR_\indI=-3\piIR_\indI-\frac{V_\indI}{H}+\xi^\covP_\indI,
    }
}
with use of the It\^o-covariant derivatives~(\ref{eq: ItoD for X})--(\ref{eq: ItoD for V}). Let us also remind the reader that the local Hubble parameter $H$ is explicitly given in terms of the IR fields and momenta through the  Friedmann constraint~\eqref{laspeIR}
\bae{
    3\Mpl^2H^2=\frac{1}{2}G^{IJ}(\phiIR)\piIR_I\piIR_J+V(\phiIR)\,,
}
without modification compared to the heuristic approach.

Eventually, let us comment on the status of these equations. As the derivation above shows, these are the semi-classical equations governing the trajectories that have a non-zero weight in the closed-time path integral, but they do not yet correspond to physical quantities: the expectation values of the quantum theory
are only recovered once taking the ensemble averages over the noises. More precisely, this statistical average exactly reproduces the quantum average only when the initial effective action $\Seff[\varphi^{\indX\indI\Kela}]$ is at most quadratic in the quantum components, resulting in the above delta functional (letting aside here the fact that we only integrated out the UV fluctuations at quadratic order in the action). It is in that sense that the stochastic equations~\eqref{eq: Ito-Langevin-final}, derived at lowest non-trivial order in the quantum components, can be qualified as ``semi-classical''.

As described in Sec.~\ref{sec: stochastic anomalies}, physical quantities derived from It\^o's SDE~\eqref{eq: Ito-Langevin-final} only depend on the auto-correlation of the noises, which is physically specified by the UV two-point correlations. The presence of It\^o-covariant derivatives also manifestly guarantees general covariance.
These equations are thus free from any stochastic anomaly.
Furthermore, as we stressed above, the reality of noises is also ensured,  
as their auto-correlations~\eqref{eq: noise stat.} derived from the CTP formalism are automatically given by the real part of the UV two-point functions.

\section{Markovian analytical approximations and phase-space Fokker-Planck equation}
\label{sec:Markovian}

As we explained in Sec.~\ref{subsec:Markovian?}, stochastic inflation is strictly speaking not described by a Markov process. Indeed, the noise amplitude is the solution of the differential equation verified by the UV modes which develop on the stochastic IR background, rather than an explicit function of the IR fields themselves. In particular, the noise amplitude \emph{a priori} depends on the whole past history of the stochastic process.
However, in some situations, the noise amplitude can be approximately expressed in terms of the instantaneous IR fields, in which case the dynamics can be thought of as Markovian
and a powerful tool becomes accessible: the Fokker-Planck equation. In this section, we deal with 
these Markovian cases.  We begin by showing the covariant Fokker-Planck equation that dictates the evolution of the one-point probability density function (PDF) for the IR fields and momenta, that can be inferred from the Langevin equations~\eqref{eq: Ito-Langevin-final} when assuming a Markovian dynamics. Then we show how to approximate the noise amplitude, first in the simpler situation in which the scalar fields are strictly massless, and then in a generic case under the assumption of slow-varying masses.

\subsection{Covariant Fokker-Planck equation in phase space}

Let us first reemphasise that throughout this work, we treat the fields as locally homogeneous, i.e. at leading order in a gradient expansion. Although this might seem very crude, following the separate universe approach, this nonetheless enables one to capture the full nonlinear dynamics on super-Hubble scales.
Hence, as described in Sec.~\ref{subsec: stochastic equations}, the Langevin equations~\eqref{eq: Ito-Langevin-final} govern the stochastic dynamics of a representative $\sigma$-Hubble patch.
In the Markovian limit, with the assumption that the noise amplitudes are well approximated as functions of the current IR field values (and momenta), these Langevin equations give rise to the corresponding Fokker-Planck (FP) equation, with use of the rule presented in Appendix~\ref{subsec:FP}, as
\bae{\label{eq: phase space fokker-planck}
	\partial_N P=&-\phasecovD_{\varphi^I}\left[\frac{G^{IJ}}{H}\varpi_J P\right]+\partial_{\varpi_I}\left[\left(3\piIR_\indI+\frac{V_\indI}{H}\right)P\right] \nonumber \\
	&+\frac{1}{2}\phasecovD_{\varphi^I}\phasecovD_{\varphi^J}(A^{QQ\indI\indJ}P)+\phasecovD_{\varphi^I}\partial_{\varpi_J}(A^{Q\covP\indI}{}_\indJ P)
	+\frac{1}{2}\partial_{\varpi_I}\partial_{\varpi_J}(A^{\covP\covP}{}_{\indI\indJ}P)\,.
}
Here we defined a last covariant derivative 
$\phasecovD_{\phiIR^\indI}$ with respect to the IR fields, the phase-space one, whose action on a rank-1 tensor is $\phasecovD_{\phiIR^\indI}\genU^\indJ=\nabla_\indI\genU^\indJ+\Gamma_{\indI\indL}^\indK\piIR_\indK\partial_{\piIR_\indL}\genU^\indJ$ and generalisation to rank-$n$ tensors is straightforward. 
As for the $A^{\covX\covY\indI\indJ}$'s, these are the noises' auto-correlations at coincident points:
\bae{
    A^{\covX\covY\indI\indJ}(N)\delta(N-N^\prime)=\braket{\xi^{\covX\indI}(N)\xi^{\covY\indJ}(N^\prime)}=\frac{\ksigma^\prime}{\ksigma}  \Re[\calP^{\covX\covY\indI\indJ}(N,\ksigma(N))]\delta(N-N^\prime),
    \label{noises-FP-coincident}
}
which are here assumed to be functions of $\phiIR^\indI(N)$ and $\piIR_\indI(N)$,
and up to the factor $\ksigma^\prime/\ksigma$ that may be approximated by unity, are nothing else than the real parts of the UV power spectra.

One should remember that in the FP equation in field space~\eqref{eq: covariant fokker-planck}, which we previously wrote for pedagogical reasons, the scalar PDF $\Ps$ is rescaled compared to the PDF $P$ that directly results from the Langevin equations. 
Here, on the contrary, it is easy to check that the phase-space PDF $P(\varphi^\indI,\varpi_\indI,N)$ (truly the transition probability given some initial state), is already a scalar quantity, without the need of any rescaling.
In this respect, although we skipped the intermediate steps of the computation, we stress that Eq.~\eqref{eq: phase space fokker-planck} is not postulated, but simply derived from the Langevin equations and Eq.~\eqref{FP-from-Langevin}, with covariant phase-space derivatives naturally emerging from the computation.
Given the important complexity of the phase-space It\^o-Langevin equations~\eqref{eq: Ito-Langevin-final}, the manifestly covariant form of the FP equation~\eqref{eq: phase space fokker-planck} is rather remarkable and provides a non-trivial consistency check of the former.
This equation generalises the FP equation that we proposed in our previous paper in a simpler setup~\cite{Pinol:2018euk}: in field space and for test scalar fields in de Sitter spacetime only. Also, the ``stochastic anomalies" were not solved there, and the form of the FP equation was simply assumed based on the requirement of general covariance.  
Not only do we present here the derivation of this phase-space FP equation, but we are also confident that it can now be used to compute correlation functions of multifield inflation with curved field space in realistic setups where the fields backreact on the geometry of spacetime.
However, it would be restrictive to consider that the virtue of this equation only concerns these situations: the It\^o-Stratonovich ambiguity was also plaguing single-field inflation, and our first-principle derivation, with emphasis on manifest covariance, enabled us to solve it in this simpler context as well.

The remaining nontrivial difficulty is now to prescribe values for the auto-correlation of the noises $A^{\covX\covY\indI\indJ}$, 
and in the next two sections, we turn to interesting particular cases where we can give analytical estimates.
Note that this will be possible because we assume from now on a slow-varying regime, which was not the case until here.
Also, because the dynamics of the UV modes in SPT is conveniently solved in terms of the conformal time $\tau$ such that $\dd N=  a H \dd \tau$, we will also make use of this time variable in what follows.
In our context in which $H$ is a stochastic quantity, conformal time is strictly speaking not a deterministic variable like the number of \efolds, but we will nonetheless make this approximation, justified as follows.
The noise auto-correlation at time $N$ only depends on the UV fluctuations with wavenumber $\ksigma(N)$, which
exited the Hubble radius $\simeq -\mathrm{ln}(\sigma)$ \efolds before $N$. As the UV fluctuations follow the Bunch-Davies behaviour until only a few \efolds before Hubble crossing (all the more so for light fields of particular relevance in the stochastic formalism), in practice it is necessary to follow the evolution of a given mode $k_\sigma(N)$ during only for a few \efolds (typically 5), a duration that is not large enough for stochastic effects to accumulate and significantly affect the local Hubble scale.

\subsection{Massless limit}
\label{sec:light}

For analytically understanding the UV fluctuations, it is particularly useful to use the projections of the mode functions on a set of parallel-transported vielbeins, the $Q^\alpha_\indA$ introduced in Eq.~\eqref{def-projected-perturbations}.
They provide independent degrees of freedom deep inside the Hubble radius, only mixing via the projected mass matrix $M^2{}^\alpha{}_\beta$, as can be seen from their EoM~\eqref{eq: UV eom mode vielbein}.
In this section, we consider that this projected mass matrix is completely negligible.
By consistency of the slow-varying approximation, we also use the zeroth-order, locally de-Sitter expression of the scale factor $a(\tau)\simeq -1/(\Hs \tau)$, where $\Hs$ denotes the Hubble scale, considered constant around Hubble crossing, i.e. in the period interpolating between the Bunch-Davies regime and the crossing of the coarse-graining scale, such that $k_\sigma(\tau) \tau \simeq -\sigma (1+{\cal O}(\epsilon))$.

Under these conditions, the mode functions $Q^\alpha_\indA$ simply provide $N_\mathrm{fields}$ independent copies $\left(Q^\alpha_\indA \propto \delta^\alpha_\indA\right)$ of the standard single-field massless mode function in quasi de Sitter spacetime, which read, with Bunch-Davies initial condition:
\bae{
    Q^\alpha_\indA(\tau,k)=-i\delta^\alpha_A\frac{\ee^{-ik\tau}}{a\sqrt{2k}}\left(1-\frac{i}{k\tau}\right) \,.
    \label{simple-mode-functions}
}
Note that we used the freedom of redefining mode functions with an arbitrary unitary matrix as explained in Sec.~\ref{subsec: classicalisation}, in order to choose a phase that leads to explicitly real values of $Q^\alpha_A$ (and $\covP^\alpha_A$) on super-Hubble scales.
From these mode functions, and using Eq.~\eqref{two-point-vacuum}, one deduces the multifield power spectrum of the UV modes at coarse-graining scale crossing:
\bae{
    \calP^{QQ\indI\indJ} 
    \left(\tau,k_\sigma(\tau)\right)&=\frac{k_\sigma(\tau)^3}{2\pi^2}e^\indI_\alpha e^\indJ_\beta Q^\alpha_\indA\left(\tau,k_\sigma(\tau)\right) Q^{\beta*}_\indA\left(\tau,k_\sigma(\tau)\right) \nonumber\\
    & = \left(\frac{H_\star}{2\pi}\right)^2G^{\indI\indJ}\left(1+\sigma^2 \right),
    \label{PQQ-simple}
}
where we recall that $\sigma \ll 1$, so that the last term should be neglected.
Notice that although mass effects are 
not taken into account in this section, the introduction of the parallel-transported vielbeins enables one to capture the geometrical effects of the curved field space at the level of UV fields $\left(\calP^{QQ\indI\indJ} \propto G^{\indI\indJ}\right)$. Then, neglecting slow-roll suppressed metric perturbations $\propto M^{2}{}_{\covP Q}$ in $\covP$ for consistency, the momentum UV modes read
\bae{
    \covP^\alpha_\indA 
    (\tau,k)&\simeq  
    \frac{\dd}{a\dd \tau} Q^\alpha_\indA(\tau,k)=-\delta^\alpha_A\frac{H^2 \ee^{-ik\tau}}{
    \sqrt{2k^3}}k^2 \tau^2\,, 
    \label{P-mode-function-simple}
}
so that using again Eq.~\eqref{two-point-vacuum}, one obtains, for the power spectra involving momenta :
\bae{
    \calP^{Q\covP\indI}{}_\indJ 
    \left(\tau,k_\sigma(\tau)\right)&=
    -\sigma^2 H_\star \left(\frac{H_\star}{2\pi}\right)^2\delta^\indI_\indJ\left(1-i\sigma\right), \\
    \calP^{\covP \covP}{}_{\indI\indJ}\left(\tau,k_\sigma(\tau)\right)&=\sigma^4 H_\star^2 \left(\frac{H_\star}{2\pi}\right)^2G_{\indI\indJ}\,.
    }
Note that the cross power-spectrum has a non-zero imaginary part $\Im\calP^{Q\covP\indI}{}_\indJ=\sg^3H_\star\left(\frac{H_\star}{2\pi}\right)^2\delta^\indI_\indJ$, remnant from the quantum nature of the scalar fields, and completely fixed by the non-vanishing commutation relation between $Q$ and $\covP$ in Eq.~\eqref{eq: commutation relations}.
Naturally, the fact that it is suppressed by the small parameter $\sigma$ is related to the highly squeezed state of the fluctuations and to the fact that they ``classicalise''
on super-Hubble scales. However, notice anyway that the Schwinger-Keldysh derivation shows that only the real parts of the power spectra appear in the statistics of the stochastic noises.

In the strict massless and ``slow-roll'' regime of this section, the mode functions of the momenta~\eqref{P-mode-function-simple} at coarse-graining scale crossing are suppressed by $\sigma^2$ compared to the ones of the fields~\eqref{simple-mode-functions}, hence the power spectra involving the former should be self-consistently set to zero in practical computations. However, this property only holds within this framework, and in general, the power spectra involving momenta, while 
``slow-roll'' suppressed, are not $\sigma$ suppressed and should be considered, as we will show in the next section.

\subsection{Slow-varying masses}
\label{sec:generic}

Let us now go one step further and consider the effects of a non-zero mass matrix $M^2{}^\alpha{}_\beta$. First we notice that at early times the mass term is negligible compared to the gradient term, i.e., $\forall\alpha,\beta, \quad M^2{}^\alpha{}_\beta\ll k^2/a^2$.
Thus initial conditions and the first stage of evolution of the perturbations are equivalent to the massless case. However, the behaviour is different around Hubble crossing. To identify  these non-trivial mass effects, we make the assumption that the projected mass matrix is approximately constant in the period interpolating between the Bunch-Davies regime and the crossing of the coarse-graining scale, a feature observed in many concrete models of inflation.
It is then possible to diagonalise the mass matrix locally, \emph{around the time of Hubble
crossing}, making use of the set of mass eigenvalues $m_\indi^2$ and corresponding
eigenvectors $\D^\alpha{}_\indi$ such that
$M^{2\alpha}{}_\beta\D^\beta{}_\indi=m_\indi^2\D^\alpha{}_\indi$\, (no sum on $\indi$).
According to our assumption, these quantities can be considered constant in the interpolating period, so that the vielbein-basis EoM~\eqref{eq: UV eom mode vielbein} result then in the simple set of diagonal equations in the mass eigenbasis:
\bae{\label{eq: UV eom mode eigenbasis}
    \partial_N^2 Q^\indi_\indA 
    +\left(3-\epsilon\right)\partial_NQ^\indi_\indA 
    +\left(\frac{k^2}{a^2H^2}+\frac{m^2_\indi}{H^2}\right)Q^\indi_A=0, \qquad \text{(no sum on $\indi$)},
}
where $Q^\indi_\indA=(\D^{-1})^\indi{}_\alpha Q^\alpha_\indA$
denotes the projected mode functions on the mass eigenbasis. 
We here note that the mass matrix $M^{2\alpha}{}_\beta$ is real and symmetric, hence the mass eigenvalue $m_i^2$ are real, and one can take the diagonalising matrix $\D^\alpha{}_i$ to be a real orthonormal matrix 
(with $(\D^{-1})^\indi{}_\alpha=(\D^T)^\indi{}_\alpha=\D^\alpha{}_\indi$). 
It is important to notice that the mass eigenvalues $m_\indi^2$ are scalars in field space, and that they also correspond to eigenvalues of the original mass matrix $M^2{}^\indI{}_\indJ$, with eigenvectors given by $e^\indI_\indi=e^\indI_\alpha\D^\alpha{}_\indi$, i.e.
\bae{
    M^{2\indI}{}_\indJ e^\indJ_\indi=m_i^2e^\indI_\indi, \qquad \text{(no sum on $\indi$)}.
    \label{originalM2-mi}
}
Moreover, taking into account the orthonormality of $\D^\alpha{}_\indi$, the set of vectors $e^\indI_\indi$, rotated from the vielbeins $e^\indI_\alpha$, constitute another set of vielbeins, hence satisfying $G_{\indI\indJ}e^\indI_\indi e^\indJ_\indj=\delta_{\indi\indj}$ and $\delta^{\indi\indj}e^\indI_\indi e^\indJ_\indj=G^{\indI\indJ}$.

The initial conditions for the $Q^\indi_\indA$ are simply given by
$Q^\indi_A(\tau,k) \underset{-k\tau\gg 1}{\to} 
\frac{-i}{a\sqrt{2k}}(\D^{-1})^\indi{}_\indA \ee^{-ik\tau}$ with $(\D^{-1})^\indi{}_\indA=(\D^{-1})^\indi{}_\alpha\delta^\alpha_\indA$, so that the corresponding solution of Eq.~\eqref{eq: UV eom mode eigenbasis} reads, at leading-order in the slow-varying approximation:
\bae{\label{eq: mode Q}
    Q^\indi_\indA(\tau,k)=
    (\D^{-1})^\indi{}_\indA Q^\indi(\tau,k), \qquad \text{(no sum on $\indi$)},
}
with $Q^\indi$ the familiar single-field mode function
\bae{
    Q^\indi(\tau,k)=\frac{\ee^{i(\nu_\indi-1/2)\pi/2}}{2a}\sqrt{-\pi\tau}H_{\nu_\indi}^{(1)}(-k\tau), \quad \text{with }
   \nu_i= \bce{
        \dps
        \sqrt{\frac{9}{4}-\frac{m_\indi^2}{H^2}}, & \text{if }
        \dps
        \frac{m_\indi^2}{H^2}<\frac{9}{4}, \\[10pt]
        \dps
        i\sqrt{\frac{m_\indi^2}{H^2}-\frac{9}{4}}, & \text{if }
        \dps
        \frac{m_\indi^2}{H^2}\geq\frac{9}{4}.
    }
    \label{def-Qi}
}
expressed in terms of $H_{\nu_\indi}^{(1)}$, the Hankel function of the first kind and of order $\nu_\indi$.
Hence, one obtains the dimensionless power spectrum of UV modes at the time when $k=\sigma aH$ as
\bae{\label{eq: UV PS massive case}
    \calP^{QQ\indI\indJ}(\tau,\ksigma(\tau))=\frac{\ksigma^3(\tau)}{2\pi^2}\sum_\indi e^\indI_\indi e^\indJ_\indi|Q^\indi(\tau,\ksigma(\tau))|^2,
}   
where we used the orthonormality of the matrix $D^\alpha{}_\indi$.
The result~\eqref{eq: UV PS massive case} is interesting because intermediate steps like the parallel-transported vielbeins or the diagonalising matrix $D^\alpha{}_\indi$ disappear altogether: to compute the right-hand side, the only requirement is to know the 
mass eigenvalues $m^2_\indi$ and the corresponding eigenvectors $e^I_i$ forming a set of vielbeins, Eq.~\eqref{originalM2-mi}, which is easy to obtain numerically from $M^{2\indI}{}_\indJ$ once a position in phase space $\left(\phiIR^\indI,\piIR_\indI\right)$ is specified. Moreover, the sum in Eq.~\eqref{eq: UV PS massive case} is nicely understood as a mass-weighted metric in field space, and indeed, the massless limit $\propto G^{\indI\indJ}$ is easily recovered by setting $\nu_\indi=3/2,\, \forall i$.
Notice also that $\calP^{QQ\indI\indJ}$ is automatically real and symmetric, as should be from first principles (see Sec.~\ref{subsec: classicalisation}).

Moving to momenta, and according to the UV EoM~\eqref{eq: UV EoM}, one obtains
\bae{\label{eq: covPi}
    \covP^\indi_\indA=H \partial_NQ^\indi_\indA 
    -\frac{\piIR^\indi\piIR_\indj}{2\Mpl^2H}Q^\indj_\indA,
}
where $\piIR_\indi=e^I_i \varpi_I$. Using the properties of Hankel functions, we note that 
the time derivative of $Q^i$ can be simply expressed, at leading-order in the slow-varying approximation, as
\footnote{For completeness, at next order one has instead $q_{\nu_\indi}(\sigma)=\left(\nu_i-\frac{3}{2}\right)+\epsilon\left(\frac{1}{2}-\nu_i \right)-\sigma\frac{H_{\nu_\indi-1}^{(1)}(\sigma)}{H_{\nu_\indi}^{(1)}(\sigma)}$, 
$\nu_i=\sqrt{9/4+3\epsilon -(1+2\epsilon)m_i^2/H^2}$, and assuming both $\epsilon$ and $m_i^2/H^2$ small, one further obtains $q_{\nu_i}(\sigma) = -m_i^2/(3H^2)+ O(\epsilon^2) + O(\epsilon \times m_i^2/H^2) + O(\sigma^2)$.
However, evaluating $q_{\nu_i}$ beyond leading-order is 
too precise compared to the rest of our computation, as we have anyway considered the projected mass matrix to be constant.}
\bae{
    \partial_NQ^\indi|_{\ksigma} 
    =q_{\nu_\indi}(\sigma)Q^\indi|_{\ksigma}, \quad \text{with} \quad q_{\nu_\indi}(\sigma)= \nu_\indi-\frac{3}{2} 
    -\sg\frac{H_{\nu_\indi-1}^{(1)}(\sg)}{H_{\nu_\indi}^{(1)}(\sg)}\,,
    \label{def-qnu}
}
where we evaluated it at the time of crossing of the coarse-graining scale. One therefore obtains
\bae{
    \covP^\indi_\indA|_{\ksigma}=\sum_{\indj}\left(Hq_{\nu_\indi}(\sigma)\delta^\indi_\indj-\frac{\piIR^\indi\piIR_\indj}{2\Mpl^2H}\right)   (\D^{-1})^\indj{}_\indA Q^\indj|_{\ksigma}  
    \equiv \sum_{\indj} \scrQ^\indi{}_\indj 
    (\D^{-1})^\indj{}_\indA Q^\indj|_{\ksigma} ,
}
so that all power spectra at coarse-graining scale crossing can be summarised as
\bae{
    \calP^{QQ\indI\indJ}&=\frac{\ksigma^3}{2\pi^2}\sum_{\indi}e^\indI_\indi e^\indJ_\indi|Q^\indi|^2_{\ksigma},
    \label{eq: power spectra massive case-QQ}\\
    \calP^{Q\covP\indI}{}_\indJ&=\frac{\ksigma^3}{2\pi^2}\sum_{\indi,\indj}e^\indI_\indi e_{\indJ\indj}\scrQ^{*\indj}{}_\indi|Q^\indi|^2_{\ksigma},
    \label{eq: power spectra massive case-QP}\\
    \calP^{\covP\covP}{}_{\indI\indJ}&=\frac{\ksigma^3}{2\pi^2}\sum_{\indi,\indj,\indk}e_{\indI\indi}e_{\indJ\indj}\scrQ^\indi{}_{\indk}\scrQ^{*\indj}{}_\indk|Q^\indk|^2_{\ksigma}.
    \label{eq: power spectra massive case-PP}
}
One knows from first principles that $\calP^{\covP\covP}{}_{\indI\indJ}$ should be real and symmetric, similarly to $\calP^{QQ\indI\indJ}$, while $\Im\calP^{Q\covP\indI}{}_\indJ=\sg^3H_\star\left(\frac{H_\star}{2\pi}\right)^2\delta^\indI_\indJ$. Because our analytical expressions are based on several approximations, these properties are not necessarily precisely verified by Eqs.~\eqref{eq: power spectra massive case-QP}--\eqref{eq: power spectra massive case-PP}, a discrepancy that can be used as a quantitative diagnostic of the quality of the approximations in practical numerical computations. However, we note once again that only the real parts of the power spectra anyway enter into the properties of the stochastic noises (see Eq.~\eqref{noises-FP-coincident}).

Although obtained for analytically estimating the noises amplitudes in stochastic multifield inflation, 
Eqs.~\eqref{eq: power spectra massive case-QQ}--\eqref{eq: power spectra massive case-PP} are of more general interest in the context of multifield inflation with slow-varying quantities, replacing $\sigma$ by $k/aH$ when necessary, and they constitute new results to the best of our knowledge.\footnote{A related formula for the trace $G_{IJ}\calP^{QQ\indI\indJ}$ has already been used without proof in Ref.~\cite{McAllister:2012am}, with excellent agreement with exact numerical computations.}
Given the number of approximations performed, it is difficult to control the degree of accuracy of the above formulae, but they constitute a proof of principle that it is possible to obtain Markovian analytical approximations, and they constitute a basis for future improvements.

The discussion has been kept quite general until now, but as is well known, the behaviour of super-Hubble fluctuations strongly depends on the mass parameter. Hence, 
from these generically applicable formulas,
in the stochastic context, two physically different regimes should be distinguished depending on the various values of $m_\indi^2$, leading either to real positive $\nu_\indi$ for ``light'' fields (the first line in Eq.~\eqref{def-Qi}), or imaginary $\nu_\indi \equiv i \mu_i$ for heavy fields (the second line there).
For heavy fields, one can write
\bae{
\frac{\ksigma^3(\tau)}{2\pi^2}|Q^\indi(\tau,\ksigma(\tau))|^2 \underset{ \frac{m_\indi^2}{H^2}\geq\frac{9}{4}}
{=} 4 \pi \ee^{-\mu_i \pi}\left(\frac{H}{2\pi}\right)^2 \left(\frac{\sg}{2}\right)^{3} |H_{i\mu_i}^{(1)}(\sg)|^2\,,
\label{mod-square-massive}
}
with the small argument expansion 
\bae{
H_{i\mu}^{(1)}(\sg) \underset{\sg \ll 1}{\simeq}  -i  \frac{\Gamma(i \mu)}{\pi}\left(\frac{\sigma}{2} \right)^{-i \mu}+\frac{1+\coth(\mu \pi)}{\Gamma(1+i \mu)}\left(\frac{\sigma}{2} \right)^{i \mu}\,.
}
The factor $|H_{i\mu_i}^{(1)}(\sg)|^2$ in Eq.~\eqref{mod-square-massive} hence describes the characteristic super-Hubble oscillations of heavy fields, but more importantly here, the power spectrum~\eqref{mod-square-massive} is suppressed by $\sigma^3$. This explicit dependence on the a priori arbitrary coarse-graining parameter $\sigma$ is not really worrisome: it simply comes from the fact that fluctuations of heavy fields are strongly suppressed on super-Hubble scales, and should simply be discarded from the stochastic description (and in the sums~\eqref{eq: power spectra massive case-QP}--\eqref{eq: power spectra massive case-PP}), whose aim is to describe the long-term dynamics generated by light scalars. Turning to them, and using
$H_\nu^{(1)}(\sg) \underset{\sg \ll 1}{\simeq} -(i/\pi) \Gamma(\nu)\left(\sg/2\right)^{-\nu}$, the last term  of $q_{\nu_i}(\sigma)$ in Eq.~\eqref{def-qnu} should be neglected as of order $\sigma^2$, and one obtains
\bae{
\frac{\ksigma^3(\tau)}{2\pi^2}|Q^\indi(\tau,\ksigma(\tau))|^2  \underset{ \frac{m_\indi^2}{H^2}< \frac{9}{4}}
{=} \left(\frac{H}{2\pi}\right)^2\left( \frac{\Gamma(\nu_\indi)}{\Gamma(3/2)} \right)^2 \left(\frac{\sg}{2}\right)^{3-2\nu_\indi}\,,
\label{mod-square-light}
}
here with only a power-law dependence on $\sigma$. 
This dependence can be neglected, and $\left(\frac{\sg}{2}\right)^{3-2\nu_\indi}$ can be approximated by unity, under the condition that the latter is taken to verify
\bae{
    \frac{\sg}{2} \gg \ee^{-\left(3-2\nu_i
    \right)^{-1}},
}
which is easily compatible with $\sigma \ll 1$ for a light enough mass (see Refs.~\cite{Starobinsky:1994bd,Grain:2017dqa} for discussions in a single-field context). 
For intermediate masses $0.1 \lesssim m_i^2/H^2 \lesssim 1$, stochastic effects are less important but may not be completely negligible (see, e.g., Ref.~\cite{Fumagalli:2019ohr}), and the resulting $\sigma$-dependence indicates that the coarse-graining procedure, made at leading-order in the gradient expansion, should be refined in order to properly treat these situations.
More precisely, let us add that in theories that are not completely scale invariant, it is expected that the Langevin equations, which describe the distribution of field values in $\sigma$-Hubble patches, do depend on $\sigma$. Yet another question is to see how $\sigma$ disappears when computing physical observables on scales much larger than the cutoff scale. It is likely that the stochastic-$\delta N$ approach needs to be modified to deal with these situations of intermediate masses, but this is largely outside the scope of this paper.

Eventually, as discussed in Sec.~\ref{subsec: classicalisation}, one can check explicitly that for light scalars with $m_i^2<9/4 H^2$, the complex mode functions $Q^\indI_\indA(N,k)$ and $\covP_{\indI\indA}(N,k)$ (or equivalently, $Q^i_A$ and $\covP^\indi_\indA$) become approximately real up to an irrelevant constant unitary matrix. This stems from the fact that, the $Q^\alpha$ being independent fields inside the Hubble radius, the variables $(\D^{-1})^\indi{}_\alpha Q^\alpha$, obtained by rotation of the former, equally provide a set of independent variables (and indeed, we have seen that the orthonormal matrix $D^\alpha{}_\indi$ drops out of all correlators). Hence, one could also have rotated the annihilation (and creation) operators $\hat{a}^{\indA}$ and absorbed in their definitions the individual phase factors $\ee^{i \nu_i \pi/2}$ of the mode functions~\eqref{def-Qi}. The corresponding transformation can be described by the relations \eqref{rotation-a-operators}--\eqref{rotation-basis} with the unitary matrix
\bae{
    \U^{\At}{}_\indB=\D^{\At}{}_\indi\,\diag\left(\ee^{i(\nu_i-3/2)\pi/2}\right){\!}^\indi{}_\indj(\D^{-1})^\indj{}_\indB\,,
}
with which one obtains the expressions
\bae{
\barQ^\indi_\At=(\D^{-1})^\indi{}_\At \ee^{- i (\nu_i-3/2) \pi/2} Q^i \quad \text{(no sum on $\indi$)}\,, \qquad
\bar{\covP}^\indi_{\At}=\scrQ^\indi{}_\indj \barQ^\indj_\At
}
that become manifestly real on super-Hubble scales.

\section{Conclusions}\label{sec: conclusions}

In this paper, we derive an effective stochastic theory for the super-Hubble, coarse-grained, scalar fields during inflation.
We do so in a phase-space approach and for the general class of nonlinear sigma models~\eqref{S-intro}, characterised by their potentials and curved field spaces.
We first give in section~\ref{sec: heuristic} a “heuristic” derivation of the corresponding Langevin equations in phase space, in order to introduce concepts and notations used all the way. We point out the limitations of the heuristic approach that uses the classical equations of motion, as well as the non-Markovian nature of the dynamics. Section~\ref{sec: stochastic anomalies} is devoted to the resolution of the “inflationary stochastic anomalies” that we pointed out in our previous paper~\cite{Pinol:2018euk}: because of the very quantum nature of the scalar fields, the theory contains a preferred frame that corresponds to the basis of independent creation and annihilation operators. 
This frame must be used to define independent noises in the Langevin equations, removing the possibility of any
ambiguity in the choice of such a frame, and the corresponding Langevin equations should be interpreted according to a Stratonovich, midpoint, discretisation scheme.
Along this discussion, we show how the classicalisation of quantum fluctuations on super-Hubble scales enables one to interpret the noises as classical random variables rather than quantum operators. We also show explicitly the transformation of the Stratonovich-Langevin equations to their It\^o version by the addition of noise-induced drifts, and explain how these terms can be combined with the usual time-derivatives to define new, covariant in It\^o calculus, time-derivatives. With the final form~\eqref{Langevin-intro}, the Langevin equations can be readily used in numerical and analytical computations.

Section~\ref{sec: effective hamiltonian action} is devoted to the rigorous derivation of the Langevin equations using a path-integral approach, which solves the remaining conceptual issues of the heuristic one. We begin by recalling that for 
the intrinsically time-dependent problems of interest in cosmology, like in other nonequilibrium situations, the relevant partition function is the one that dictates “in-in” correlation functions and causal equations of motions, and that it is defined by a closed-time path of integration. Equivalently, in this also called Schwinger-Keldysh formalism, the degrees of freedom are doubled along the conventional path, and we pay particular attention to the boundary conditions that connect them. In accordance with first principles, we also use the Hamiltonian action rather than the Lagrangian one, which is conceptually clearer for our phase-space study and in a stochastic context in which fields and momenta are not time-differentiable in the ordinary sense. Eventually, to deal with the UV parts of the fields and momenta, we identify phase-space covariant Vilkovisky-DeWitt variables, a crucial step to maintain the general covariance of the stochastic theory under redefinitions of the scalar fields. Because we are only interested in the super-Hubble dynamics, we integrate out explicitly the UV fields
from the path integral, and find the influence action that describes the deviation of the IR dynamics from the background one of Standard Perturbation Theory. The final result is the Hamiltonian, coarse-grained effective action for the IR fields at first order in quantum corrections, which after a final manipulation consisting of introducing auxiliary classical variables $\xi$, can be shown to give rise to the noises in the Stratonovich-Langevin equations. The statistics of the noise at a given time is found as the real parts of the UV power spectra at
the coarse-graining scale of that time. The fact that the noises are explicitly real is one of the improvements from the path-integral approach compared to the heuristic one. In section~\ref{sec:Markovian} we consider cases where the Markovian approximation is valid, and derive the covariant, phase-space Fokker-Planck (FP) equation corresponding to our Langevin equations. Thanks to the resolution of the anomalies, this equation is free from the ambiguities previously present in the literature, even in the single-field case. We also provide explicit analytical formulae for the noises correlations in multifield contexts, for massless scalar fields, as well as in generic situations under a slow-varying approximation.

We are confident that the formalism presented in this paper can be used in many interesting applications, both theoretical and phenomenological.
First, the It\^o-Langevin equations coupled to the UV EoM could be fully solved numerically without resorting to the Markovian hypothesis. That would however require following the evolutions of as many modes as time steps in the computation, in order to predict the correct noise amplitude at any time, and depending on the previous realisations of the noises and IR dynamics. These simulations should be done a large number of times, in order to compute statistical averages.
Although being the most rigorous approach, it may be simpler to first consider the Markovian approximation, replacing the commonly approximated noise amplitude $(H/2\pi)^2$ by the formulae that we give in Eqs.~\eqref{eq: power spectra massive case-QQ}--\eqref{eq: power spectra massive case-PP}, and only then determine the IR dynamics, either numerically or analytically. 
Observationally relevant quantities, such as the power spectrum and the full PDF of the curvature perturbation, as well as the mass distribution of PBHs in relevant models, can then be computed by use of the stochastic-$\delta N$ formalism, either applied to the result of many stochastic simulations in separate universes, or readily working at the level of the FP equation~\eqref{eq: phase space fokker-planck}.
We stress that due to the generality of our formalism, such computations can be made, not only in single-field contexts, but in the very large class of multifield models with curved field space, where qualitatively new phenomena can be expected.
It would also be interesting to compare the computations of correlation functions made with the stochastic formalism to pure QFT calculations, notably in de Sitter, or to determine equilibrium PDFs in phase space, as well as to study the eigenvalues and eigenvectors of the FP operator in simple multifield contexts.

Eventually, this paper not only provides one with a useful formalism that can be used from now on, but it also paves the way for going further. First, thanks to the rigorous path-integral derivation, corrections to the present stochastic formalism can be in principle computed. Technically, that would require going at next order in the expansion of the Hamiltonian coarse-grained effective action in the quantum components of the fields and momenta. Another interesting avenue is to unveil the effect of non-Gaussianities on the stochastic formalism, by expanding the Hamiltonian action up to cubic order as we do in Eqs.~\eqref{eq: cubic order hamiltonian action}--\eqref{eq: cubic order hamiltonian action-end}, and considering the effect of non-linear mode couplings. We leave these interesting possibilities for future works.

\acknowledgments

We are grateful to
Thibaut Arnoulx de Pirey Saint Alby,
Camille Aron,
Cliff Burgess,
Guillaume Faye,
Tomohiro Fujita,
Jacopo Fumagalli,
Vivien Lecomte,
Jer\^ome Martin,
Cyril Pitrou,
Gerasimos Rigopoulos, 
Julien Serreau,
Takahiro Tanaka,
Junsei Tokuda,
Andrew J. Tolley,
Vincent Vennin, 
and Lukas Witkowski
for helpful discussions.  We also thank the anonymous referee for insightful comments.
L.P and S.RP are supported by the European Research Council under the European Union's Horizon 2020 research and innovation programme (grant agreement No 758792, project GEODESI).
Y.T. is supported by JSPS KAKENHI Grants 
No. JP18J01992 and No. JP19K14707, and was supported by a grant from R\'egion \^Ile-de-France at the initial stage of this work.

\appendix 

\section{Stochastic calculus}
\label{appendix: stochastic calculus}

In this Appendix, we describe the exotic features
of stochastic calculus that are intimately due to the 
non-differentiability of stochastic random variables. 
We begin by explaining explicitly how discretisation schemes are defined, and how moving from one to another changes the continuous description, contrary to what happens for ordinary calculus.
Then, we review some properties of It\^o calculus, which is very different from the ordinary one as the standard chain rule for the derivative of composite functions does not hold. In particular, we explain how to define covariant derivatives compatible with this stochastic calculus. We eventually give the form of the Fokker-Planck equation corresponding to Langevin equations in any discretisation scheme.

\subsection{Going from any scheme $\discAlpha$ to It\^o}

Let us consider multiple stochastic processes $\genS^\indn(N)$ that are characterised by a multidimensional Langevin equation of the form
\bae{
    \dif{\genS^\indn}{N}=h^\indn(N,\genS)+g^\indn_\indA(N,\genS)\xi^\indA, \qquad \braket{\xi^\indA(N)\xi^\indB(N^\prime)}=\delta^{\indA\indB}\delta(N-N^\prime), 
}
where $\xi^\indA$ are independent normalised Gaussian white noises. 
Here for notational simplicity we do as if the processes were Markovian and the noise amplitudes $g^\indn_\indA$ were functions of the stochastic processes $\genS^\indn$ at the same time $N$.
However, there could be more general cases for which the stochastic processes affect their own noise amplitudes in a more complicated way. As explained in Sec.~\ref{subsec:Markovian?}, this is actually the case in stochastic inflation for which the amplitude of the noise is dictated by differential equations that simply include the stochastic processes themselves, a situation with implicit dependence that we sometimes refer to by using the expression ``Langevin-type" equations.
The noise amplitude can also depend explicitly on some subset of the stochastic processes $\genS^\indn$ and implicitly on others, which is the case in Eq.~\eqref{eq: Langevin independent noise vielbein} for example. Keeping this in mind, we will still use the notation $g^\indn_\indA(N,\genS)$ for simplicity in these situations.

The time integration of such stochastic process is first described with use of a discrete time step $\Delta N$, and then evaluated in the continuous limit $\Delta N\to0$, like for the ordinary Riemann integral.
In this discrete description, the noise term at the step $i$ is first replaced by $\Delta W^\indA_i$, which are defined as independent random Gaussian variables with variance $\Delta N$:
\bae{\label{eq: DwDw}
    \Delta W_i^\indA\Delta W_j^\indB=\delta^{\indA\indB} \delta_{ij}\Delta N.
}
The fact that $\Delta W_i^\indA / \Delta N \sim (\Delta N)^{-1/2}$ has the important consequence that the noise, and thus the whole stochastic process, is not differentiable in the mathematical sense. All complexities of stochastic calculus arise from this simple fact. Still, it is possible to write the evolution of $\genS^\indn$ from the step $i$ to $i+1$ as
\bae{
    \Delta\genS^\indn_i=\genS^\indn_{i+1}-\genS^\indn_i=h^\indn(N_{i+\discAlpha},\genS_{i+\discAlpha})\Delta N+g^\indn_\indA(N_{i+\discAlpha},\genS_{i+\discAlpha})\Delta W^\indA_i.
}
In general, the coefficients $h^\indn$ and $g^\indn_\indA$ can be functions of time $N$ and of the stochastic process $\genS^\indn$ itself. In this case of so-called multiplicative noise, the particular time at which they are evaluated between $i$ and $i+1$, parametrized by $0\leq\discAlpha\leq1$ as
\bae{
    N_{i+\discAlpha}=N_i+\discAlpha\Delta N, \qquad \genS^\indn_{i+\discAlpha}=(1-\discAlpha)\genS^\indn_i+\discAlpha\genS^\indn_{i+1},
}
can matter, contrary to the ordinary Riemann integral for differentiable function.
Indeed, let us find the transformation law between an arbitrary $\discAlpha$ discretisation and the so-called It\^o~\cite{ito1944109}, or prepoint, discretisation defined by $\discAlpha=0$:
\bae{
    \Delta\genS^\indn_i&=h^\indn(N_{i+\discAlpha},\genS_{i+\discAlpha})\Delta N+g^\indn_\indA(N_{i+\discAlpha},\genS_{i+\discAlpha})\Delta W_i^\indA \nonumber \\
    &=\left[h^\indn+\pdif{h^\indn}{N}\discAlpha\Delta N+\pdif{h^\indn}{\genS^\indm}\discAlpha\Delta\genS_i^\indm\right]_{N_i,\genS_i}\!\!\!\!\!\!\Delta N+\left[g^\indn_\indA+\pdif{g^\indn_\indA}{N}\discAlpha\Delta N+\pdif{g^\indn_\indA}{\genS^\indm}\discAlpha\Delta\genS^\indm_i\right]_{N_i,\genS_i}\!\!\!\!\!\!\Delta W_i^\indA+\cdots \nonumber \\
    &=\left[h^\indn+\discAlpha g^\indm_\indA\pdif{g^\indn_\indA}{\genS^\indm}\right]_{N_i,\genS_i}\!\!\!\!\!\!\Delta N+g^\indn_\indA(N_i,\genS_i)\Delta W_i^\indA+\calO\left((\Delta N)^{3/2}\right).
}
Crucially, because of the $(\Delta N)^{1/2}$-dependence of the discrete noise $\Delta W_i^\indA$, the term $\discAlpha(\partial_\indm g^\indn_\indA)\times \Delta\genS^\indm_\indi\Delta W^\indA_\indi$ 
in the second line, contains a piece linear in $\Delta N$ that should be kept for consistency and results in a noise-induced drift $\discAlpha g^\indm_\indA\partial_\indm g^\indn_\indA$. In terms of the continuous description, the Langevin equation is thus properly defined only once a given discretisation scheme $\discAlpha$ is chosen. However it is always possible to relate it to a different Langevin equation corrected by the noise-induced drift and interpreted in the It\^o, $\discAlpha=0$, discretisation as
\bae{\label{eq: scheme conversion}
    \boxed{
    \text{$\discAlpha$-scheme:} \qquad
    \dif{\genS^\indn}{N}=h^\indn+g^\indn_\indA\circ_\discAlpha\xi^\indA=\left(h^\indn+\discAlpha g^\indm_\indA\pdif{g^\indn_\indA}{\genS^\indm}\right)+g^\indn_\indA\xi^\indA,
    }
}
where the implicit sum on $\indm$ only runs on indices that denote stochastic processes entering \emph{explicitly} in the noise amplitudes $g^\indn_\indA$. We have introduced here the $\circ_\discAlpha$ symbol to make the $\discAlpha$ discretisation explicit, while no circle should be understood as the It\^o discretisation. Another particular choice is the so-called Stratonovich, or midpoint, $\discAlpha=1/2$, scheme~\cite{stratonovich1966new}, which we will represent by the circle $\circ$ without the subscript $\discAlpha$.
Different discretisation schemes can thus be connected by the induced drift term $\discAlpha g^\indm_\indA\partial_\indm g^\indn_\indA$ in the continuous description. 
The relation between a Langevin equation interpreted in the Stratonovich scheme, and the equivalent one in the It\^o scheme is used in Sec.~\ref{sec: stochastic anomalies}.

\subsection{It\^o calculus and covariance}\label{subsec: Ito calculus}

In this part we derive useful relations in the continuous description of stochastic processes when the underlying time discretisation is understood in the It\^o scheme, taking into account the fact that so-called It\^o calculus differs from ordinary calculus. 
First, the differential form of a function $f(N,\genS^\indn)$ of time and of the stochastic processes $\genS^\indn$, should be consistently expanded at ``quadratic" order as
\bae{
    \dd f(N,\genS^\indn)=\pdif{f}{N}\dd N+\pdif{f}{\genS^\indn}\dd\genS^\indn+\frac{1}{2}\frac{\partial^2f}{\partial\genS^\indn\partial\genS^\indm}\dd\genS^\indn\dd\genS^\indm.
}
Suppose the stochastic processes $\genS^\indn$ verify the following It\^o's Langevin equation:
\bae{
    \dd\genS^\indn=h^\indn\dd N+g^\indn_\indA\dd W^\indA.
    \label{SDE-Ito}
}
Then, the differential form $\dd f$ can be written with the so-called It\^o's lemma~\cite{ito1944109},
\bae{\label{eq: Ito lemma}
    \boxed{
        \text{It\^o's lemma:} \qquad
        \dd f(N,\genS^\indn)=\left(\pdif{f}{N}+\frac{1}{2}\frac{\partial^2f}{\partial\genS^\indn\partial\genS^\indm}A^{\indn\indm}\right)\dd N+\pdif{f}{\genS^\indn}\dd\genS^\indn,
    }
}
where $A^{\indn\indm}=g^\indn_\indA g^\indm_\indA$ is the auto-correlation of the noises.
It shows that a function of stochastic variables does not follow the ordinary chain rule of differentiation in It\^o calculus.
It should be noted that the scheme conversion~(\ref{eq: scheme conversion}) reveals on the contrary that the Stratonovich discretisation does follow the standard chain rule, as the noise-induced drift compensates for the quadratic correction in It\^o's lemma:
\bae{
    \dd f(N,\genS^\indn)&=\left(\pdif{f}{N}+\frac{1}{2}\frac{\partial^2f}{\partial\genS^\indn\partial\genS^\indm}A^{\indn\indm}\right)\dd N+\left(\pdif{f}{\genS^\indn}\circ\dd\genS^\indn-\frac{1}{2}\frac{\partial^2f}{\partial\genS^\indn\partial\genS^\indm}\dd\genS^\indn\dd\genS^\indm\right) \nonumber \\
    &=\pdif{f}{N}\dd N+\pdif{f}{\genS^\indn}\circ\dd\genS^\indn.
}
That is why the Stratonovich discretisation is useful in 
physical contexts where changes of variables are ubiquitous.

Motivated by our will to develop a manifestly covariant theory of stochastic multifield inflation, let us now consider a manifold with metric $G_{\indI\indJ}(\genX^\indK)$, with coordinates $\genX^\indI$ subject to SDE of the type~\eqref{SDE-Ito}. It\^o's discretisation does not follow the standard chain rule and therefore it also breaks covariance under 
coordinate transformation in the ordinary sense. 
It is however possible to define stochastic covariant derivatives by adding suitable counter terms to ordinary derivatives.
For the coordinates $\genX^\indI$, we define 
\bae{
    \boxed{
        \text{It\^o-covariant derivative for coordinates:} \qquad
	    \ItoD\genX^\indI=\dd\genX^\indI+\frac{1}{2}\Gamma^\indI_{\indJ\indK}A^{\genX\genX\indJ\indK}\dd N,
	}
}
where $\Gamma^\indI_{\indJ\indK}$ denote the Christoffel symbols of the metric, 
and $A^{\genX\genX\indI\indJ}=g^{\genX\indI}_\indA g^{\genX\indJ}_\indA$ represents the auto-correlation of the noise $g^{\genX\indI}_\indA$,  
assumed to be a vector at $\genX^\indI$.
Under the coordinate transformation $\genX^\indI\to\Xt^\indIt(\genX)$,
$\ItoD\genX^\indI$ indeed transforms itself as a 
vector:
\bae{
	\ItoD\Xt^{\indIt}=\pdif{\Xt^{\indIt}}{\genX^\indI}\ItoD\genX^\indI,
}
as can be shown with use of the It\^o's lemma~(\ref{eq: Ito lemma}) for $\dd\Xt^\indIt(\genX)$ and of the transformation law for the Christoffel symbol:
\bae{\label{eq: Chris trs}
	\pdif{\Xt^{\indIt}}{\genX^\indI}\Gamma^\indI_{\indJ\indK}=\pdif{\Xt^{\indJt}}{\genX^\indJ}\pdif{\Xt^{\indKt}}{\genX^\indK}\Gammat^{\indIt}_{\indJt\indKt}
	+\frac{\partial^2\Xt^{\indIt}}{\partial\genX^\indJ\partial\genX^\indK}.
}
We then also define covariant It\^o derivatives for tangent vectors $\genU^\indI$ and covectors $\genV_\indI$. In the new coordinate system $\Xt^{\indIt}(\genX)$, their components read by definition:
\bae{
    \Ut^{\indIt}(\genX,\genU)=\pdif{\Xt^{\indIt}(\genX)}{\genX^\indI}\genU^\indI\,, \qquad
	\Vt_{\indIt}(\genX,\genV)=	\pdif{\genX^\indI(\Xt)}{\Xt^{\indIt}}   \genV_\indI\,.
}
After some algebra, involving derivatives of these transformations as well as the one of the Christoffel symbols~(\ref{eq: Chris trs}), and using It\^o's lemma for $\dd\Ut^\indIt(\genX,\genU)$ and $\dd\Vt_\indIt(\genX,\genV)$, one finds the following suitable covariant It\^o derivatives:
\begin{empheq}[box=\fbox]{align}
    &\text{It\^o-covariant derivative for vectors:} \nonumber \\
	&\quad \ItoD\genU^\indI=\covD\genU^\indI+\frac{1}{2}\left(\Gamma^\indI_{\indJ\indS,\indK}-\Gamma^\indM_{\indJ\indS}\Gamma^\indI_{\indM\indK}\right)\genU^\indS A^{\genX\genX\indJ\indK}\dd N+\Gamma^\indI_{\indJ\indK}A^{\genX\covU\indJ\indK}\dd N, \\
	&\text{It\^o-covariant derivative for covectors:} \nonumber \\
	&\quad 	\ItoD\genV_\indI=\covD\genV_\indI-\frac{1}{2}\left(\Gamma^\indS_{\indI\indJ,\indK}+\Gamma^\indM_{\indI\indJ}\Gamma^\indS_{\indK\indM}\right)\genV_\indS A^{\genX\genX\indJ\indK}\dd N
	-\Gamma^\indK_{\indI\indJ}A^{\genX\covV\indJ}{}_\indK\dd N.
\end{empheq}
Here $A^{\genX\covU\indI\indJ}=g^{\genX\indI}_\indA g^{\covU\indJ}_\indA$ and $A^{\genX\covV\indI}{}_\indJ=g^{\genX\indI}_\indA g^\covV_{\indJ\indA}$ are the cross-correlations between the coordinate noise $g^{\genX\indI}_\indA$ and the covariant combinations of (co)vector noise:
\bae{
    g^{\covU\indI}_\indA=g^{\genU\indI}_\indA+\Gamma^\indI_{\indJ\indK}\genU^\indJ g^{\genX\indK}_\indA, \qquad g^\covV_{\indI\indA}=g^\genV_{\indI\indA}-\Gamma_{\indI\indJ}^\indK\genV_\indK g^{\genX\indJ}_\indA.
}
It can be checked that they indeed transform as tensors under It\^o calculus, i.e. with:
\bae{
    \ItoD\Ut^{\indIt}=\pdif{\Xt^{\indIt}}{\genX^\indI}\ItoD\genU^\indI\,, \qquad	\ItoD\Vt_{\indIt}=	\pdif{\genX^\indI}{\Xt^{\indIt}}   \ItoD\genV_\indI\,.
}	
	
\subsection{From Langevin to Fokker-Planck}\label{subsec:FP}

When the noise amplitude $g_\indA^\indn$ depends only explicitly on the stochastic processes $\genS^\indn$, the whole process is said to be Markovian. Indeed, the distribution from which the noise is drawn at a time $N$ is then fixed by the values of the processes $\genS^\indn$ \emph{at this time $N$ only}, and the past history (the path in the space of the processes that led to this particular point) is irrelevant. Thus, a new useful tool becomes accessible: the Fokker-Planck (FP) equation. 

The FP equation is a partial differential equation (PDE) for the transition probability of the stochastic processes $\genS^\indn$. This transition probability 
$P(\genS^n,N | \genS^\indn_\mathrm{ini}, N_\mathrm{ini})$ is defined as the probability that the stochastic processes take the values $\genS^\indn$ at the time $N$ knowing that they initially had the value $\genS^\indn_\mathrm{ini}$ at time $N_\mathrm{ini}$. If the processes $\genS^\indn$ verify the Langevin equation~\eqref{eq: scheme conversion} with a given choice of discretisation $\alpha$, it can be shown that the transition probability (that can be thought of as a one-point Probability Density Function (PDF) for the processes $\genS^\indn$) verifies the Fokker-Planck equation:
\bae{
\frac{\partial P}{\partial N}=-\frac{\partial}{\partial\genS^\indn}\left[\left(h^\indn+\discAlpha g_\indA^\indm \frac{\partial  g_\indA^\indn}{\partial \genS^\indm}\right) P \right]+\frac{1}{2}\frac{\partial^2}{\partial \genS^\indn \partial\genS^\indm}\left[g_\indA^\indn g_\indA^\indm P \right].
\label{FP-from-Langevin}
}
This is nothing but a convection-diffusion equation for the PDF $P(\genS^\indn,N)$, where the diffusion is due to the stochastic noise while the drift is due to the deterministic force in the Langevin equation (plus a noise-induced drift for discretisations different from It\^o, $\discAlpha=0$).

\section{Covariant expansion of the Hamiltonian}
\label{appendix: covP}

In this appendix, we revisit in more detail the choice of covariant UV perturbations. 
In the main body of the paper, we connect the finite displacements $\dphi^\indI$ and $\dpi_\indI$ and the covariant ``initial velocity" $Q^\indI=\dd\phifull^\indI/\dd\lambda|_{\lambda=0}$ and $\covP_\indI=\covD_\lambda\pifull_\indI|_{\lambda=0}$ through the 
geodesic-type interpolation $\covD_\lambda^2\phifull^\indI=0$ and $\covD_\lambda^2\pifull_\indI=0$.
However, the requirement of covariance does not by itself prohibit any curvature invariant term in the corresponding momentum equation.
For example, let us consider the transportation of the on-shell solution. Considering the Minkowski limit for simplicity in this discussion, the on-shell momenta are given by $\pi_\indI=G_{\indI\indJ}\dot{\phi}^\indJ$ where a dot represents a generic time derivative (see Eq.~(\ref{eq: rescaled phi EoM})).
Therefore, along the field-space geodesic defined
by 
$\covD_\lambda^2\phifull^\indI=0$, the on-shell $\pifull$ follows
\bae{\label{eq: DDpi with Riemann}
    \covD_\lambda^2\pifull_\indI&=G_{\indI\indJ}\covD_\lambda^2\dot{\phifull}^\indJ=G_{\indI\indJ}\covD_\lambda^2\covD_t\phifull^\indJ
    =G_{\indI\indJ}\covD_\lambda\covD_t\covD_\lambda\phifull^\indJ=G_{\indI\indJ}[\covD_\lambda,\covD_t]\covD_\lambda\phifull^\indJ \nonumber \\
    &=R_{\indI\lambda\lambda t}=R_{\indI\indJ\indK}{}^\indS\pifull_\indS\dif{\phifull^\indJ}{\lambda}\dif{\phifull^\indK}{\lambda}\neq 0 \text{ a priori},
}
where $R^\indS{}_{\indI\indJ\indK}$ is the Riemann tensor of the field space:
\bae{
    R^\indS{}_{\indI\indJ\indK}=\Gamma^\indS_{\indI\indK,\indJ}-\Gamma^\indS_{\indI\indJ,\indK}
    +\Gamma^\indR_{\indI\indK}\Gamma^\indS_{\indJ\indR}-\Gamma^\indR_{\indI\indJ}\Gamma^\indS_{\indK\indR}.
}
To derive Eq.~(\ref{eq: DDpi with Riemann}), we made use of the general fact $[\covD_\lambda,\covD_t]\phifull^\indI=0$ as well as the geodesic equation $\covD_\lambda^2\phifull^\indI=0$. This new interpolation law obviously yields an additional quadratic term proportional to the Riemann tensor, in the expression of $\dpi_\indI$ in terms of $(Q^\indI,\covP_\indI)$.
However the momenta $\pifull_\indI$ are \emph{a priori} off-shell in the path integral that we perform in Sec.~\ref{sec: effective hamiltonian action}. Thus, they have no particular reason to follow 
the interpolation law~\eqref{eq: DDpi with Riemann}, which is nothing but one of the possible covariant deviations from the geodesic interpolation, and any such covariant deviation is allowed. The point is that the quadratic correction due to the non-trivial ``acceleration'' $\covD_\lambda^2\pifull_\indI$ can always be 
absorbed into the definition of the covariant momentum UV perturbation. When the interpolation law is given by Eq.~\eqref{eq: DDpi with Riemann}, we will denote the former as
$\covP_{I,1/2}$ for a reason that will soon become clear, and one finds the modification compared to the case where it is $\covD_\lambda^2 \pifull_\indI=0$ that defines $\covP_\indI$:
\bae{
    \covP_{I,1/2}=\covP_I+\frac{1}{2} R_{\indI\indJ\indK}{}^\indL \piIR_L Q^\indJ Q^\indK.
}
Note that this other
choice would leave unchanged, both the commutation relation
$[Q^\indI(N,\mathbf{x}), \\
\covP_{\indJ,1/2}(N,\mathbf{x}^\prime)]=i\delta^\indI_\indJ\delta^{(3)}(\mathbf{x}-\mathbf{x}^\prime)$ 
and the linear UV equations of motion. 
Interestingly, $\covP_{\indI,1/2}$ is the only choice that makes completely disappear the term in ``background EoM'' $\times$ 
Riemann $\times$ $QQ$ in $S^{(2)}$, Eq.~\eqref{eq: cov S2}, and that we discarded in the main text. As already explained, any covariant deviation from the geodesic interpolation law
is allowed by the requirement to define a covariant  momentum perturbation, so this leaves the possibility of defining for instance a family of covectors
\bae{\label{eq: Pkappa}
    \covP_{I,\kappa}=\covP_I+\kappa R_{\indI\indJ\indK}{}^\indL \piIR_L Q^\indJ Q^\indK,
}  
that all verify the necessary conditions: covariance, standard commutation relation and linear UV equations of motion.
We acknowledge that this ambiguity remains in our treatment of
stochastic inflation, but the difference between two choices of covariant perturbations goes beyond the usual approximations of stochastic inflation, and in particular the Gaussianity of the noise. However, there might be potentially interesting geometrical effects induced by the curved field space beyond this approximation.

To see the apparent effect of the $\kappa$ ambiguity, let us explicitly show the cubic Hamiltonian action in terms of the variables $\covP_{I,\kappa}$.
To this end, a geometric approach is more useful than directly expanding the differences $\dphi^\indI=\phifull^\indI-\phiIR^\indI$ and $\dpi_\indI=\pifull_\indI-\piIR_\indI$. We first note that the action for perturbations $S^\mathrm{(ptb)}$ is given by the difference between the full action and the IR action as
\bae{
    S^\mathrm{(ptb)}=S[\phifull,\pifull,\lapse,\shift]-S[\phiIR,\piIR,\lapseIR,0].
}
The non-trivial aspect comes from the fact that the full variables and their IR parts live at the different field-space points $\phifull^\indI(\lambda=1)$ and $\phiIR^\indI=\phifull^\indI(\lambda=0)$. That is why their direct differences are not covariant objects.
However the action itself is a scalar in field space and thus it is invariant under the parallel transport from $\lambda=1$ to $\lambda=0$:
\bae{
    S^{(\mathrm{ptb})}=S_\parallel[\phifull,\pifull,\lapse,\shift]_{\lambda=0}-S[\phiIR,\piIR,\lapseIR,0].
}
These transported variables can be directly compared to the IR (co)vectors, and the covariant expansion of the Hamiltonian action is easily derived
that way.

\bfe{width=0.9\hsize}{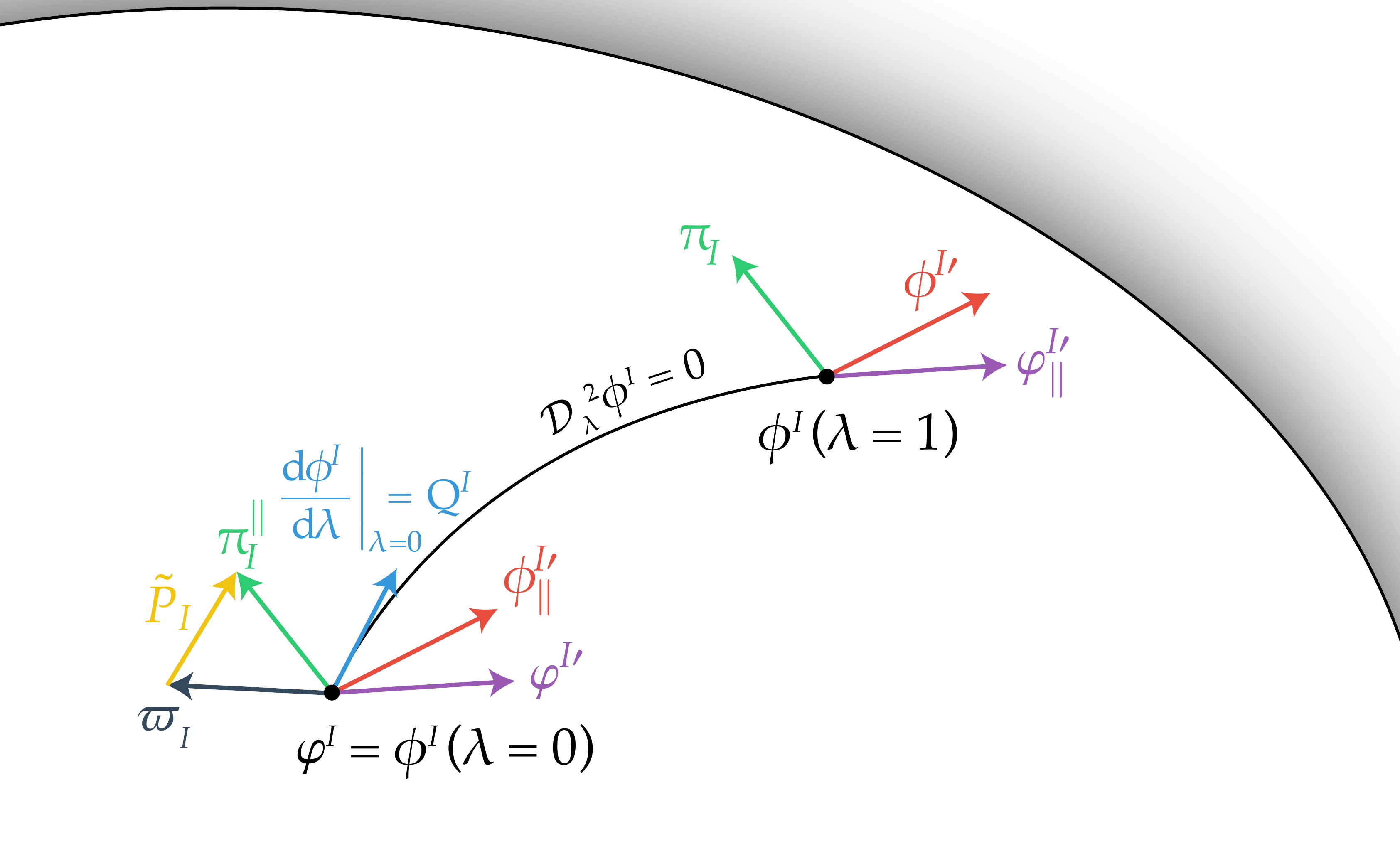}{
Schematic representation of the Hamiltonian variables on a curved field space. The full (co)vectors $\phi^{\indI\prime}$ and $\pifull_\indI$ are defined at $\phifull^\indI$, while their IR parts $\phiIR^{\indI\prime}$ and $\piIR_\indI$ live at $\phiIR^\indI$. These two points can be connected by the geodesic $\lambda$-line $\covD_\lambda^2\phifull^\indI(\lambda)=0$, characterised by the contravariant UV perturbation $Q^\indI=\dd\phifull^\indI/\dd\lambda|_{\lambda=0}$ which is a vector at $\phiIR^\indI$. Along this line, vectors living at each point can be parallel transported to each others (vectors in the same color). This procedure enables one to efficiently expand the action in terms of (co)vectors at the IR point $\phiIR^\indI$.}
{fig: transport method} 

Let us describe this approach in detail (see Fig.~\ref{fig: transport method} for a schematic representation).
We first recall that the full 
field $\phifull^\indI$ (at $\lambda=1$) and its IR part $\phiIR^\indI$ (at $\lambda=0$) are connected by a unique geodesic line
\bae{\label{eq: lambda line}
    \covD_\lambda^2\phifull^\indI=0, \quad \Leftrightarrow \quad \dif{^2\phifull^\indI}{\lambda^2}=-\Gamma^\indI_{\indJ\indK}\dif{\phifull^\indJ}{\lambda}\dif{\phifull^\indK}{\lambda},
}
with the affine parameter $\lambda$. This geodesic line can be characterised by the contravariant vector at $\lambda=0$
\bae{
    Q^\indI:=\left.\dif{\phifull^\indI}{\lambda}\right|_{\lambda=0}.
}
Then one can parallel transport, e.g., the full momentum $\pifull_\indI$, 
a covector at $\lambda=1$, to the IR point $\lambda=0$ along this $\lambda$-line. The transport condition is given by $\covD_\lambda\pifull^\parallel_\indI=0$ with the initial condition $\pifull^\parallel_\indI|_{\lambda=1}=\pifull_\indI$.
The transported momentum $\pifull^\parallel_\indI|_{\lambda=0}$ 
is a covector at $\lambda=0$, so that it can be directly compared to the other covector there, $\piIR_\indI$. Thus one can define the covariant UV momentum by their difference, i.e.: 
\bae{\label{eq: transport P}
    \covP_\indI \equiv \pifull^\parallel_\indI|_{\lambda=0}-\piIR_\indI.
}
This new definition coincides with the previous one 
$\covP_\indI=\covD_\lambda \pifull_\indI|_{\lambda=0}$.

As the scalar product (computed with the field-space metric) between two parallel-transported quantities is constant, the kinetic term in the Hamiltonian, for example, can be easily expressed as\footnote{The position of the parallel symbol $\parallel$ is arbitrarily raised and lowered without any specific meaning. We put it on the opposite side of the field index $\indI$ just for clarity.}
\bae{
    \pifull_\indI\pifull^\indI|_{\lambda=1}=\pifull_\indI^\parallel\pifull^\indI_\parallel|_{\lambda=0}=(\piIR_\indI+\covP_\indI)(\piIR^\indI+\covP^\indI).
}
Similarly, all scalar products in the Hamiltonian action such as $\pifull_\indI\pifull^\indI$, $\pifull_\indI\dotphi^\indI$, $\pifull_\indI\partial_i\phifull^\indI$, and $G_{\indI \indJ} \partial_i\phifull^\indI\partial_i\phifull^\indJ$ can be expanded at $\lambda=0$ as $\pifull_\indI^\parallel\pifull^\indI_\parallel|_{\lambda=0}$, $\pifull_\indI^\parallel\dotphi^\indI_\parallel|_{\lambda=0}$, and so on.
Using the geodesic equation~\eqref{eq: lambda line} as well as expansions around $\lambda=0$ of the Christoffel symbols and their derivatives, it
is then straightforward, although tedious, to find the following expressions:
\bae{
    \dotphi^\indI_\parallel|_{\lambda=0}=&\,\dotphiIR^\indI+\covD_N Q^\indI+\frac{1}{2}R^\indI{}_{\indJ\indK\indL}\dotphiIR^\indL Q^\indJ Q^\indK \nonumber \\
    &+\frac{1}{6}R^\indI{}_{\indJ\indK\indM;\indL}\dotphiIR^\indM Q^\indJ Q^\indK Q^\indL+\frac{1}{6}R^\indI{}_{\indJ\indK\indL}Q^\indJ Q^\indK\covD_N Q^\indL+\cdots, \\
    (\partial_i\phifull^\indI)_\parallel|_{\lambda=0}=&\partial_iQ^\indI+\frac{1}{6}R^\indI{}_{\indJ\indK\indL}Q^\indJ Q^\indK\partial_iQ^\indL+\cdots\,.
}
The scalar potential $V$ can be also expanded in a covariant way as
\bae{
    V|_{\lambda=1}&=V+V_\indI Q^\indI+\frac{1}{2}V_{;\indI\indJ}Q^\indI Q^\indJ+\frac{1}{6}V_{;\indI\indJ\indK}Q^\indI Q^\indJ Q^\indK+\cdots.
}
The ADM variables $\lapse$ and $\shift^i$ are also scalars in field space but they should be considered independent of $\phifull^\indI$ and $\pifull_\indI$ before one takes into account their constraint equations. Thus, one can define their UV parts by the simple differences
\bae{
    \lapseUV \equiv \lapse-\lapseIR, \qquad a^{-2}\partial_i\shiftUV \equiv \shift^i-0.
}

Up to cubic order in the covariant UV perturbations
$Q^\indI$ and $\covP_\indI$, the Hamiltonian action then reads\footnote{Making use of the linear constraints~\eqref{lapseUV} and \eqref{eq: perturbation energy const}, one can check that our cubic Hamiltonian is consistent with results in the literature (see, e.g. Refs.~\cite{Elliston:2012ab,Butchers:2018hds}), albeit importantly in our stochastic context, no background equation of motion has been used here.
Note that, as can be seen in the linear perturbation EoM~(\ref{eq: UV EoM}), our 
definition of $\covP$ differs, even at linear order, from the ordinary notation where the on-shell momentum perturbation is defined as $\covP\propto\covD_N Q$.}
\bae{\label{eq: cubic order hamiltonian action}
    \So&=\int\dd^4x\,a^3\left[\piIR_\indI\dotphiIR^\indI-\frac{1}{H}\left(\frac{1}{2}\piIR_\indI\piIR^\indI+V+3\Mpl^2H^2\right)\right], \displaybreak[0] \\
    \Sl&=\int\dd^4x\,a^3\left[\covP_{\indI,\kappa}\left(\dotphiIR^\indI-\frac{\piIR^\indI}{H}\right)-Q^\indI\left(\covD_N\piIR_\indI+3\piIR_\indI+\frac{V_\indI}{H}\right)-\lapseUV\left(\frac{1}{2}\piIR_\indI\piIR^\indI+V-3\Mpl^2H^2\right)\right], \displaybreak[0] \\
    \Sz&=\int\dd^4x\,a^3\left[\covP_{\indI,\kappa}\covD_N Q^\indI-3\Mpl^2H^3\lapseUV^2-\lapseUV\left(\piIR_\indI\covP^\indI_\kappa+V_\indI Q^\indI+2\Mpl^2H^2\frac{\nabla^2}{a^2}\shiftUV\right)+\piIR_\indI Q^\indI\frac{\nabla^2}{a^2}\shiftUV \right. \nonumber \\
    &\qquad-\frac{1}{H}\left(\frac{1}{2}\covP_{\indI,\kappa}\covP^\indI_\kappa-\frac{1}{2}Q_\indI\frac{\nabla^2}{a^2}Q^\indI+\frac{1}{2}V_{;\indI\indJ}Q^\indI Q^\indJ-\frac{1}{2}R_\indI{}^{\indK\indL}{}_\indJ\piIR_\indK\piIR_\indL Q^\indI Q^\indJ\right)\nonumber \\
    &\qquad\left.+\left(\frac{1}{2}-\kappa\right)\left(\dotphiIR^\indI-\frac{\piIR^\indI}{H}\right)R_{\indI\indJ\indK}{}^\indL\piIR_\indL Q^\indJ Q^\indK\right], \label{eq: kappa S2} \displaybreak[0] \\
    \SE&=\int\dd^4x\,a^3\left[3\Mpl^2H^4\lapseUV^3+2\Mpl^2H^3\lapseUV^2\frac{\nabla^2}{a^2}\shiftUV+\frac{1}{2}\Mpl^2H^2\lapseUV\left(\left(\frac{\shiftUV_{ii}}{a^2}\right)^2-\frac{\shiftUV_{ij}}{a^2}\frac{\shiftUV_{ij}}{a^2}\right)
    \right. \nonumber \\
    &\qquad
    -\frac{1}{a^2}\shiftUV_i\covP_{\indI,\kappa}\partial_iQ^\indI   -\lapseUV\left(\frac{1}{2}\covP_{\indI,\kappa}\covP^\indI_\kappa+\frac{1}{2a^2}\partial_iQ^\indI\partial_iQ_\indI+\frac{1}{2}V_{;\indI\indJ}Q^\indI Q^\indJ-\kappa R_\indI{}^{\indK\indL}{}_\indJ\piIR_\indK\piIR_\indL Q^\indI Q^\indJ\right) \nonumber \\
    &\qquad+\lapseIR\left(\left(\frac{1}{2}+\kappa\right)R^\indL{}_{\indI\indJ}{}^\indK\piIR_\indL Q^\indI Q^\indJ\covP_{\indK,\kappa}-\frac{1}{6}V_{;\indI\indJ\indK}Q^\indI Q^\indJ Q^\indK+\frac{1}{6}R^\indL{}_{\indI\indJ}{}^\indM{}_{;\indK}\piIR_\indL\piIR_\indM Q^\indI Q^\indJ Q^\indK\right) \nonumber \\
    &\qquad +\frac{1}{6}\left(\dotphiIR^\indI-\frac{\piIR^\indI}{H}\right)\left(3R_{\indI\indJ\indK}{}^\indL Q^\indJ Q^\indK\covP_{\indL,\kappa}+R_{\indI\indJ\indK}{}^\indM{}_{;\indL}\piIR_\indM Q^\indJ Q^\indK Q^\indL\right) \nonumber \\
    &\qquad\left.+\left(\frac{1}{6}-\kappa\right)R^\indL{}_{\indI\indJ\indK}\piIR_\indL Q^\indI Q^\indJ\covD_N Q^\indK\right].
    \label{eq: cubic order hamiltonian action-end}
}
Here we use the generalised momentum $\covP_{\indI,\kappa}$~(\ref{eq: Pkappa}), allowing a shift from the original momentum~(\ref{eq: transport P}). 
As mentioned before, $\kappa=1/2$ eliminates the term proportional to the
``background-like'' EoM $\dotphiIR^\indI-\piIR^\indI/H$ in the quadratic action $\Sz$. On the other hand, if one chooses $\kappa=1/6$,
there is no term containing $\covD_N Q^\indI$ in the cubic action, which indicates that $\covP_{\indI,1/6}$ might enjoy interesting properties. Although not directly used in this paper, we hope these results can be useful for future works.

\section{Covariant notations}\label{appendix: notations}

We gather in this appendix the various covariant notations that we use in the paper.

\subsection{Manifolds, indices and related metrics}

To use compact and covariant notations, we had to introduce several types of indices denoting coordinates on various manifolds. To help the reader, here we show a comprehensive list of these notations together with the corresponding metrics that must be used to raise and lower indices.

\begin{table}[H]
\large
\renewcommand{\arraystretch}{1.5} 
\hspace{-0.5cm}
\begin{tabular}{| m{7cm} | m{1.4cm} | m{5cm} | m{1.8cm} |}
    \hline
    Manifold & Index & Metric & Dimension  \\
    \hline
    \hline
    Spacetime 
    & $\mu$ & $g_{\mu\nu}$ & $4$ \\
    \hline
    $3$-dimensional spatial hypersurfaces 
    & $\indi$ & $\gamma_{\indi\indj}=a^2\delta_{\indi\indj}$ in flat gauge & $3$ \\
    \hline
    Field space & $\indI$ & $G_{\indI\indJ}$ evaluated at $\phifull^\indI$ or $\phiIR^\indI$  & $N_\mathrm{fields}$ \\
    \hline
    Phase space & 
    $X$ or $\covX$ & 
    $\dps-i\sigma_{2\indX\indY}=\bigl(\begin{smallmatrix}0&-1\\1&0\end{smallmatrix}\bigr)$
    & 
    $2$ 
    \\
    \hline
    Set of creation-annihilation operators & $\indA$ & $\delta_{\indA\indB}$  & $N_\mathrm{fields}$\\
    \hline
    Vielbeins' frame & $\alpha$ & $\delta_{\alpha\beta}$  & $N_\mathrm{fields}$\\
    \hline
    Mass squared matrix eigenvalues & $\indi$ & $\delta_{\indi\indj}$  & $N_\mathrm{fields}$\\
    \hline
    Set $\{+,-\}$ of CTP branches & $\inda$ & $\sigma_{3 \inda\indb}=\diag(1,-1)_{\inda\indb}$  & $2$\\
    \hline
    Keldysh basis $\{\KelC,\KelD\}$ & 
    $\Kela$ &   
    $\dps\sigma_{1\Kela\Kelb}=\bigl(\begin{smallmatrix}0&1\\1&0\end{smallmatrix}\bigr)$
    & 
    $2$ 
    \\
    \hline
\end{tabular}
\end{table}

\subsection{Covariant derivatives}

Motivated by our will to have a manifestly covariant theory of multifield stochastic inflation, we have to use a certain number of covariant derivatives. Here we list all of them and show their actions on an IR stochastic vector in phase space $\genU^\indI\left(\phiIR^\indI,\piIR_\indI\right)$ from which their actions on any IR stochastic tensor can be deduced. Note that $A^{Q\covU\indJ\indK}=A^{Q\genU\indJ\indK}+\Gamma^\indK_{\indL\indM}\genU^\indL A^{QQ\indJ\indM}$.

\begin{table}[H]
\large
\renewcommand{\arraystretch}{1.5} 
\hspace{-1cm}
\begin{tabular}{| m{4cm} | m{1.6cm} | m{11.5cm} |}
    \hline
    Name & Notation & Action on a field-space vector  \\
    \hline
    \hline
    Field space & $\N_\indJ$ & $\N_\indJ \genU^\indI=\partial_\indJ \genU^\indI + \Gamma^\indI_{\indJ\indK} \genU^\indK$ \\
    \hline
    Phase space & $D_{\phiIR^\indJ}$ & $D_{\phiIR^\indJ} \genU^\indI=\N_\indJ \genU^\indI + \Gamma^\indK_{\indJ\indL} \piIR_\indK \partial_{\piIR_\indL}\genU^\indI$ \\
    \hline
    Time (deterministic or Stratonovich scheme) & $\covD_N$ & $\covD_N \genU^\indI=\partial_N \genU^\indI + \Gamma^\indI_{\indJ\indK}  \left(\partial_N\phiIR^\indJ\right) \genU^\indK $  \\
    \hline
    \vspace{3pt} Time (It\^o scheme) \vspace{3pt} & \vspace{3pt} $\ItoD_N$ \vspace{3pt} & 
    \vspace{3pt}
    $\dps
    \ItoD_N \genU^\indI=\covD_N \genU^\indI + \frac{1}{2}\left(\Gamma^\indI_{\indJ\indS,\indK}-\Gamma^\indM_{\indJ\indS}\Gamma^\indI_{\indM\indK}\right) \genU^\indS A^{QQ\indJ\indK} + \Gamma^\indI_{\indJ\indK}A^{Q\covU\indJ\indK}$ 
    \vspace{3pt}
    \\
    \hline
\end{tabular}
\end{table}

\subsection{Covariant perturbations?}

Under redefinitions of the scalar fields, naive perturbations of the full fields $\phifull^\indI$ and $\pifull_\indI$ around a classical background (or IR) value,  do not transform covariantly
beyond linear order. This subtlety was discussed in Sec.~\ref{subsec: covariant perturbations} and in Appendix~\ref{appendix: covP}. Here we summarise our notations for ``perturbations" including the naive ones, $\dphi^\indI=\phifull^\indI-\phiIR^\indI$ and $\dpi_\indI=\pifull_\indI-\piIR_\indI$, as well as the true covariant objects such as $Q^\indI$ and $\covP_\indI$. Their covariance (``Yes", ``No", or at ``Linear order" only) is also explicitly shown.

\begin{table}[H]
\large
\renewcommand{\arraystretch}{1.5}
\hspace{-1cm}
\begin{tabular}{| m{2cm} | m{11cm} | m{3cm} |}
    \hline
    Notation & Definition & Covariant?   \\
    \hline
    \hline
    \vspace{3pt} $\delta\phifull^\indI$ \vspace{3pt} & 
    \vspace{3pt}
    $\dps\phifull^\indI-\phiIR^\indI =Q^\indI-\frac{1}{2}\Gamma^\indI_{\indJ\indK}Q^\indJ Q^\indK+\cdots$
    \vspace{3pt} & \vspace{3pt} Linear order \vspace{3pt} \\
    \hline
    \vspace{3pt} $\delta\pifull_\indI$ \vspace{3pt} & 
    \vspace{3pt}
    $\begin{aligned}
        \pifull_\indI-\piIR_\indI=&\,\covP_\indI+\Gamma_{\indI\indJ}^\indK\piIR_\indK Q^\indJ+\Gamma^\indK_{\indI\indJ}Q^\indJ\covP_\indK \\ 
        &+\frac{1}{2}(\Gamma^\indS_{\indI\indJ,\indK}-\Gamma^\indS_{\indI\indR}\Gamma^\indR_{\indJ\indK}+\Gamma^\indR_{\indI\indJ}\Gamma^\indS_{\indR\indK})\piIR_\indS Q^\indJ Q^\indK+\cdots
    \end{aligned}$
    \vspace{3pt}
    & \vspace{3pt} No \vspace{3pt} \\
    \hline
    \vspace{3pt} $Q^\indI$ \vspace{3pt} & 
    \vspace{3pt}
    $\dps \left.\dif{\phifull^\indI}{\lambda}\right|_{\lambda=0}$
    \vspace{3pt}
    & \vspace{3pt} Yes \vspace{3pt}
    \\
    \hline
    \vspace{3pt} $P_\indI$ \vspace{3pt} &
    \vspace{3pt}
    $\dps\left.\dif{\pifull_\indI}{\lambda}\right|_{\lambda=0}$
    \vspace{3pt}   
    & \vspace{3pt} No \vspace{3pt}  \\
    \hline
    $\covP_\indI$ & $\covD_\lambda \pifull_\indI|_{\lambda=0}=P_\indI-\Gamma^\indK_{\indI\indJ}\piIR_\indK Q^J$ & Yes   
    \\
    \hline
    $\covP_{\indI,\kappa}$ & $\covP_I+\kappa R_{\indI\indJ\indK}{}^\indL \piIR_L Q^\indJ Q^\indK$ & Yes   
    \\
    \hline
\end{tabular}
\end{table}

\bibliographystyle{JHEP}
\bibliography{main}
\end{document}